\documentclass[a4paper,11pt]{article}
\pdfoutput=1 

\usepackage{jcappub} 
\usepackage{lineno} 

\usepackage[T1]{fontenc} 

\usepackage{booktabs} 
\usepackage{multirow} 
\usepackage{threeparttable}
\usepackage{rotating}
\usepackage{orcidlink}

\usepackage{comment}

\title{\boldmath Strong Lensing Tomography: Double and pseudo multi-source plane strong gravitational lensing to constrain dark energy}

\author[a,1]{Paras Sharma\orcidlink{0000-0002-6909-205X},\note{Corresponding author.}}
\author[a]{Simon Birrer\orcidlink{0000-0003-3195-5507},}
\author[a,b,c]{Narayan Khadka\orcidlink{0000-0001-5512-2716},}
\author[i]{Timo Anguita\orcidlink{0000-0003-0930-5815},}
\author[c]{Adam S. Bolton\orcidlink{0000-0002-9836-603X},}
\author[b, c]{Sydney Erickson\orcidlink{0000-0001-5717-2688},}
\author[g]{Philip Holloway\orcidlink{0009-0002-8896-6100},}
\author[d]{Tian Li\orcidlink{0009-0005-5008-0381},}
\author[b, c]{Phil Marshall\orcidlink{0000-0002-0113-5770},}
\author[h]{Dieu D. Nguyen\orcidlink{0000-0002-5678-1008},}
\author[j]{Graham P. Smith\orcidlink{0000-0003-4494-8277},}
\author[f]{Crescenzo Tortora\orcidlink{0000-0001-7958-6531},}
\author[e]{Bryce Wedig\orcidlink{0000-0002-0748-7312},}
\author[]{the Strong Lensing Science Collaboration, and}
\author[]{the LSST Dark Energy Science Collaboration}

\affiliation[a]{Department of Physics and Astronomy, Stony Brook University, Stony Brook, NY 11794, USA}
\affiliation[b]{Kavli Institute for Particle Astrophysics and Cosmology, Department of Physics, Stanford University}
\affiliation[c]{SLAC National Accelerator Laboratory, 2575 Sand Hill Rd., Menlo Park, CA 94025, USA}
\affiliation[d]{Institute of Cosmology and Gravitation, University of Portsmouth, Burnaby Rd, Portsmouth PO1 3FX, UK}
\affiliation[e]{Department of Physics and McDonnell Center for the Space Sciences, Washington University, St. Louis, MO 63130, USA}
\affiliation[f]{INAF -- Osservatorio Astronomico di Capodimonte, Salita Moiariello 16, I-80131, Napoli, Italy}
\affiliation[g]{Department of Physics, Oxford University, Keble Road, Oxford, OX1 3RH, UK}
\affiliation[h]{University of Michigan, 500 S State St, Ann Arbor, MI 48109, USA}
\affiliation[i]{Instituto de Astrofísica, Departamento de Física y Astronomia, Facultad de Ciencias Exactas, Universidad Andres Bello, Fernández Concha 700, Santiago, Chile}
\affiliation[j]{School of Physics and Astronomy, University of Birmingham, Birmingham, B15 2TT, United Kingdom}

\emailAdd{paras.sharma@stonybrook.edu}
\emailAdd{simon.birrer@stonybrook.edu}

\abstract{Tomographic measurements of gravitational lensing with different lens and source redshift distributions contain crucial information about the universe's relative expansion rate, and hence dark energy.
While this technique is well-established in weak lensing, its application to strong lensing has traditionally focused on Double Source Plane Lenses (DSPLs). However, DSPLs are exceedingly rare and fundamentally limited by the Mass-Sheet Degeneracy (MSD), a systematic uncertainty underexplored in previous literature.
To overcome these challenges, we introduce Pseudo Double-Source Plane Lenses (PDSPLs): pairs of independent single-source plane lenses with self-similar deflectors. This generalizes the DSPL formalism to the $\sim 10^5$ galaxy-galaxy lenses expected from upcoming surveys like LSST, Euclid, and Roman. Unlike true DSPLs, PDSPLs are free from the intermediate source mass problem by construction, eliminating the associated secondary MSD and the need for multi-plane ray tracing. We incorporate the deflector galaxy's MSD into a hierarchical population model and use it to forecast cosmological constraints, demonstrating that this degeneracy severely degrades the constraining power of small DSPL samples and thereby motivating our PDSPL statistical approach.
We forecast constraints on the dark energy equation of state under a Flat $w_0w_a$CDM cosmology. Assuming unbiased Gaussian photo-z uncertainties of $\sigma_z/(1+z)=0.03$, the mock LSST 10-year photometric sample is forecast to achieve $\sigma(w_0)\sim0.45$ while simultaneously constraining the population-mean MSD parameter and deflector power-law slope to $\sim 2\%$. Adding a prior $\Omega_{\rm m}\sim\mathcal{N}(0.3, 0.05)$, representing combination with external cosmological probes such as CMB, BAO, or SNe Ia, improves the forecast to $\sigma(w_0)\sim0.29$, comparable to current Stage III weak-lensing results.
Under the adopted unbiased redshift-error model, the larger LSST Y10 photometric sample provides greater statistical precision than the smaller simulated spectroscopic subsets (e.g., from 4MOST). This statistical advantage does not establish comparable systematic accuracy. Application to real data will nevertheless require spectroscopic calibration of the photo-z distributions and observational validation of how residual differences in the mass profiles of paired deflectors affect their Einstein-radius ratios.
}

\begin{document}
\maketitle
\flushbottom


\section{Introduction}
\label{sec:intro}
Understanding the fundamental nature of dark energy remains one of the most pressing challenges in modern cosmology. Recent observations—including evidence for potential dynamical dark energy \citep{DESI:BAO:2025, Zhang:2026, deCruz:2026}, along with persistent tensions with $\Lambda$CDM \citep{DiValentino:2021, Abdalla:2022}—have intensified interest in exploring dark energy physics beyond the standard cosmological constant \citep{Chevallier:2001, Linder:2003, Caldwell:2005, Singh:2016, Zhao:2017}. These developments motivate the search for robust, late-time geometric probes. Tomographic measurements of gravitational lensing—which track how the deflection of light evolves across sources at varying distances—provides a powerful, purely geometric probe of the Universe's expansion history and the dark energy equation of state. By analysing the relative deflection of light from sources at varying distances, we can constrain these cosmological parameters and break the degeneracies often inherent to single-plane measurements \citep{Jain:2003, Hu:2002}.

This tomographic approach has been prominently applied to weak lensing shape measurements with multiple redshift bins in the lens and source population \citep{DES+KiDS:2023}. In the strong lensing regime, the technique relies on the rare alignment of multiple sources behind a single deflector. Galaxy clusters, acting as dominant deflectors for multiple background sources at distinct redshifts, have proven to be powerful laboratories for this method \citep{Jullo:2010, Caminha:2022}. While the alignment of two distinct sources behind a single galaxy-scale deflector—forming a double-source plane lens (DSPL)—is exceptionally rare, these systems offer unique value. When identified, the ratio of their Einstein radii yields a robust cosmological observable independent of the Hubble constant ($H_0$) that has been successfully used to constrain the dark energy equation of state \citep{Gavazzi:2008, Collett:2014, Johnson:2025, Sahu:2025, Bowden:2025}.

Despite their conceptual appeal, DSPLs remain exceedingly rare, with only a small number of confirmed systems to date \citep[e.g.,][]{Gavazzi:2008, Tanaka:2016, Wong:2017, Tanaka:2020, Dux:2025, Li:2025}. Moreover, DSPLs present unique modeling challenges; standard analyses often neglect the "compound lensing" effect—where the intermediate source acts as a secondary deflector—an approximation that degrades when the sources are widely separated in redshift \citep{Johnson:2025}. These limitations motivate the search for alternative strategies that retain the geometric sensitivity of DSPLs while mitigating their observational sparsity and modelling complexity.

We are currently entering the era of next-generation wide-field surveys. Facilities such as the Vera C. Rubin Observatory, which will conduct the Legacy Survey of Space and Time \citep[LSST;][]{Rubin:Ivezic:2019}, Euclid \cite{Euclid:Laureijs:2011}, and the Nancy G. Roman Space Telescope \citep[formerly WFIRST; ][]{Roman:Spergel:2015} are expected to discover $\mathcal{O}(10^5)$ galaxy-scale strong lenses \citep{Collett:2015, Shajib:2024, Ferrami:2024, Wedig:2025}. This dramatic increase in sample size opens new avenues for extracting cosmological signals from the population statistics of strong lenses, rather than relying on rare individual alignments.

To capitalize on these upcoming catalogs, we propose \emph{pseudo} double-source plane lenses (PDSPLs): pairs of independent strong lenses, each with its own deflector and source, selected such that the deflectors are approximately self-similar. With these considerations, the systems can be treated as an approximate DSPL, enabling the extraction of a generalized Einstein radius ratio observable. This approach effectively transforms strong lensing tomography from a search for rare alignments into a statistics-driven science, capitalizing on the vast catalogs of single-source plane lenses anticipated from upcoming surveys \citep{Collett:2015, Shajib:2024}. Crucially, this method does not rely on expensive velocity dispersion measurements (although representative spectroscopy remains important for calibrating the photo-z distributions) and possesses systematic uncertainties distinct from those of weak lensing tomography.

The cosmological constraining power of the Einstein-radius ratios of DSPLs and PDSPLs is fundamentally limited by uncertainties in the deflector mass profiles. This is primarily due to the well-known mass-sheet degeneracy \citep[MSD;][]{Falco:1985}, which allows for transformations of the lensing mass distribution that preserve image positions but alter inferred mass scales. As demonstrated by \cite{Schneider:2014}, this degeneracy limits the precision with which one can extract cosmological information from DSPLs unless it is explicitly broken by external datasets (e.g., stellar kinematics) or statistically constrained \citep{Birrer:2020, Khadka:2024}.

In this paper, we derive how the observables in (P)DSPL transform under an MSD and what quantities remain invariant. We perform an updated cosmological forecast based on a hierarchical population model of the DSPL deflectors. The system-level mass-profile parameters are analytically marginalized, allowing their population means and intrinsic scatters to be inferred jointly with the dark energy equation-of-state parameters.

Further, to assess the cosmological utility of PDSPLs, we generate a realistic, population-level simulation of galaxy-galaxy lenses expected to be observed by LSST, pair near-identical deflectors, and apply the same hierarchical population model with its approximate marginalized likelihood. We further project the capabilities of upcoming surveys in leveraging either PDSPLs or DSPLs for precision cosmology.



This paper is structured as follows: In Section~\ref{sec:multi_source_plane} we provide the basic formalism for single- and double-source plane lensing. Section~\ref{sec:msd} details the formalism of the mass-sheet degeneracy in both single- and double-source plane lensing. Specifically, we derive how the observable ratio of two Einstein radii transforms under an MSD for a power-law deflector mass density. In Section~\ref{sec:pseudo_multi_plane_formalism}, we generalize the DSPL method to near-identical pairs of lenses with different sources and derive the criteria under which two single-source plane lenses can be combined into an approximate double-source plane lens. In Section~\ref{sec:forecast} we provide a forecast for the expected constraints on the dark energy equation of state from the expected set of DSPLs and PDSPLs that LSST can provide. 
We provide further discussions of the newly proposed method in Section~\ref{sec:discussion} and conclude in Section~\ref{sec:conclusion}.

\section{Multi-source plane lensing formalism}\label{sec:multi_source_plane}

\subsection{Single plane lensing}
In general, we denote $\boldsymbol{\beta}$ as the unlensed angular coordinate of the source, $\boldsymbol{\theta}$ as the observed angular coordinate of the lensed image on the sky, and $\boldsymbol{\alpha}$ as the reduced angular deflection vector mapping the two. The lens equation (Equation~\ref{eqn:lens_equation}) describes this geometric distortion:

\begin{equation}\label{eqn:lens_equation}
    \boldsymbol{\beta} = \boldsymbol{\theta} - \boldsymbol{\alpha}(\boldsymbol{\theta}).
\end{equation}

In case of a single deflector, the physical deflection angle $\hat{\boldsymbol{\alpha}}$ is related to the reduced deflection angle $\boldsymbol{\alpha}$ as

\begin{equation} \label{eqn:physical_reduced_deflection}
    \boldsymbol{\alpha}(\boldsymbol{\theta}) = \frac{D_{\rm ds}}{D_{\rm s}} \boldsymbol{\hat{\alpha}}(D_{\rm d}\boldsymbol{\theta}),
\end{equation}
where $D_{\rm s}$ is the angular diameter distance from the observer to the source, and $D_{\rm ds}$ is the angular diameter distance from the deflector to the source, respectively.



\subsection{Double source plane lensing}
For a single deflector at redshift $z_{\rm d}$ lensing two background sources at redshifts $z_{\rm s1}$ and $z_{\rm s2}$, measuring the resulting Einstein radii $\theta_{\rm E, 1}$ and $\theta_{\rm E, 2}$ yields a distance ratio constraint.

For two photons passing through the same point in the lens plane, but originating on different source planes, the ratio of scaled deflection angles is

\begin{equation} \label{eqn:deflection_ratio_dsp}
 \frac{\alpha_1}{\alpha_2} = \frac{D_{\rm ds1}D_{\rm s2}}{D_{\rm ds2}D_{\rm s1}} \equiv \beta,
\end{equation}

where $D_{\rm si}$ is the angular diameter distance from the observer to the i'th source, and $D_{\rm dsi}$ is the angular diameter distance to the i'th source when observed from the deflector.

For Einstein radii that are not formed at the same angular position, the interpretation of the comparison between two or more Einstein radii becomes model dependent. The primary cosmology-relevant observable in a DSPL is the ratio of Einstein radii
\begin{equation}
    \beta_{\rm E} \equiv \frac{\theta_{\rm E, 1}}{\theta_{\rm E, 2}}.
\end{equation}

For a singular isothermal sphere (SIS) deflector, the deflection angle is constant as a function of radius, i.e.
\begin{equation}\label{eqn:sis_beta}
 \alpha_{\rm SIS}(\theta) = \theta_{\rm E},  
\end{equation}
hence, the ratio of Einstein radii, $ \beta_{\rm E}$ is identical to the quantity $\beta$ in Equation~\ref{eqn:deflection_ratio_dsp}, $\beta_{\rm E, SIS} = \beta$.

The deflection angle of a power-law mass density with logarithmic three-dimensional density profile slope $\gamma_{\rm pl}$ is 
\begin{equation} \label{eqn:power_law_deflection}
 \alpha_{\rm PL}(\theta) = \theta_{\rm E}  \left(\frac{\theta}{\theta_{\rm E}}\right)^{\left(2 - \gamma_{\rm pl}\right)}.
\end{equation}

Given two Einstein radii measurements, we can calculate the ratio of deflection angles at any point, such as at the angle of the two Einstein radii, which is given by

\begin{equation} \label{eqn:beta_gamma_scaling}
    \beta = \frac{\alpha_{\rm{pl}, 1}(\theta_{\rm{E}, 1})}{\alpha_{\rm{pl}, 2}(\theta_{\rm{E}, 1})} = \frac{\alpha_{\rm{pl}, 1}(\theta_{\rm{E}, 2})}{\alpha_{\rm{pl}, 2}(\theta_{\rm{E}, 2})} = \left(\beta_{\rm E, pl} \right)^{\gamma_{\rm{pl}}-1}.
\end{equation}
We recover the expression of $\beta$ for a SIS profile (Eqn.~\ref{eqn:sis_beta}) with the isothermal power-law slope $\gamma_{\rm pl} = 2$.


\section{Mass-sheet degeneracy and multi-source-plane lensing}\label{sec:msd}

\subsection{The MST with a single lens and source plane}

The mass-sheet transform (MST) is a multiplicative transform of the lens equation preserving image positions (and thus any higher order differentials too) under a linear source displacement $\boldsymbol{\beta} \rightarrow \lambda\boldsymbol{\beta}$ and was introduced by \cite{Falco:1985, Gorenstein:1988} as such

\begin{equation} \label{eqn:mst_deflection}
    \lambda_{\rm MST} \boldsymbol{\beta} = \boldsymbol{\theta} - \lambda_{\rm MST} \boldsymbol{\alpha}(\boldsymbol{\theta}) - (1 - \lambda_{\rm MST}) \boldsymbol{\theta}.
\end{equation}

The term $(1 - \lambda_{\rm MST}) \boldsymbol{\theta}$ in Equation~\ref{eqn:mst_deflection} describes an infinite sheet of convergence, $\kappa$ (defined as the dimensionless surface mass density), hence the name Mass-Sheet Transform (MST). The MST is the mathematical transformation that gives rise to the Mass-Sheet Degeneracy (MSD).
Thus, the corresponding transform of the convergence profile is given by

\begin{equation}
  \kappa_{\rm MST}(\boldsymbol{\theta}) =  \lambda_{\rm MST} \kappa(\boldsymbol{\theta}) + (1 - \lambda_{\rm MST}).    
\end{equation}

The MST can be described as a global transform of the convergence and hence it can lead to physical solutions for a wide range of values of $\lambda_{\rm MST}$. A fact that makes the MST a prominent and relevant degeneracy for many applications, in particular the measurement of the Hubble constant with time-delay cosmography \citep{SchneiderSluse:2013, Birrer:2016, Sonnenfeld:2018, Kochanek:2020}.
Only observables related to the absolute source size, intrinsic magnification of the lensed source, the absolute lensing potential, or the relative time delay when imposing a known cosmology with absolute distances, are able to break this degeneracy.

\subsection{The MST with two source planes}

The strict mathematical form of the MST is broken when introducing two or more source planes. Let us perform an MST to a first source plane, $z_{\rm s1}$, with corresponding $\lambda_{\rm MST, 1}$ (Eqn. \ref{eqn:mst_deflection}).
The physical deflection angle at the deflector plane is then (Eqn. \ref{eqn:physical_reduced_deflection})

\begin{equation}
    \boldsymbol{\hat{\alpha}}_{\lambda_{1}}(\boldsymbol{\theta}) = \lambda_{1} \frac{D_{\rm s1}}{D_{\rm ds1}} \boldsymbol{\alpha}_1(\boldsymbol{\theta}) + (1 - \lambda_{1}) \frac{D_{\rm s1}}{D_{\rm ds1}} \boldsymbol{\theta},
\end{equation}
where we referred to $\boldsymbol{\alpha}_1$ as the reduced deflection field to $z_{\rm s1}$ prior to an MST.
When evaluating the reduced deflection field for the second source plane $z_{\rm s2}$, the reduced deflection field results in

\begin{equation}
    \boldsymbol{\alpha}_{2, \lambda_1}(\boldsymbol{\theta}) = \lambda_{1} \frac{D_{\rm s1} D_{\rm ds2}}{D_{\rm ds1} D_{\rm s2}} \boldsymbol{\alpha}_1(\boldsymbol{\theta}) + (1 - \lambda_{1}) \frac{D_{\rm s1} D_{\rm ds2}}{D_{\rm ds1} D_{\rm s2}} \boldsymbol{\theta}.
\end{equation}

We can now substitute the relative apparent deflection scaling $\beta$ (Eqn. \ref{eqn:deflection_ratio_dsp}) in the equation above, and convert the apparent deflection field $\boldsymbol{\alpha}_2 = \boldsymbol{\alpha}_1 /\beta$

\begin{equation}
    \boldsymbol{\alpha}_{2, \lambda_1}(\boldsymbol{\theta}) = \lambda_{1} \boldsymbol{\alpha}_2(\boldsymbol{\theta}) + (1 - \lambda_{1}) \frac{1}{\beta} \boldsymbol{\theta}.
\end{equation}

The equation above is not an invariant MST anymore. Re-writing the equation above to the closest form of the MST is

\begin{equation} \label{eqn:approx_mst_dsp}
    \boldsymbol{\alpha}_{2, \lambda_1}(\boldsymbol{\theta}) = \lambda_1 \boldsymbol{\alpha}_2(\boldsymbol{\theta}) + (1 - \lambda_{1}) \boldsymbol{\theta} + \left(\frac{1}{\beta} - 1 \right)\left(1 - \lambda_1\right)\boldsymbol{\theta}.
\end{equation}

In short, an MST for one source plane, $z_{\rm s1}$, with $\lambda_1$, leads to a transform at a different source plane, $z_{\rm s2}$ that is the combination of an MST with the same $\lambda$ with an additional convergence term $(\beta^{-1} -1)(1-\lambda)$. Consequently, the second source experiences a modified deflection field that physically alters its corresponding Einstein radius—a measurable shift that we quantify in the following section.


\subsection{The MST in DSP with a power-law density profile}

There exists no general relation between a sheet of mass/convergence and the change of the Einstein radius, as it depends on the radial projected mass density profile of the deflector.
For a power-law density profile (Eqn.~\ref{eqn:power_law_deflection}), a mass sheet with convergence $\kappa_{\rm mst}$ changes the Einstein radius by
\footnote{The new Einstein radius is defined demanding $\alpha_{{\rm pl}+\kappa}(\theta_{\rm E, pl+\kappa}) = \theta_{\rm E, pl+\kappa}$.}

\begin{equation}\label{eqn:theta_E_pl+mass_sheet}
    \theta_{{\rm E, pl}+\kappa} = \theta_{\rm E, pl} \left(1 -\kappa_{\rm } \right)^{\frac{1}{1 -\gamma_{\rm pl}}}.
\end{equation}

For a power-law density profile that is invariantly transformed under an MST at the source plane $z_{\rm s1}$ with $\lambda_1$ (meaning, among other lensing observables, that $\theta_{\rm E1}$ remains unchanged), this transformation will change the Einstein radius at the second source plane $z_{\rm s2}$, $\theta_{\rm E2}$, as (Eqns.~\ref{eqn:approx_mst_dsp}+\ref{eqn:theta_E_pl+mass_sheet}):

\begin{equation}\label{eqn:mst_pl_theta_E}
    \left( \frac{\theta_{\rm E1}}{\theta_{\rm E2}} \right)_{\lambda} = \left(\frac{\theta_{\rm E1}}{\theta_{\rm E2}} \right)_{\lambda = 1} \left(1 - (1 - \lambda) \left(\frac{1}{\beta} -1\right) \right)^{\frac{1}{\gamma_{\rm pl} - 1}}.
\end{equation}

Putting it all together, combining the result of the MST of the equation above (Eqn.~\ref{eqn:mst_pl_theta_E}) with the scaling of the power-law slope excluding the MST (Eqn.~\ref{eqn:beta_gamma_scaling}), the ratio of two Einstein radii $\theta_{\rm E1}$ and $\theta_{\rm E2}$ for two source redshifts $z_{\rm s1}$ and $z_{\rm s2}$ scales with power-law slope $\gamma_{\rm pl}$ and mass-sheet term $\lambda$ as

\begin{equation}\label{eqn:mst_epl}
    \beta_{\rm E,pl} = \left( \frac{\theta_{\rm E1}}{\theta_{\rm E2}} \right)_{\lambda, \gamma_{\rm pl}} = \beta^{\frac{1}{\gamma_{\rm pl} - 1}}\left(1 - (1 - \lambda) \left(\frac{1}{\beta} -1\right) \right)^{\frac{1}{\gamma_{\rm pl}-1}}
    = \left(\beta - (1-\lambda)(1-\beta) \right)^{\frac{1}{\gamma_{\rm pl} - 1}}.
\end{equation}

The observable, the ratio of two Einstein radii, is hence a function of not only the cosmology (encompassed in $\beta$), but also of the MST parameter $\lambda$ and the power-law density slope $\gamma_{\rm pl}$. 
Solving Equation~\ref{eqn:mst_epl} for $\beta$ results in

\begin{equation}\label{eqn:beta_mst_pl}
    \beta = \frac{1}{2 - \lambda} \left[ \left( \frac{\theta_{\rm E1}}{\theta_{\rm E2}} \right)^{\gamma_{\rm pl} -1} + (1 - \lambda) \right].
\end{equation}

It is important to note that the above formalism represents an \emph{idealized} Double Source Plane Lens (DSPL) scenario, where the intermediate source (at $z_{\rm s1}$) is treated purely as a luminous background object and its own mass is neglected. In true, physical DSPL systems, the intermediate source acts as a secondary deflector, introducing a complex compound lensing effect and a secondary, independent mass-sheet degeneracy associated with the first source plane \cite{Johnson:2025}. By neglecting this intermediate mass, we isolate the fundamental geometric distance scaling and any associated MST-like degrees of freedom. Crucially, this idealized, mathematically simpler configuration is exactly the regime that our Pseudo-DSPL (PDSPL) framework emulates (Section~\ref{sec:pseudo_multi_plane_formalism}).

\section{Generalized pseudo-multi-source plane lensing}\label{sec:pseudo_multi_plane_formalism}
A \emph{Pseudo Multi-Source Plane Lens} (PMSPL) may be constructed by pairing multiple single-source plane lens systems, under the assumption that the deflectors exhibit sufficiently similar mass distributions. The primary motivation for this approach stems from the observational reality: while genuine double- or multi-source plane lenses (DSPLs or MSPLs) are intrinsically rare and challenging to identify in current datasets, the number of confirmed single-source plane lenses is already substantial and poised to increase dramatically. Next-generation wide-field surveys are expected to uncover $\sim\mathcal{O}(10^5)$ strong lensing systems \citep{Shajib:2024}, greatly enhancing the likelihood of identifying deflector pairs with near-identical properties.

A key consideration in this framework is quantifying the required degree of similarity between deflectors for the PMSPL approximation to remain valid. Equally important is the practical challenge of identifying such near-identical systems within large, structurally diverse datasets. These two aspects—theoretical tolerance and observational identifiability—represent distinct but related challenges. In both cases, one can define a characteristic noise floor that captures the uncertainty introduced by residual differences between deflectors.

Despite these challenges, the observational advantage of this approach is clear. The ability to form PMSPLs from the vastly more numerous population of single-source plane lenses opens the door to tomographic cosmological analyses at a scale that would be infeasible relying solely on genuine DSPL or MSPL systems, which are expected to remain sparse even in future surveys.

\subsection{Pseudo Double Source Plane Lenses}
\begin{figure} 
  \includegraphics[width=0.7\textwidth]{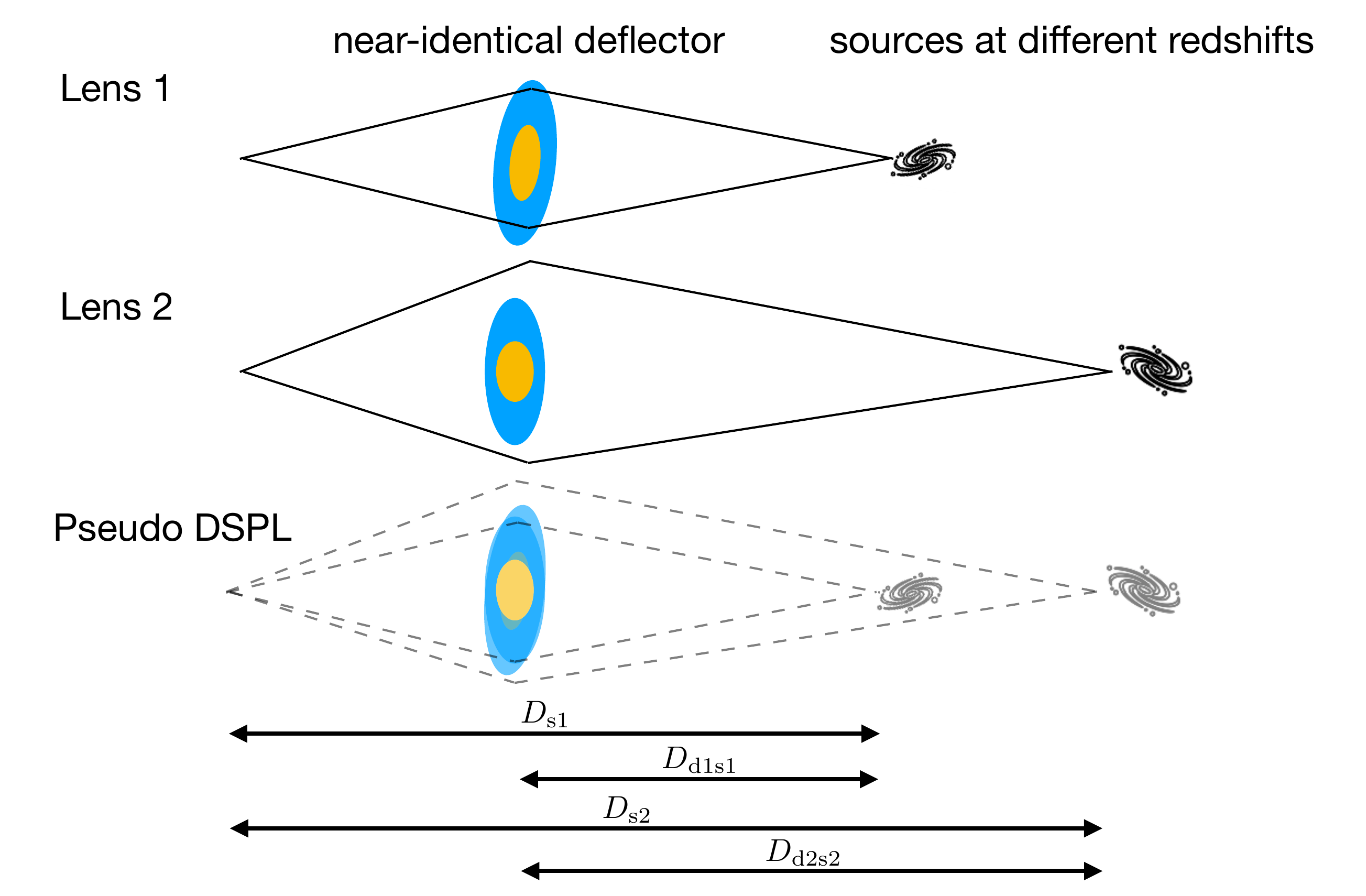} 
  \centering
  \caption{Illustration of two separate lenses with near-identical deflectors and sources at different redshifts. Near-identical pairs of deflectors can be used as pseudo double source plane lenses. \label{fig:pseudo_dsp_illustration} 
}
\end{figure}

A pseudo double-source plane lens (PDSPL) represents the simplest realization of the broader PMSPL framework, constructed from a pair of single-source plane lenses whose deflectors exhibit near-identical mass profiles. An illustrative example of such a configuration is shown in Figure~\ref{fig:pseudo_dsp_illustration}. Importantly, the mathematical formalism developed for conventional double-source plane lenses (DSPLs) remains directly applicable to PDSPLs, provided the pair of lenses possess self-similar deflectors.
In what follows, we focus exclusively on the PDSPL case, which captures the essential features of the approach while offering a tractable starting point for both analytical development and observational implementation.

\begin{figure}
    \centering
    \includegraphics[width=.8\linewidth]{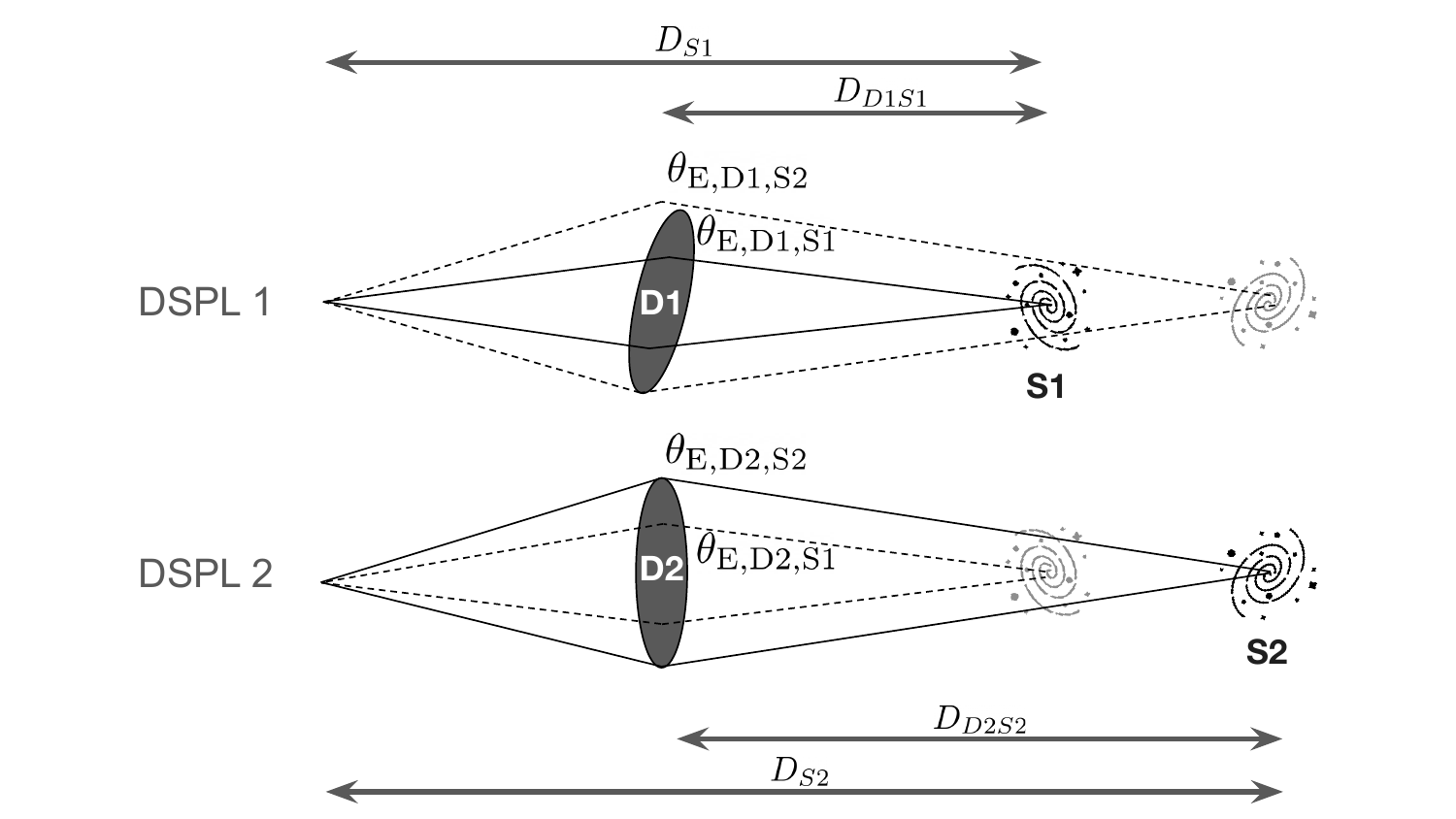}
    \caption{Illustration of the two possible DSPL configurations obtained by placing one source behind the deflector of the other lensing system. In the ``DSPL 1'' configuration, a virtual source S2 is positioned behind deflector D1 in addition to the existing source S1. Conversely, in the ``DSPL 2'' configuration, a virtual source S1 is placed behind deflector D2 along with the existing source S2. The corresponding Einstein radii for each configuration are indicated in the figure.}
    \label{fig:dspl_virt_illustration}
\end{figure}

\subsection{Noise in interpreting PDSPL as DSPL}\label{sec:noise_interpreting_dspl}
In the context of this paper, self-similarity of deflectors is closely tied to whether a pair of single-source plane lenses can mimic the Einstein-radius ratio of a DSPL. To answer this question, we refer to Figure~\ref{fig:pseudo_dsp_illustration}. Consider two single-source-plane lens systems: the first has a deflector at $z_{D1}$ and a source at $z_{S1}$, while the second has a deflector at $z_{D2}$ and a source at $z_{S2}$. The observable quantity relevant for the PDSPL analysis is
\begin{equation}
    \beta_{\rm E, pseudo} \equiv \frac{\theta_{\rm E, D1, S1}}{\theta_{\rm E, D2, S2}}
\end{equation}
where $\theta_{\rm E, Di, Sj}$ is the Einstein radius for the lens configuration with deflector $\rm Di$ and source $\rm Sj$. Note that in practice, $\rm i=j$ corresponds to the observed single-source plane lenses where we directly measure $\theta_{\rm E, D1, S1}$ and $\theta_{\rm E, D2, S2}$ from observed images.

Unlike in a DSPL, the mass normalization does not cancel identically in a PDSPL because the two Einstein radii are produced by physically distinct deflectors. To make this dependence explicit, consider a power-law density profile for the $j$th deflector,
\begin{equation}
    \rho_j(r)
    =
    \rho_{0,j}
    \left(\frac{r}{r_{0,j}}\right)^{-\gamma_{\rm pl}}.
\end{equation}
The corresponding Einstein radius scales as \cite{Agnello:2013}
\begin{equation}
\label{eqn:einstein_radius_mass_normalization}
    \theta_{{\rm E},j}^{\gamma_{\rm pl}-1}
    \propto
    \left(\rho_{0,j}r_{0,j}^{\gamma_{\rm pl}}\right)
    D_{{\rm d},j}^{\,2-\gamma_{\rm pl}}
    \frac{D_{{\rm d}_j{\rm s}_j}}{D_{{\rm s},j}}
    \equiv
    A_j
    D_{{\rm d},j}^{\,2-\gamma_{\rm pl}}
    \frac{D_{{\rm d}_j{\rm s}_j}}{D_{{\rm s},j}},
\end{equation}
where $A_j\equiv\rho_{0,j}r_{0,j}^{\gamma_{\rm pl}}$ denotes the physical mass normalization of the $j$th deflector. For illustration, if the two deflectors have a common density slope, their PDSPL Einstein-radius ratio is
\begin{equation}
\label{eqn:pdspl_mass_normalization}
    \beta_{\rm E,pseudo}
    =
    \left[
        \frac{A_1}{A_2}
        \left(
            \frac{D_{{\rm d},1}}
                 {D_{{\rm d},2}}
        \right)^{2-\gamma_{\rm pl}}
        \frac{
            D_{{\rm d}_1{\rm s}_1}/D_{{\rm s},1}
        }{
            D_{{\rm d}_2{\rm s}_2}/D_{{\rm s},2}
        }
    \right]^{1/(\gamma_{\rm pl}-1)}.
\end{equation}
Thus, $\beta_{\rm E,pseudo}$ contains the relative physical mass normalization and deflector-distance scaling of the two lenses in addition to the desired dependence on the geometric distance ratio $\beta$. These extra factors cancel for a true DSPL, where both sources are lensed by the same deflector, but not for a PDSPL. Differences in density slope and other mass-profile properties may introduce further corrections. The relevant normalization quantity is therefore not the absolute mass normalization of either deflector, but their ratio $A_1/A_2$. A broad range of absolute galaxy masses is not problematic by itself, provided that the pairing procedure selects deflectors whose relative normalization is sufficiently close to unity. The remaining departure of $A_1/A_2$ from unity contributes to the PDSPL residual calibrated below.

We do not assume that these additional contributions vanish. Instead, we quantify their combined effect by comparing $\beta_{\rm E,pseudo}$ with the Einstein-radius ratio that would be obtained if the two sources were placed behind the same deflector. We construct this reference using the two virtual DSPL configurations shown in Figure~\ref{fig:dspl_virt_illustration}.

For the first virtual configuration, we place source S2 behind deflector D1 such that $z_{D1}<z_{S1}<z_{S2}$. The resulting Einstein-radius ratio for the DSPL configuration with deflector D1 is
\begin{equation}
    {\beta}_{\rm E, DSPL, D1} \equiv \frac{\theta_{\rm E, D1, S1}}{\theta_{\rm E, D1, S2}}.
\end{equation}
For the second virtual configuration, we place source S1 behind deflector D2 such that $z_{D2}<z_{S1}<z_{S2}$. The corresponding Einstein-radius ratio is
\begin{equation}
    {\beta}_{\rm E, DSPL, D2} \equiv \frac{\theta_{\rm E, D2, S1}}{\theta_{\rm E, D2, S2}}.
\end{equation}

We combine the two virtual configurations by defining
\begin{equation}
    {\beta}_{\rm E,DSPL}
    \equiv
    \frac{1}{2}
    \left(
        {\beta}_{\rm E,DSPL,D1}
        +
        {\beta}_{\rm E,DSPL,D2}
    \right).
\end{equation}
The fractional residual incurred when interpreting a PDSPL as an approximate DSPL is then
\begin{equation}\label{eqn:beta_E_relative_scatter}
    \frac{\Delta\beta_{\rm E}}{\beta_{\rm E}} = \frac{\beta_{\rm E, pseudo} - {\beta}_{\rm E, DSPL}}{\beta_{\rm E, DSPL}}.
\end{equation}
One of the main goals of this paper is to determine how small this residual becomes for approximately matched deflectors. This quantity is distinct from the measurement uncertainty on $\beta_{\rm E,pseudo}$ propagated from the measured Einstein radii. For a population of paired lens systems, we characterize the fractional residual by its standard deviation,
\begin{equation}\label{eqn:beta_E_relative_scatter_population}
    \left.\frac{\Delta\beta_{\rm E}}{\beta_{\rm E}} \right|_{\rm pop.} = \text{std}\left(\frac{\beta_{\rm E, pseudo} - {\beta}_{\rm E, DSPL}}{\beta_{\rm E, DSPL}}\right)
\end{equation}

Ideally, identifying near-identical deflectors in observed systems would involve minimizing $\Delta\beta_{\rm E}/\beta_{\rm E}$; however, this quantity is not directly observable. Consequently, in the following subsection, we define a deflector pair dissimilarity parameter as an observational proxy that correlates directly with the relative scatter in $\beta_{\rm E}$.

\subsection{Deflector Self-Similarity}
In order for a pair of single-source plane lenses to be considered a PDSPL (or an approximate DSPL) we need to determine the criterion for two deflectors to be near-identical (or self-similar). Assuming deflectors to be elliptical galaxies, a non-exhaustive list of deflector properties to describe them can be: their redshift ($z_D$), velocity dispersion (${\sigma_{v, D}}$), effective radius (${R_{\rm e}}$), observed magnitudes or fluxes (${m_{X,D}}$ or ${f_{X,D}}$), ellipticity (${\varepsilon}_{D}$), or their average surface brightness ($\mu_e$) within ${R_{\rm e}}$.

The relative differences (${\Delta z_D / z_D}$, ${\Delta \sigma_{v, D}/\sigma_{v, D}}$, $\cdots$) in these parameters between the two deflectors set the criteria of self-similarity. Here we define these quantities as
\begin{equation}\label{eqn:rel_scatter_deflector_params}
    \frac{\Delta z_D}{z_D} \equiv \frac{|z_{D1} - z_{D2}|}{\text{mean}(z_{D1}, z_{D2})} \ \ ; \ \ \frac{\Delta \sigma_{v, D}}{\sigma_{v, D}} \equiv \frac{|\sigma_{v, D1} - \sigma_{v, D2}|}{\text{mean}(\sigma_{v, D1}, \sigma_{v, D2})} \ \ ; \ \ \cdots
\end{equation}
For a population of deflector pairs indexed with i, the relative scatter can be defined as
\begin{equation}\label{eqn:rel_scatter_deflector_params_population}
    \left.\frac{\Delta z_D}{z_D}\right|_{\rm pop.} \equiv \text{std}\left(\frac{\Delta z_{D\rm i}}{z_{D\rm i}}\right) \ \ ; \ \ \left.\frac{\Delta \sigma_{v, D}}{\sigma_{v, D}}\right|_{\rm pop.} \equiv \text{std}\left(\frac{\Delta \sigma_{v, D\rm i}}{\sigma_{v, D\rm i}}\right)\ \ ; \ \ \cdots
\end{equation}
where $\Delta z_{D\rm i}$ represents the redshift difference and $z_{D\rm i}$ represents the mean redshift for the two deflectors in i'th pair. The same convention is followed for $\sigma_{v, D}$ and other deflector properties as well. 

To quantify how similar or dissimilar deflectors can get, we define the deflector pair dissimilarity parameter for the i'th deflector pair as
\begin{equation}\label{eqn:dissimilarity_parameter}
    \mathcal{D}_{\rm i}(z_{D}, R_e, \cdots) = \sqrt{\frac{1}{N_{\rm pp}}\left(\left|\frac{\Delta z_{D\rm i}}{z_{D\rm i}}\right|^2 + \left|\frac{\Delta R_{e\rm i}}{R_{e\rm i}}\right|^2 + \cdots\right)}.
\end{equation}
which depends on the parameters used for pairing the deflectors and $N_{\rm pp}$ specifies the total number of these ``pairing parameters''. A smaller value of $\mathcal{D}_{\rm i}(z_{D}, R_e, \cdots)$ indicates near-identical deflectors. We note that while a fractional difference (e.g., $\Delta X/X$) is physically meaningful for linear quantities, computing it for logarithmic quantities like magnitudes ($m_i$) and colors is physically unorthodox due to arbitrary zero-points. Thus we convert magnitudes to fluxes and colors to flux ratios and use them as pairing parameters. While this unweighted Euclidean metric provides a functional baseline for identifying pairs in our simulated forecasts, the observational parameters do not impact the lensing observables equally. A weighted distance metric is a natural and necessary extension for real survey data. We discuss this optimization further in Section \ref{sec:discussion:defl_pairing}.

\subsection{Photometric vs Spectroscopic Observables for Pairing Deflector Galaxies}
\label{subsec:photo_vs_spectro_params_selection}
Upcoming wide-field imaging surveys, including LSST, Euclid, and Roman, will provide a wealth of photometric data for an unprecedented number of deflector galaxies, yielding an estimated sample of $\sim$120,000 galaxy-scale strong lenses (predominantly massive elliptical deflectors)~\citep{Shajib:2024, Collett:2015, Ferrami:2024, Wedig:2025} from LSST alone. For each system, these surveys will deliver multi-band photometric magnitudes, morphological parameters derived from the galaxy's light profile (e.g., effective radius), and photo-z estimates with a target precision of $\sigma_z/(1+z) < 0.05$~\cite{LSSTScienceBook:2009}.

Complementary spectroscopic follow-up campaigns with facilities like the 4-metre Multi-Object Spectroscopic Telescope (4MOST) are projected to secure precise spectroscopic redshifts for a subset of $\sim$10,000 of these galaxy-galaxy lenses \cite{Collett_4MOST_SLSLS:2023}. Furthermore, these observations will yield central stellar velocity dispersion ($\sigma_v$) measurements---a key dynamical probe---for approximately 5,000 of these deflector galaxies.

The vast majority of deflector galaxies will have well-measured photometric properties but will lack direct spectroscopic redshift and velocity dispersion measurements. This significant disparity in sample sizes between the photometric and spectroscopic datasets motivates us to investigate whether photometric observables alone can identify sufficiently similar deflectors. Massive elliptical galaxies obey empirical scaling relations connecting their luminosity, size, surface brightness, and stellar velocity dispersion. These include the Faber--Jackson relation \citep{Faber:1976}, the Kormendy relation \citep{Kormendy:1977}, and the Fundamental Plane \citep{Djorgovski:1987,Dressler:1987}. These relations motivate the use of photometric and structural observables as statistical proxies for deflector similarity. However, their intrinsic scatter means that photometrically similar galaxies need not have identical total mass
profiles. The effectiveness of such matching must therefore be quantified rather than assumed.

We therefore compare three observationally motivated pairing strategies: one using photo-z estimates, effective radii, fluxes, and flux ratios; one replacing the deflector photo-z estimates with spectroscopic redshifts; and one additionally including measured stellar velocity dispersions. In the following sections, we construct simulated lens samples representative of these observing scenarios and directly quantify the residual scatter in the PDSPL Einstein-radius ratio produced by each strategy.

\subsection{Simulated galaxy-galaxy lenses from \texttt{SLSim}}\label{sec:slsim_lenses}
To investigate how the scatter in the properties of paired deflectors propagates to the observed scatter in $\beta_{\rm E, pseudo}$, we simulate single-source-plane (galaxy--galaxy) lenses using the \texttt{SLSim} (Strong-Lensing Simulation)\footnote{\href{https://github.com/LSST-strong-lensing/slsim}{https://github.com/LSST-strong-lensing/slsim}} pipeline \cite{Khadka:2026}. The simulations are performed assuming a \texttt{Flat-wCDM} cosmology with $H_0 = 70~\mathrm{km\,s^{-1}\,Mpc^{-1}}$, $\Omega_{\mathrm{m}} = 0.3$, and $w = -1$. 

The deflector population is modelled as elliptical (red) galaxies with a power-law mass density profile having $\langle\gamma_{\rm pl}\rangle = 2.078$ with an intrinsic scatter of $\sigma(\gamma_{\rm pl}) = 0.16$ \cite{Auger:2010}, spanning a redshift range of $z_D \in [0.01, 2.5]$. The background sources are modelled as spiral (blue) galaxies with redshifts $z_S \in [0.1, 5.0]$. In \texttt{SLSim} elliptical galaxies are modelled using a double Schechter function \citep{Schechter:1976} using the \texttt{skypy} \cite{skypy:2021} pipeline to encompass both massive and dwarf populations. The velocity dispersions for these galaxies are drawn by abundance matching with the observed SDSS velocity dispersion distribution \citep{bernadietal2010}. For spiral galaxies, a single Schechter function is used with parameters including characteristic mass, normalization, and power law index specific to spiral galaxies.

For each galaxy--galaxy lens system, the Einstein radius was calculated using the \texttt{SLSim} function \texttt{theta\_E\_from\_vel\_disp\_epl}\footnote{\href{https://slsim.readthedocs.io/en/stable/_modules/slsim/Deflectors/MassLightConnection/velocity_dispersion.html}{\texttt{SLSim} velocity dispersion module}} with \texttt{sis\_convention=True}. Under this convention, the assigned velocity dispersion is interpreted as an SIS-equivalent velocity dispersion and
\begin{equation}
    \theta_{\rm E}
    =
    4\pi
    \left(\frac{\sigma_v}{c}\right)^2
    \frac{D_{\rm ds}}{D_{\rm s}}.
\end{equation}
The same convention is used for every observed and virtual lens--source configuration entering the calibration (Section~\ref{sec:noise_interpreting_dspl}). The Einstein radius therefore does not depend on the individual power-law density slope of the deflector. The Einstein radii of the two PDSPL members are calculated independently from their respective velocity dispersions and lens--source geometries, so differences in their SIS-equivalent mass normalizations remain present. This contribution is included in the calibrated deflector-mismatch scatter $\sigma_{\beta_{\rm E},\mathcal D}$ (Equation~\ref{eqn:model_beta_E_dissimilarity}). The population mean and intrinsic scatter of $\gamma_{\rm pl}$ are introduced separately through the hierarchical population model (see Equation~\ref{eqn:sigma_beta_E_pop_expanded} in Section~\ref{sec:likelihood_fnc}).

For the purposes of estimating the scatter in $\beta_{\rm E}$ solely from the deflector dissimilarity, we neglect line-of-sight (LoS) perturbations from intervening halos, as these primarily contribute an additional stochastic scatter of $\sim 1\%$ to $\beta_{\rm E}$ and can be treated separately \cite{Johnson:2025} while performing the cosmological forecast (see Section~\ref{sec:forecast}).

To ensure a realistic sample of discoverable gravitational lenses, we impose a set of detectability criteria on the simulated population following the methodology of Collett 2015 \cite[Eq. 6-9]{Collett:2015}. We simulate a survey area of 20,000 deg$^2$, corresponding to the full LSST footprint. A lens system is considered detectable if it satisfies three primary conditions: multiple imaging, resolvability, and sensitivity. First, the system must produce multiple source images, requiring the unlensed source position ($r_{\rm pos} = \sqrt{x_s^2 + y_s^2}$) to be within the Einstein radius ($\theta_E > r_{\rm pos}$). Second, the images must be resolved from the deflector and each other. We require that the Einstein radius exceeds the effective seeing limit (FWHM), $s$, such that $\theta_E^2 > r_{\rm s}^2 + (s/2)^2$, and that the tangential extension of the arcs is resolvable ($\mu_{\rm tot} r_{\rm s} > s$, where $\mu_{\rm tot}$ is the total magnification). To ensure identifiable arc morphology, we further require a total magnification of $\mu_{\rm tot} > 3$. 
We adopt an optimistic seeing of $s_{\rm opt} = 0.5''$ for LSST, assuming that strong lenses will be discovered using stacks of the best-seeing epochs (``optimal stacking''; \cite{Collett:2015}) rather than the median survey seeing of $s_{\rm med} \approx 0.7''$ \cite{LSSTScienceBook:2009}; this value enters exclusively the geometric resolvability cuts described above. Finally, we impose a sensitivity cut requiring the total signal-to-noise ratio of the lensed arcs to be ${\rm SNR} > 20$ in $g, r, i$ bands following the prescription of \cite{Holloway:2023} as implemented in \texttt{SLSim}, which uses the default \texttt{lenstronomy}\footnote{\href{https://github.com/lenstronomy/lenstronomy}{https://github.com/lenstronomy/lenstronomy}}\cite{Birrer:2018, Birrer:2021} LSST observational configuration for per-band seeing and sky background. We further require the $i$-band contrast ratio $> 2$ (ratio of mean surface brightness of lensed source to the surface brightness of the deflector at image position) for at least 2 lensed source images. These cuts ensure that the lensed arcs are both detectable and visible against deflector light. Applying these criteria yields a parent sample (\emph{LSST Y10}) of $\sim 116,000$ detectable galaxy-galaxy lenses across the LSST footprint, a figure consistent with the estimate of \cite{Collett:2015}. We note that utilizing the median seeing would instead yield $\sim 55,000$ lenses, which is slightly smaller than the conservative estimate of \cite{Collett:2015}, attributable to our additional contrast ratio cut. 

We further derive the \emph{LSST Y1} sample from the full \emph{LSST Y10} parent sample by modeling the accumulation of detectable lenses as a function of survey duration, specifically assuming a $\sqrt{t}$ dependence~\cite{Shajib:2024}. This 
scaling approximates the growth in survey depth with time: as the survey accumulates exposures, the effective SNR of stacked images grows as $\sqrt{t}$, progressively bringing fainter arcs above the detection threshold. The Y1 sample is therefore constructed by selecting the top $\sim 37,000$ lenses ($116,000 / \sqrt{10}$) with the highest SNR from the Y10 parent sample, which naturally corresponds to the brightest, most easily detectable systems that would be found first.

We define the spectroscopic follow-up sample anticipated for the 4MOST Strong Lensing Spectroscopic Legacy Survey (4SLSLS, \cite{Collett_4MOST_SLSLS:2023}) by selecting targets from the simulated population (over 20,000 deg$^2$) that satisfy both spectroscopic feasibility and high-resolution imaging detectability. First, to ensure spectroscopic success, we impose hard selection cuts: a lensed source magnitude limit of $m_{r} < 24.0$ and redshift $z_S < 1.5$ to ensure that key emission lines (e.g., [OII]) from the source galaxy fall within the 4MOST wavelength coverage (370–950 nm) \cite{Collett_4MOST_SLSLS:2023, Li:2024}. Additionally, we require the deflector galaxy to be brighter than $m_{r} < 20.0$, ensuring sufficient continuum signal-to-noise for potential dynamical measurements. 
We then assess the detectability of these candidates assuming high-resolution imaging (effectively mimicking the joint discovery space of LSST, Euclid, and Roman) by applying the \cite{Collett:2015} detectability criteria with an optimistic seeing of $0.2''$, a signal-to-noise ratio threshold of ${\rm SNR} > 20$, and no contrast ratio cut.
Finally, to match the survey's fiber allocation while maximizing scientific yield, we rank the detectable systems by their deflector $r$-band magnitude. We define the \emph{4MOST ($z^{\rm spec}$)} sample as the top 10,000 systems from this ranked list. For the brightest subset, the high deflector flux allows for the measurement of stellar velocity dispersions ($\sigma_{v, D}$) in addition to redshifts; thus, we define the \emph{4MOST ($z^{\rm spec}$ + $\sigma_{v, D}$)} sample as the top 5,000 systems from the same ranked list.

\begin{figure}
    \centering
    \includegraphics[width=\linewidth]{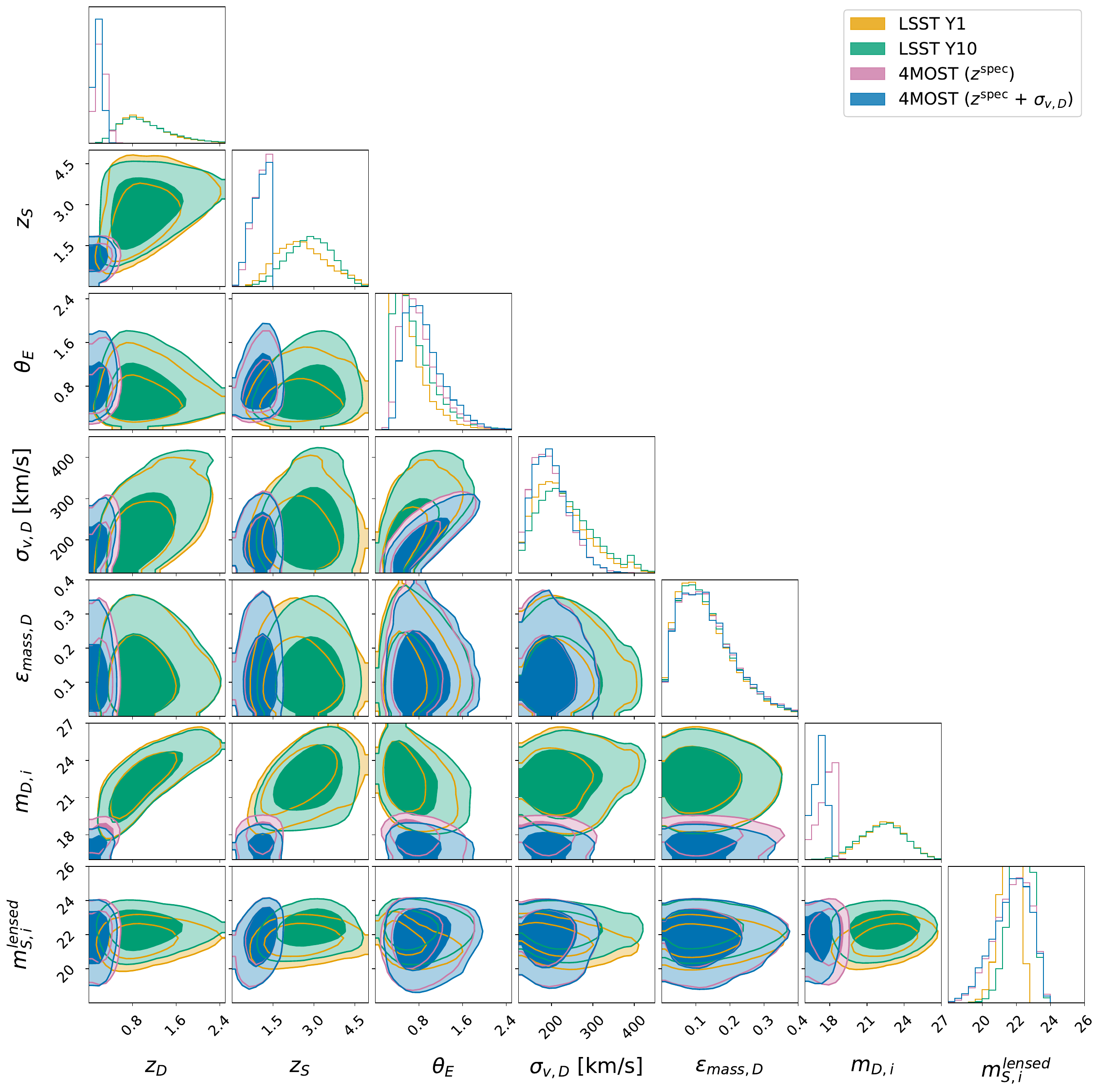}
    \caption{Population of galaxy-galaxy lenses simulated using \texttt{SLSim} with elliptical (red) galaxies as deflectors and spiral (blue) galaxies as sources within a sky area of 20000 deg$^2$. $\sigma_{v, D}$, $z_D$, $\varepsilon_{mass, D}$, and $m_{D, i}$ represent the velocity dispersion, redshift, mass ellipticity, and i-band magnitude of the deflector galaxies, respectively, while $z_S$ and $m_{S, i}^{lensed}$ denote the redshift and i-band magnitude of the lensed sources. $\theta_E$ represents the Einstein radius of the lens system.}
    \label{fig:slsim_corner_GGL}
\end{figure}

All generated galaxy galaxy lens samples are summarized in Table~\ref{tab:slsim_ggl_samples} along with the histograms of their key properties shown in Figure~\ref{fig:slsim_corner_GGL}.

\begin{table}[ht]
    \centering
    \caption{Description of the galaxy-galaxy lens samples generated with \texttt{SLSim}, drawn over an area of 20000 deg$^2$ by placing appropriate observability cuts on the lens systems. Pairing parameters for each sample are also provided below.}\vspace{1em}
    \label{tab:slsim_ggl_samples}
    \renewcommand{\arraystretch}{1.4} 
    \begin{tabular}{l c p{8.6cm}}
    \hline
    \textbf{Sample} & \textbf{\# Lenses} & \textbf{Description} \\
    \hline
    
    \multirow{2}{*}{LSST Y10} & 
    \multirow{2}{*}{$\sim$116000} & 
    \textbf{Selection:} Collett 2015 \cite{Collett:2015} cuts with $s_{\rm opt} = 0.5''$: \newline
    $\theta_E > r_{\rm pos} = \sqrt{x_s^2 + y_s^2}$, \quad $\theta_E > \sqrt{r_{\rm s}^2 + (s_{\rm opt}/2)^2}$, \quad $\mu_{\rm tot} r_{\rm s} > s_{\rm opt}$, \quad
    $\mu_{\rm tot} > 3$, \quad ${\rm SNR_{g, r, i}} > 20$, \quad $i$-band contrast ratio $> 2$ for at least 2 images \\
     & & \textbf{Pairing:} $z_{\mathrm{lens}}^{\rm photo}$, $R_e$ [arcsec], $f_{D,i}$, $f_{D,g}/f_{D,r}$, $f_{D,r}/f_{D,i}$ \\
    \hline

    \multirow{2}{*}{LSST Y1} & 
    \multirow{2}{*}{$\sim$37000} & 
    \textbf{Selection:} Best SNR subset ($N_{Y1} \approx N_{Y10}/\sqrt{10}$) \\
     & & \textbf{Pairing:} $z_{\mathrm{lens}}^{\rm photo}$, $R_e$ [arcsec], $f_{D,i}$, $f_{D,g}/f_{D,r}$, $f_{D,r}/f_{D,i}$ \\
    \hline

    \multirow{3}{*}{{4MOST ($z^{\rm spec}$)}} & 
    \multirow{3}{*}{10,000} & 
    \textbf{Selection:} High-Res ($s=0.2''$) detectable candidates (from LSST, Euclid and Roman) with: \newline
    $m_{r, \rm src} < 24.0$, \quad $m_{r, \rm def} < 20.0$, \quad $z_S < 1.5$. \newline
    Selected as the \textbf{Top 10,000} ranked by deflector brightness ($m_{r, \rm def}$). \\
     & & \textbf{Pairing:} $z_{\mathrm{lens}}^{\rm spec}$, $R_e$ [arcsec], $f_{D,i}$, $f_{D,g}/f_{D,r}$, $f_{D,r}/f_{D,i}$ \\
    \hline

    \multirow{3}{*}{{4MOST ($z^{\rm spec} + \sigma_{v, D}$)}} & 
    \multirow{3}{*}{5,000} & 
    \textbf{Selection:} The \textbf{Top 5,000} brightest systems from the same ranked list as the 4MOST ($z^{\rm spec}$) sample (effectively the brightest 50\%). \\
     & & \textbf{Pairing:} $z_{\mathrm{lens}}^{\rm spec}$, $R_e$ [arcsec], $\sigma_{v, D}$ [km/s], $f_{D,i}$, $f_{D,g}/f_{D,r}$, $f_{D,r}/f_{D,i}$ \\
    \hline
    \end{tabular}
\end{table}

\subsection{Finding Self-Similar Deflector Pairs within \texttt{SLSim} lenses}\label{sec:deflector_pairing}
As established in Section~\ref{subsec:photo_vs_spectro_params_selection}, photometric data constitute the most prevalent observational modality for gravitational lenses in upcoming surveys. Accordingly, our primary strategy for identifying self-similar deflectors is predicated upon photometric observables and their derivative quantities. Here we evaluate the performance of three pairing strategies. The first relies exclusively on photometric parameters—redshift, size, magnitude, and colors. The second incorporates spectroscopic redshifts, where available from the 4MOST sample, alongside photometric size, magnitude, and colors. Finally, the third method combines spectroscopic redshifts and velocity dispersions with photometric size for systems in the 4MOST sample where both spectroscopic measurements are available. The list of pairing parameters used to pair deflectors for each sample is shown in Table~\ref{tab:slsim_ggl_samples}. Note that we convert magnitudes/colors to fluxes or flux ratios before using them for pairing so that all pairing parameters are on a linear scale.

Prior to pairing the deflectors, it is necessary to account for uncertainties in the pairing parameters as expected from the respective surveys. According to the LSST Science Book \cite{LSSTScienceBook:2009}, the target photo-z uncertainty for deflector and source galaxies is $\sigma_{z}/(1+z) < 0.05$, with an aspirational goal of 0.02. For simplicity, we adopt a fixed redshift uncertainty of $\sigma_{z}/(1+z) = 0.03$ for both deflectors and sources in all LSST Y10 lenses. This is an idealized Gaussian baseline rather than a demonstrated precision for lensed sources. Blending with the foreground deflector may increase their redshift uncertainties or introduce systematic biases. We investigate both effects in Appendix~\ref{app:photoz_precision_bias}. For spectroscopic redshifts, we assume an uncertainty of $\sigma_{z} \sim 10^{-4}$, applied to both 4MOST-based samples. Additionally, for deflector velocity dispersions, we adopt an uncertainty of 10 km s$^{-1}$ following \cite{Li:2024}. The size ($R_e$) of deflector galaxies also contains some measurement uncertainty, which we assume to be 5\% here across all samples. 
Finally, we incorporate a photometric flux uncertainty of 1\% (corresponding to $\sim$10 millimags) for the deflector photometry.

The implementation of deflector pairing begins by converting photometric magnitudes into pseudo-fluxes ($f_{D,i}$) and flux ratios ($f_{D,g}/f_{D,r}$, $f_{D,r}/f_{D,i}$), defined as $f = 10^{-0.4m}$. We apply a logarithmic transformation to all strictly positive pairing features, allowing us to employ a \texttt{kD-Tree} data structure \citep{2020SciPy} for efficient nearest-neighbor searches in this transformed high-dimensional space. This approach is well motivated: a Euclidean distance calculated in logarithmic space mathematically approximates the root-mean-square (RMS) fractional difference. Consequently, our \texttt{kD-Tree} algorithm directly optimizes for the dissimilarity metric ($\mathcal{D}_{\rm deflector}$, Equation~\ref{eqn:dissimilarity_parameter}) that characterizes the scatter in our downstream cosmological likelihood. Finally, we note that as the \texttt{kD-Tree} assigns each lens its nearest neighbor independently, the resulting pairing is not a mutually exclusive 1:1 mapping, and individual lenses may appear in multiple pairs.

As previously established, identifying near-identical deflectors requires a metric based on observed properties that effectively minimizes $\Delta\beta_{\rm E}/\beta_{\rm E}$. While various definitions for such a metric are possible, we adopt a simplified formulation suitable for forecasting purposes. Accordingly, to quantify the degree of self-similarity between paired deflector galaxies, we employ the deflector dissimilarity parameter, defined in Equation~\ref{eqn:dissimilarity_parameter}, which takes a specific functional form depending on the sample:
\begin{equation}\label{eqn:dissmilarity_parameter_diff_samples}
\mathcal{D}_{\rm deflector} \equiv \mathcal{D}_i(\text{pairing parameters}),
\end{equation}
where the specific pairing parameters adopted for each sample are summarized in Table~\ref{tab:slsim_ggl_samples}. As we will show, pairs with higher dissimilarity exhibit a proportionally larger scatter in their Einstein radius ratio. We parameterize this "noise floor" using a simple linear scatter model:
\begin{equation}\label{eqn:model_beta_E_dissimilarity}
    \sigma_{\beta_{\rm E}, \mathcal{D}} = \sigma_{\beta_{\rm E},\rm \mathcal{D}}^{(0)} + \sigma_{\beta_{\rm E},\rm \mathcal{D}}^{(1)} \ \mathcal{D}_{\rm deflector}
\end{equation}
where $\sigma_{\beta_{\rm E},\rm \mathcal{D}}^{(0)}$ is the baseline uncertainty for perfectly matched pairs, and $\sigma_{\beta_{\rm E},\rm \mathcal{D}}^{(1)}$ captures the degradation as pair dissimilarity increases. Importantly, $\sigma_{\beta_{\rm E},\mathcal D}$ characterizes the conditional scatter in the complete PDSPL residual, rather than only the measurement uncertainties in the quantities entering $\mathcal D_{\rm deflector}$. Because the Einstein radii are calculated independently from the SIS-equivalent velocity dispersions and the redshifts of the two deflectors, this scatter includes residual differences in their relative effective mass normalizations, deflector distances, and other properties not fully predicted by the pairing observables. The residual distributions are approximately Gaussian and centered on zero in the populated dissimilarity bins (Figure~\ref{fig:beta_E_vs_D}), motivating their treatment as a zero-mean stochastic contribution. Density-slope uncertainty is included separately through $\sigma_{\beta_{\rm E},\mathrm{pop}}$ in the cosmological likelihood (see Section~\ref{sec:likelihood_fnc}). This decomposition is conditional on the adopted \texttt{SLSim} population model and must ultimately be validated using observed lenses.

The contribution included in $\sigma_{\beta_{\rm E},\mathcal D}$ describes the residual physical mismatch produced by selecting pairs using noisy observables, whereas $\sigma_{\beta_{\rm E},\mathrm{photo-}z}$ propagates uncertainty in the distance-ratio prediction for an already selected system.

We adopt a linear relation because it adequately describes the dependence of the simulated scatter on $\mathcal D_{\rm deflector}$ over the range populated by each sample. The resulting calibrated relation provides an estimate of the expected pair-mismatch uncertainty for each PDSPL system and is incorporated directly into the cosmological likelihood in Section~\ref{sec:forecast_samples}.

To robustly calibrate this model and quantify the overall pairing efficiency, we performed the pairing simulation 500 times for each sample. In each realization, we added random Gaussian noise to the pairing parameters consistent with the uncertainties described above and executed the kD-Tree matching procedure using the parameters specified in Table~\ref{tab:slsim_ggl_samples}. The resulting pairing statistics—including the mean number of pairs, the mean all-pairs fractional scatter in $\beta_{\rm E}$, and the best-fit coefficients for our linear scatter model—are summarized in Table~\ref{tab:slsim_all_samples_pairing_statistics}.

\begin{figure*}
    \centering
    \includegraphics[width=\linewidth]{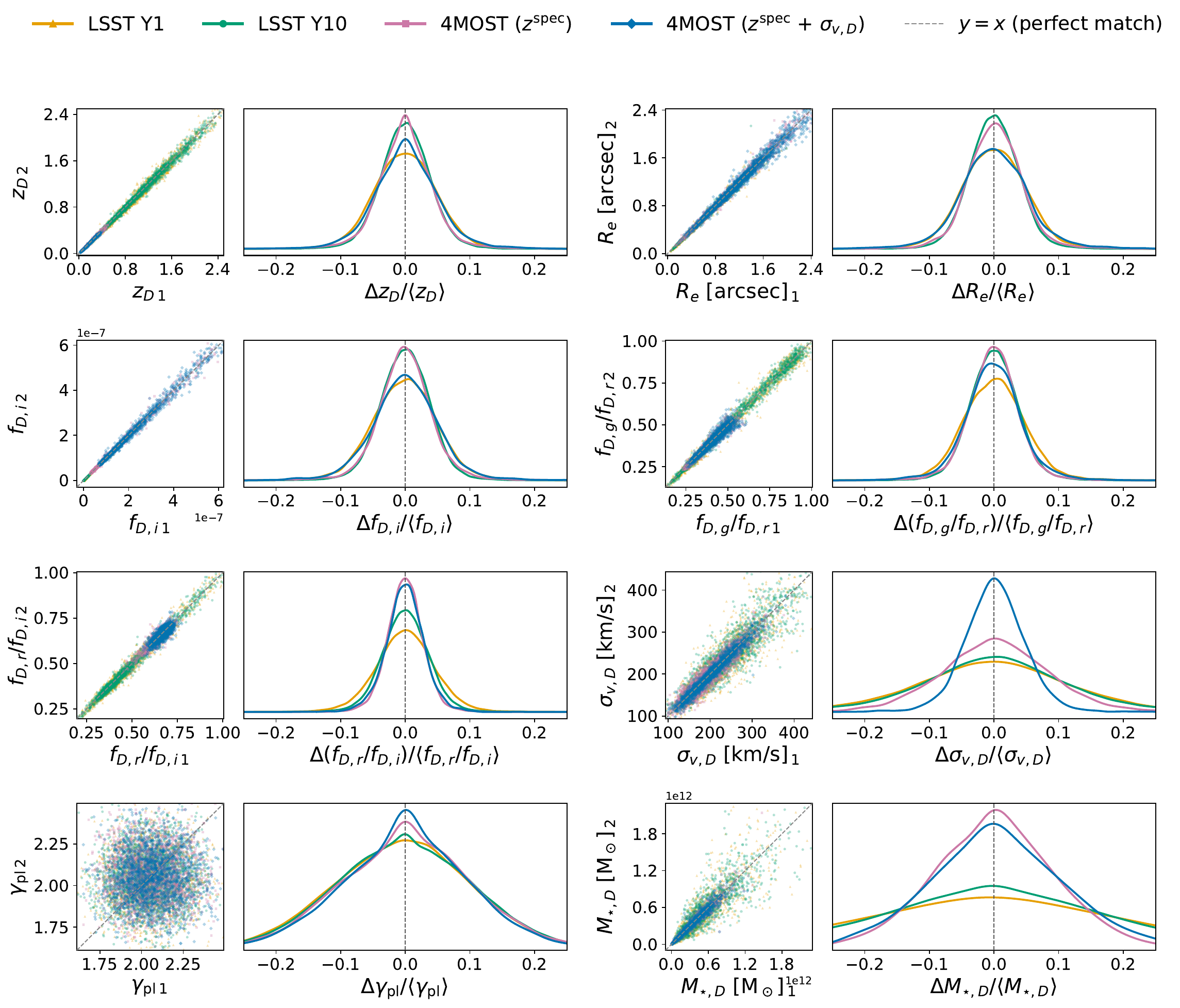}
    \caption{Pairing fidelity of deflector properties across lens samples. For each parameter/property: left panels compare values between paired deflectors (dashed gray 1:1 line); right panels show the distribution of relative fractional differences, $\Delta X / \langle X \rangle = 2(X_2 - X_1)/(X_2 + X_1)$. Rows 1--3 display pairing parameters: $z_D$, $R_e$, $f_{D,i}$, flux ratios, and $\sigma_{v, D}$. The bottom row shows residuals for $\gamma_{\rm pl}$ and stellar mass, which are not used for pairing but illustrate the matching fidelity of the physical mass models. Kinematic constraints significantly narrow the $\sigma_{v, D}$ distribution. Results represent a single representative Monte Carlo realization, similar to what's shown in left panel of Figure~\ref{fig:beta_E_vs_D}.}
    \label{fig:pairing_scatter_quality}
\end{figure*}

To visualize these results, Figure~\ref{fig:pairing_scatter_quality} details the relative differences among key quantities for the paired deflectors. Furthermore, Figure~\ref{fig:beta_E_vs_D} illustrates the resulting distribution of $\mathcal{D}_{\rm deflector}$ across all samples and its correlation with the Einstein radius ratio scatter, $\sigma_{\beta_{\rm E}, \mathcal{D}}$. By binning the deflector pairs at every 10$^{\rm th}$ percentile in $\mathcal{D}_{\rm deflector}$, the plot clearly demonstrates that more dissimilar pairs exhibit a proportionally higher scatter in $\beta_{\rm E}$.

\begin{table}
    \centering

    \caption{Summary of deflector pairing statistics for different samples. The middle three columns list the total number of lenses, the mean number of resulting deflector pairs and the mean of the standard deviation of $\Delta\beta_{\rm E}/\beta_{\rm E}$ (evaluated across all pairs in each realization), over 500 Monte Carlo realizations. The parameters $\sigma_{\beta_{\rm E},\rm \mathcal{D}}^{(0)}$ and $\sigma_{\beta_{\rm E},\rm \mathcal{D}}^{(1)}$ describe a linear fit to the relation given in Equation~\ref{eqn:model_beta_E_dissimilarity}
which quantifies how the fractional scatter in $\beta_{\rm E}$ varies with deflector pair dissimilarity, $\mathcal{D}_{\rm deflector}$. 
}\vspace{1em}
    \label{tab:slsim_all_samples_pairing_statistics}

\begingroup
\small
\setlength{\tabcolsep}{4pt}
\begin{tabular}{l | c c c | c c}
\hline
\multirow{2}{*}{\textbf{Sample}} & \multicolumn{3}{c|}{\textbf{Pairing Statistics}} & \multicolumn{2}{c}{\textbf{Linear Fit}} \\
 & \textbf{\# Lenses} & \textbf{$\langle$\# Pairs$\rangle$} & \textbf{$\langle{\texttt{std}{(\Delta\beta_{\rm E}/\beta_{\rm E})}}\rangle$} & \textbf{$\sigma_{\beta_{\rm E},\rm \mathcal{D}}^{(0)}$} & \textbf{$\sigma_{\beta_{\rm E},\rm \mathcal{D}}^{(1)}$} \\
\hline
LSST Y1 & 36831 & 27141 & 0.198 & 0.168 $\pm$ 0.002 & 0.63 $\pm$ 0.05 \\
LSST Y10 & 116471 & 85819 & 0.193 & 0.180 $\pm$ 0.002 & 0.34 $\pm$ 0.05 \\
4MOST ($z^{\rm spec}$) & 10000 & 7324 & 0.078 & 0.029 $\pm$ 0.002 & 1.25 $\pm$ 0.05 \\
4MOST ($z^{\rm spec}$ + $\sigma_{v, D}$) & 5000 & 3711 & 0.084 & 0.030 $\pm$ 0.001 & 1.09 $\pm$ 0.02 \\
\hline
\end{tabular}
\endgroup
\end{table}

\begin{figure}
    \centering
    \includegraphics[width=\linewidth]{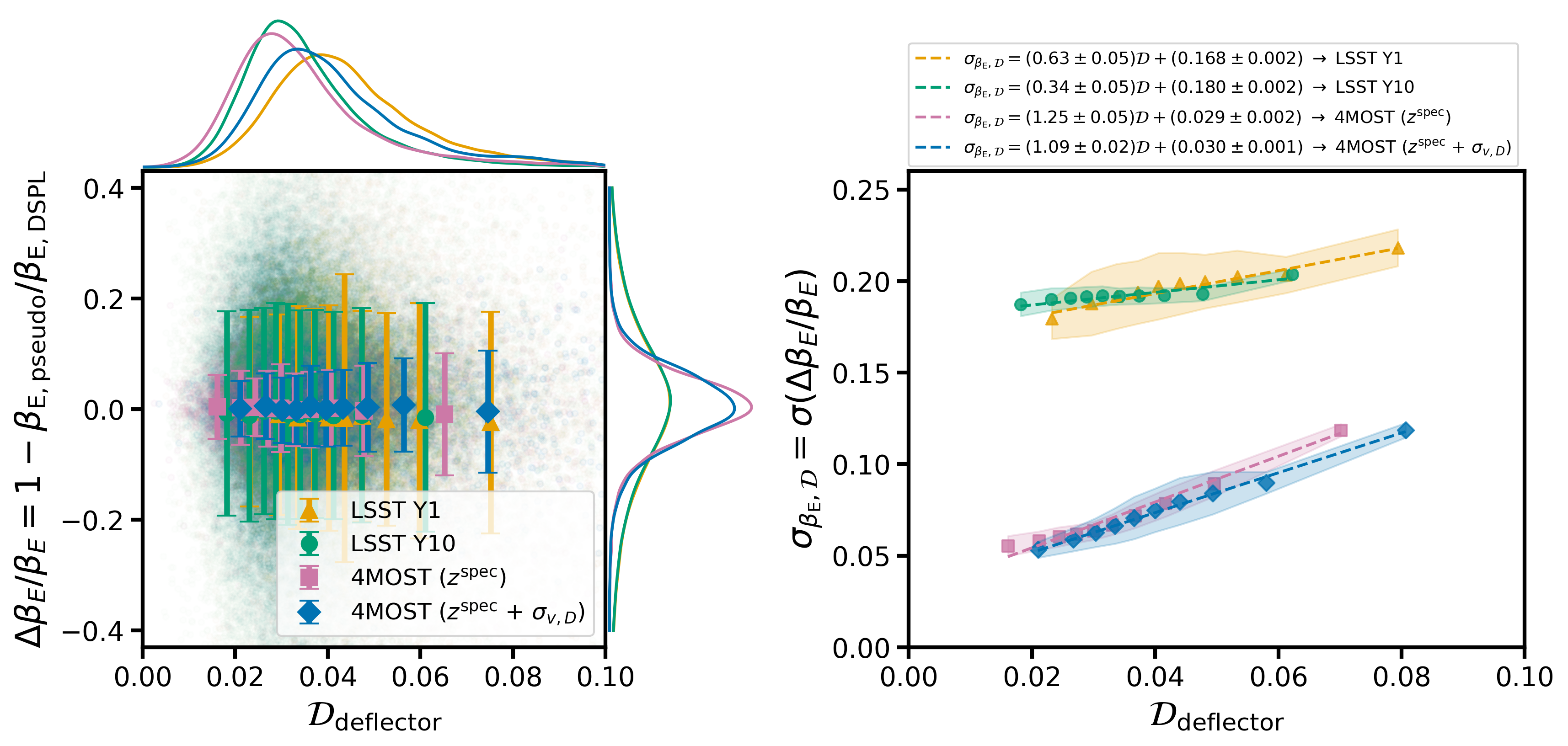}
    \caption{Simulated scatter in $\beta_{\rm E}$ as a function of the deflector pair dissimilarity parameter $\mathcal{D}_{\rm deflector}$ for PDSPLs across different samples, generated with \texttt{SLSim}. The results summarize 500 Monte Carlo realizations accounting for measurement uncertainties on $z_D$, $R_e$, $\sigma_{v, D}$, magnitudes/fluxes, and colors/flux-ratios. \emph{Left panel:} Individual fractional residuals and binned averages (with error bars showing the standard deviation within each bin) corresponding to a single representative realization. The residual distributions are approximately Gaussian and centered on zero across the populated bins. The marginal histograms show the distributions for this single run. \emph{Right panel:} The mean conditional scatter ($\sigma_{\beta_E, \mathcal{D}}$) calculated in bins corresponding to every 10$^{\rm th}$ percentile, averaged over the 500 realizations. The shaded regions represent the standard deviation of the scatter across the realizations. Weighted fits to these mean scatter values are shown, with the fitted equations included in the legend. Although the definition of $\mathcal{D}_{\rm deflector}$ (see Equation~\ref{eqn:dissimilarity_parameter} and~\ref{eqn:dissmilarity_parameter_diff_samples}) varies by sample depending on the pairing parameters, it effectively represents the percentage dissimilarity between the paired deflectors.
}
    
    \label{fig:beta_E_vs_D}
\end{figure}

Comparing the survey subsets in Figure~\ref{fig:beta_E_vs_D} reveals distinct behaviors driven by their respective observational uncertainties. The photometry-based \textit{LSST Y10} and \textit{LSST Y1} samples are primarily limited by photo-z precision. This establishes a high baseline "noise floor" ($\sigma_{\beta_{\rm E},\rm \mathcal{D}}^{(0)} \approx 0.17$--$0.18$), meaning even mathematically identical pairs ($\mathcal{D}_{\rm deflector} \to 0$) still exhibit significant scatter. Because this baseline error is dominant, the scatter exhibits a relatively shallow dependence on the dissimilarity parameter itself. Furthermore, this photo-z smearing results in a general scarcity of highly similar deflector pairs ($\mathcal{D}_{\rm deflector} \lesssim 0.02$). However, comparing the two photometric subsets highlights the direct benefit of survey duration: the vastly larger parent sample of \textit{LSST Y10} allows for the identification of more near-identical pairs than \textit{LSST Y1}. This statistical advantage is evident in the marginal distributions of Figure~\ref{fig:beta_E_vs_D}, where the $\mathcal{D}_{\rm deflector}$ distribution for \textit{LSST Y10} extends to notably lower values than its Y1 counterpart.

Conversely, the \textit{4MOST} samples demonstrate a tighter average residual scatter in $\beta_{\rm E}$ (dropping to $\sim 8\%$). With the redshift-induced geometric uncertainty virtually eliminated by exact spectroscopic redshifts, the remaining scatter in $\beta_{\rm E}$ becomes much more strongly correlated with the physical and photometric dissimilarity of the paired deflectors, resulting in a steeper linear dependence ($\sigma_{\beta_{\rm E},\rm \mathcal{D}}^{(1)} > 1.0$). This higher pairing fidelity may also partly reflect the selection of the brightest and most massive simulated deflectors, which occupy a narrower region of photometric and structural parameter space and are expected to follow relatively tight scaling relations such as the Fundamental Plane and the lensing mass--fundamental plane \citep{Djorgovski:1987,Dressler:1987,Mozumdar:2025}. We use these relations only as qualitative motivation; the pairing performance is measured directly from the simulated residuals. This precision nevertheless comes at the cost of a much smaller overall sample size.

\section{Forecasts for DSPL and PDSPL with MST}\label{sec:forecast}

\subsection{Analytical error propagation}
For an individual DSPL, we can do an analytical error propagation on the cosmological parameters stemming from the uncertainties in the measured Einstein radii ratio $\beta_{\rm E}$, the power-law slope $\gamma_{\rm pl}$ and the MST parameter $\lambda$. In particular, a change in the inferred dark energy equation of state parameter, $\delta w$, can be written to first order as

\begin{equation}
    \delta w =  \frac{\partial w}{\partial \beta} \left( \frac{\partial \beta}{\partial \beta_{\rm E}} \delta \beta_{\rm E} + \frac{\partial \beta}{\partial \lambda} \delta \lambda + \frac{\partial \beta}{\partial \gamma_{\rm pl}} \delta \gamma_{\rm pl} \right).
\end{equation}

The partial derivatives as a function of $\beta$ can be calculated from Equation~\ref{eqn:beta_mst_pl}. The partial derivatives around $\gamma_{\rm pl} \approx 2$ and $\lambda \approx 1$ result in

\begin{equation}
    \frac{\partial \beta}{\partial \beta_{\rm E}} \approx 1 \qquad
    \frac{\partial \beta}{\partial \lambda} = \frac{\partial \beta}{\partial \beta_{\rm E}}\frac{\partial \beta_{\rm E}}{\partial \lambda}\approx 1 - \beta_{\rm E} \qquad
    \frac{\partial \beta}{\partial \gamma_{\rm pl}} =\frac{\partial \beta}{\partial \beta_{\rm E}}\frac{\partial \beta_{\rm E}}{\partial \gamma_{\rm pl}} \approx -\beta_{\rm E} \log(\beta_{\rm E}).
\end{equation}

The partial change in $w$ resulting in a change from $\beta$ is generally dependent on the redshift of the lens and the two sources. For typical deflector and source redshift configurations (i.e. $z_{d} \sim 0.5$, $z_{\rm s1} \sim 1$, $z_{\rm s2}\sim 2$), we get

\begin{equation}
    \frac{\partial w}{\partial \beta} \approx - 40.
\end{equation}

Uncertainties in either $\lambda$ or $\gamma_{\rm pl}$ directly translate to uncertainties in $\beta$ and hence $w$. The power-law slope uncertainty is a strong function of $\beta_{\rm E}$ and highly suppressed for Einstein radii ratios close to one. This is expected as the change in radial density profile is a function of radial separation of the Einstein radii. The MST component $\lambda$ impacts $\beta$ of order unity. Constraining the MST parameter $\lambda$ to one percent will enable a measurement of $w$ to $0.4$ precision.

\subsection{Hierarchical population model and the likelihood function}\label{sec:likelihood_fnc}
To forecast cosmological constraints from the simulated DSPL and PDSPL samples, we use a hierarchical population model in which the cosmological parameters are inferred jointly with the hyperparameters describing the deflector mass-profile population. We consider a Flat $w_0w_a$CDM cosmology characterized by $\Omega_{\rm m}$, $w_0$, and $w_a$, with $\Omega_{\Lambda}=1-\Omega_{\rm m}$. The Einstein-radius ratio depends only on ratios of angular-diameter distances and is therefore independent of the absolute distance scale. We consequently fix $H_0=70\,\mathrm{km\,s^{-1}\,Mpc^{-1}}$ rather than including it among the sampled parameters.

We separate the sampled global parameters into cosmological and deflector-population subsets,
\begin{align}
    \Theta &= \{\Theta_c,\Theta_{\rm pop}\}, \\
    \Theta_c &= \{\Omega_{\rm m},w_0,w_a\}, \\
    \Theta_{\rm pop} &= \{\overline{\lambda}_{\rm MST},\sigma(\lambda_{\rm MST}),
    \overline{\gamma}_{\rm pl},\sigma(\gamma_{\rm pl})\}.
\end{align}
Here $\overline{\lambda}_{\rm MST}$ and $\overline{\gamma}_{\rm pl}$ are the population means, while $\sigma(\lambda_{\rm MST})$ and $\sigma(\gamma_{\rm pl})$ describe the intrinsic system-to-system scatter. In the baseline PDSPL forecasts, the two coefficients (Equation~\ref{eqn:model_beta_E_dissimilarity}) of the deflector-mismatch relation $\sigma_{\beta_{\rm E},\mathcal D}$ are fixed to their simulation-calibrated values. Appendix~\ref{app:inferred_scatter} considers the extended model in which these coefficients are also inferred.

For both DSPLs and PDSPLs, the measured observable for system $i$ is the Einstein-radius ratio
\begin{equation}
    \beta_{\text{E, meas.}, i}
    \equiv \frac{\theta_{{\rm E}1,i}}{\theta_{{\rm E}2,i}}.
\end{equation}
For a DSPL, the two Einstein radii correspond to two sources behind the same deflector. For a PDSPL, they are measured from two distinct single-source-plane lenses. The PDSPL implementation represents the two approximately matched deflector redshifts by their mean, $z_{{\rm d},i}=(z_{{\rm d}1,i}+z_{{\rm d}2,i})/2$. Following Section~\ref{sec:multi_source_plane}, the geometric distance ratio is
\begin{equation}
\label{eqn:beta_i}
    \beta_i(\Theta_c)
    \equiv
    \frac{D_{{\rm d}s1,i}(\Theta_c)D_{s2,i}(\Theta_c)}
    {D_{{\rm d}s2,i}(\Theta_c)D_{s1,i}(\Theta_c)}.
\end{equation}

For given effective system-level mass-profile parameters
$\phi_i=\{\lambda_{{\rm MST},i},\gamma_{{\rm pl},i}\}$, Equation~\ref{eqn:mst_epl} gives the forward model
\begin{equation}
\label{eqn:beta_E_individual_definition}
    m_i(\Theta_c,\phi_i)
    =
    \left[
        \beta_i(\Theta_c)
        -\left(1-\lambda_{{\rm MST},i}\right)
        \left(1-\beta_i(\Theta_c)\right)
    \right]^{1/(\gamma_{{\rm pl},i}-1)}.
\end{equation}
The central model prediction used in the marginalized likelihood is this expression evaluated at the population means:
\begin{align}
\label{eqn:beta_E_pl_i_definition}
    \beta_{{\rm E,pl},i}(\Theta)
    &\equiv m_i(\Theta_c,\overline{\phi}) \nonumber\\
    &=
    \left[
        \beta_i(\Theta_c)
        -\left(1-\overline{\lambda}_{\rm MST}\right)
        \left(1-\beta_i(\Theta_c)\right)
    \right]^{1/(\overline{\gamma}_{\rm pl}-1)},
\end{align}
where $\overline{\phi}=\{\overline{\lambda}_{\rm MST},\overline{\gamma}_{\rm pl}\}$. This explicitly defines the quantity $\beta_{{\rm E,pl},i}(\Theta)$ used below.

The distinction between the global hyperparameters $\Theta$ and the system-level parameters $\phi_i$ defines the hierarchical population model. The latter do not belong to $\Theta$; conceptually, they are latent draws from the shared population distributions
\begin{equation}
\label{eqn:population_distributions}
    \lambda_{{\rm MST},i}\sim
    \mathcal{N}\!\left(\overline{\lambda}_{\rm MST},\sigma^2(\lambda_{\rm MST})\right),
    \qquad
    \gamma_{{\rm pl},i}\sim
    \mathcal{N}\!\left(\overline{\gamma}_{\rm pl},\sigma^2(\gamma_{\rm pl})\right),
\end{equation}
which are assumed to be mutually independent. For a DSPL, $\phi_i$ describes the common deflector. The present PDSPL likelihood instead uses one effective pair-level $\phi_i$; it does not introduce separate latent mass profiles for the two physical deflectors. Their residual pair-differential mass-profile mismatch, including mass-normalization differences not predicted by the pairing observables, is represented statistically by $\sigma_{\beta_{\rm E},\mathcal D}$.

Let $\mathbf X_i$ denote the fixed inputs for system $i$: its measured redshifts, redshift uncertainties and, for a PDSPL, its observed deflector dissimilarity $\mathcal D_{{\rm deflector},i}$. A full system-level likelihood under the hierarchical population model would marginalize the latent parameters,
\begin{equation}
\label{eqn:formal_hierarchical_integral}
    p(\beta_{\text{E, meas.}, i}\mid\mathbf X_i,\Theta)
    =\int
    p(\beta_{\text{E, meas.}, i}\mid\mathbf X_i,\phi_i,\Theta_c)
    p(\phi_i\mid\Theta_{\rm pop})\,d\phi_i.
\end{equation}
The forecast is conditional on $\mathbf X_i$ and does not model the population distribution or selection function of these inputs.

Directly sampling $\lambda_{{\rm MST},i}$ and $\gamma_{{\rm pl},i}$ for every system would add $2N$ parameters. Instead, our implementation analytically marginalizes their effect after a first-order expansion of the forward model about the population means,
\begin{equation}
\label{eqn:first_order_population_expansion}
    m_i(\Theta_c,\phi_i)
    \simeq
    m_i(\Theta_c,\overline{\phi})
    +\mathbf J_i(\phi_i-\overline{\phi}),
\end{equation}
where $\mathbf J_i$ is the Jacobian with respect to $\lambda_{\rm MST}$ and $\gamma_{\rm pl}$, evaluated at $\overline{\phi}$. Since $\mathbb E[\phi_i-\overline{\phi}]=0$, the central prediction is, to first order, $m_i(\Theta_c,\overline{\phi})=\beta_{{\rm E,pl},i}(\Theta)$. The corresponding intrinsic population variance is
\begin{equation}
\label{eqn:sigma_beta_E_pop_expanded}
    \sigma_{\beta_{\rm E},{\rm pop},i}^2(\Theta)
    =
    \left[
        \left.\frac{\partial m_i}{\partial\lambda_{\rm MST}}
        \right|_{\overline{\phi}}
        \sigma(\lambda_{\rm MST})
    \right]^2
    +
    \left[
        \left.\frac{\partial m_i}{\partial\gamma_{\rm pl}}
        \right|_{\overline{\phi}}
        \sigma(\gamma_{\rm pl})
    \right]^2.
\end{equation}
Thus the latent parameters are represented implicitly through the predictive distribution rather than sampled explicitly. This is an exact analytic marginalization of the \emph{linearized Gaussian model}, but only a first-order approximation to the nonlinear model in Equation~\ref{eqn:beta_E_individual_definition}. Both population means and both intrinsic scatters remain free hyperparameters and are varied jointly with cosmology in the MCMC inference.

Under this approximation, the marginalized likelihood for system $i$ is Gaussian in the observable,
\begin{equation}
\label{eqn:marginalized_system_likelihood}
    \mathcal{L}_i(\beta_{\text{E, meas.}, i}\mid\mathbf X_i,\Theta)
    =\frac{1}{\sqrt{2\pi\sigma_{\beta_{\rm E},i}^2(\Theta)}}
    \exp\!\left[
    -\frac{\left(
        \beta_{\text{E, meas.}, i}
        -\beta_{{\rm E,pl},i}(\Theta)
    \right)^2}
    {2\sigma_{\beta_{\rm E},i}^2(\Theta)}
    \right].
\end{equation}

The total predictive variance is decomposed into statistically independent contributions,
\begin{equation}
\label{eqn:total_beta_E_variance}
    \sigma_{\beta_{\rm E},i}^2(\Theta)
    =\sigma_{\beta_{\rm E},{\rm meas},i}^2
    +\sigma_{\beta_{\rm E},{\rm los},i}^2
    +\sigma_{\beta_{\rm E},\mathcal D,i}^2(\Theta)
    +\sigma_{\beta_{\rm E},{\rm pop},i}^2(\Theta)
    +\sigma_{\beta_{\rm E},{\rm photo-}z,i}^2(\Theta).
\end{equation}
The first term describes uncertainty in the measured Einstein radii, while the second represents stochastic line-of-sight perturbations. In the forecasts, each is assigned a fractional width of $1\%$ \citep{Collett:2014, Johnson:2025, Smith:2021}. During mock construction these fractional widths are converted to absolute uncertainties using the fiducial prediction and are then held fixed during inference.

For PDSPLs, the deflector-mismatch uncertainty generally dominates the error budget, with a fractional scatter of approximately $8$--$20\%$ in the simulated samples (Figure~\ref{fig:beta_E_vs_D}). It describes the aggregate residual mismatch between the two distinct deflectors after pairing on the observables in Section~\ref{sec:deflector_pairing}. Using the linear relation calibrated in Equation~\ref{eqn:model_beta_E_dissimilarity}, its absolute contribution at a proposed model point is
\begin{equation}
\label{eqn:pair_scatter_likelihood}
    \sigma_{\beta_{\rm E},\mathcal D,i}(\Theta)
    =\beta_{{\rm E,pl},i}(\Theta)
    \left[
        \sigma_{\beta_{\rm E},\mathcal D}^{(0)}
        +\sigma_{\beta_{\rm E},\mathcal D}^{(1)}
        \mathcal D_{{\rm deflector},i}
    \right].
\end{equation}
For a true DSPL, no deflector pairing is required and we set $\sigma_{\beta_{\rm E},\mathcal D,i}=0$. The mass-profile population term in Equation~\ref{eqn:sigma_beta_E_pop_expanded} is included separately. Its fiducial injected values are $\sigma(\gamma_{\rm pl})=0.16$ \citep{Auger:2010} and $\sigma(\lambda_{\rm MST})=0.05$, but both are allowed to vary during inference.

Finally, photo-z errors are propagated into the predicted Einstein-radius ratio. They are not represented by explicitly sampled latent redshifts. For $x\in\{z_{{\rm d},i},z_{s1,i},z_{s2,i}\}$, the first-order propagated variance is
\begin{equation}
\label{eqn:sigma_photo-z}
    \sigma_{\beta_{\rm E},{\rm photo-}z,i}^2(\Theta)
    =\sum_{x}
    \left(
        \frac{\partial\beta_{{\rm E,pl},i}}{\partial\beta_i}
        \frac{\partial\beta_i}{\partial x}
        \sigma_{x,i}
    \right)^2.
\end{equation}

Thus the marginalized likelihood for the sample and the posterior for the global parameters are
\begin{align}
\label{eqn:sample_marginal_likelihood}
    \mathcal{L}_{\rm marg}({\boldsymbol{\beta}}_{\rm E, meas.}
    \mid\mathbf X,\Theta)
    &=\prod_{i=1}^{N}
    \mathcal{L}_i(\beta_{\text{E, meas.}, i}\mid\mathbf X_i,\Theta), \\
    p(\Theta\mid{\boldsymbol{\beta}}_{\rm E, meas.},\mathbf X)
    &\propto
    \mathcal{L}_{\rm marg}({\boldsymbol{\beta}}_{\rm E, meas.}
    \mid\mathbf X,\Theta)p(\Theta).
\end{align}
For these forecasts, we approximate the PDSPL pairs as statistically independent, neglecting correlations induced when an individual lens contributes to more than one pair.
Unless otherwise specified, we adopt the priors in Table~\ref{tab:forecast_priors}. For the numerical forecasts, we evaluate the expected (Asimov) form of this marginalized log-likelihood rather than drawing one noisy realization; Appendix~\ref{app:numerical_compression} gives the corresponding expression and the numerical-compression scheme.

\subsection{Simulated PDSPLs and DSPLs for forecasting}\label{sec:forecast_samples}

Figure~\ref{fig:beta_E_vs_D} illustrates the scatter of $\beta_{\rm E, pseudo}$ around $\beta_{\rm E, DSPL}$ for \texttt{SLSim} simulated galaxy–galaxy lenses paired according to the photometric and/or spectroscopic properties of their deflectors. The observed scatter from these simulations, $\sigma_{\beta_{\rm E}, \mathcal{D}} \sim 8\text{--}20\%$, reflects intrinsic variations between the paired deflectors and the finite precision of the pairing parameters, rather than measurement uncertainties in the Einstein radii themselves, which are subdominant here. The fractional residual, $\Delta\beta_{\rm E}/\beta_{\rm E}$, follows an approximately Gaussian distribution, indicating that these variations are well described by random statistical fluctuations around the true relation.

The dependence of $\sigma_{\beta_{\rm E}, \mathcal{D}}$ on the deflector dissimilarity parameter $\mathcal{D}_{\rm deflector}$ is modeled using a linear scatter model with coefficients ($\sigma_{\beta_{\rm E},\rm \mathcal{D}}^{(0)}$ and $\sigma_{\beta_{\rm E},\rm \mathcal{D}}^{(1)}$) calibrated to the simulated sample (see Equation~\ref{eqn:model_beta_E_dissimilarity} and Figure~\ref{fig:beta_E_vs_D}). We use these coefficients to dynamically estimate the pairing uncertainty $\sigma_{\beta_{\rm E}, \mathcal{D}, i}$ for each PDSPL system.

We forecast the constraining power of the full PDSPL population expected from LSST and 4MOST. As detailed in Table~\ref{tab:slsim_all_samples_pairing_statistics}, the LSST Y10 sample is expected to yield tens of thousands of deflector pairs. Evaluating the approximate marginalized likelihood of the hierarchical population model for every pair is computationally expensive. We therefore use a compressed representation in which a representative subset of systems is assigned the sample down-sampling factor (DSF) as a log-likelihood weight. This approximates the expected statistical weight of the full catalog. We detail this numerical compression implementation in Appendix~\ref{app:numerical_compression}.

To avoid random statistical fluctuations obscuring the true parameter constraining power, we evaluate the likelihood using Asimov datasets \citep{Cowan:2011}. In an Asimov dataset, the observed mock variables ($\beta_{\text{E, meas.}}$) are set exactly to the theoretical expectation value of the true model (no stochastic Gaussian noise is added, see Appendix~\ref{app:numerical_compression}), while the likelihood variance properly incorporates the full theoretical error budget. The underlying ground truth for these mocks assumes a Flat $w_0w_a$CDM cosmology with $\Omega_{\rm m} = 0.3, w_0 = -1.0, w_a = 0.0$. For the mass profile population, we inject true parameters of $\overline\gamma_{\rm pl} = 2.078$, $\sigma(\gamma_{\rm pl}) = 0.16$ \cite{Auger:2010}, $\overline\lambda_{\rm MST} = 1.0$, and $\sigma(\lambda_{\rm MST}) = 0.05$. The corresponding lens and source redshifts, alongside their observational uncertainties (e.g., precise spectroscopic redshifts or $3\%$ photo-z errors), are adopted directly from the \texttt{SLSim} generated samples.

Finally, to directly compare the tomographic leverage of PDSPLs against DSPLs\footnote{We perform a simple forecast for idealized DSPLs (with no mass for the intermediate source)}, we generate an independent mock catalog of 500 genuine DSPLs—consistent with optimistic expectations for LSST Y10 \cite{Shajib:2024}. Each DSPL system incorporates a joint measurement and line-of-sight precision of $\sim 1.4\%$ on the Einstein radius ratio, adopts the identical true cosmological and deflector population parameters as the PDSPLs, and utilizes exact spectroscopic redshifts.

\subsection{Numerical forecast with the hierarchical population model}\label{sec:numerical_forecast}

Cosmological inference utilizing DSPL or PDSPL observables necessitates a simultaneous joint fit of the cosmological parameters and the hyperparameters of the deflector population. For the PDSPL and DSPL samples outlined above, we sample the approximate marginalized likelihood of the hierarchical population model defined in Section~\ref{sec:likelihood_fnc} with Markov Chain Monte Carlo (MCMC). To efficiently evaluate these massive mock datasets without introducing random stochastic biases, we utilized the compressed Asimov log-likelihood framework detailed in Appendix~\ref{app:numerical_compression}.

In our baseline analysis, we treat the fundamental structural diversity of the deflector population---parameterized by the intrinsic scatters $\sigma(\lambda_{\rm MST})$ and $\sigma(\gamma_{\rm pl})$---as free hyperparameters. Accounting for this system-to-system variation is essential in strong-lensing cosmography to avoid artificially tight and potentially biased cosmological constraints. Conversely, for the PDSPL forecasts presented in the main text, we fix the linear dissimilarity-scatter coefficients ($\sigma_{\beta_{\rm E},\mathcal D}^{(0)}$ and $\sigma_{\beta_{\rm E},\mathcal D}^{(1)}$) to their simulation-calibrated values. Appendix~\ref{app:inferred_scatter} relaxes this assumption and infers these coefficients jointly with the cosmological and population parameters. We adopt broad uniform priors for $\Omega_{\rm m}$, $w_0$, $w_a$, $\overline\gamma_{\rm pl}$, $\overline\lambda_{\rm MST}$, and their associated intrinsic scatters, while keeping $H_0$ fixed (see Table~\ref{tab:forecast_priors}).

\begin{figure}
\centering
  \includegraphics[width=\textwidth]{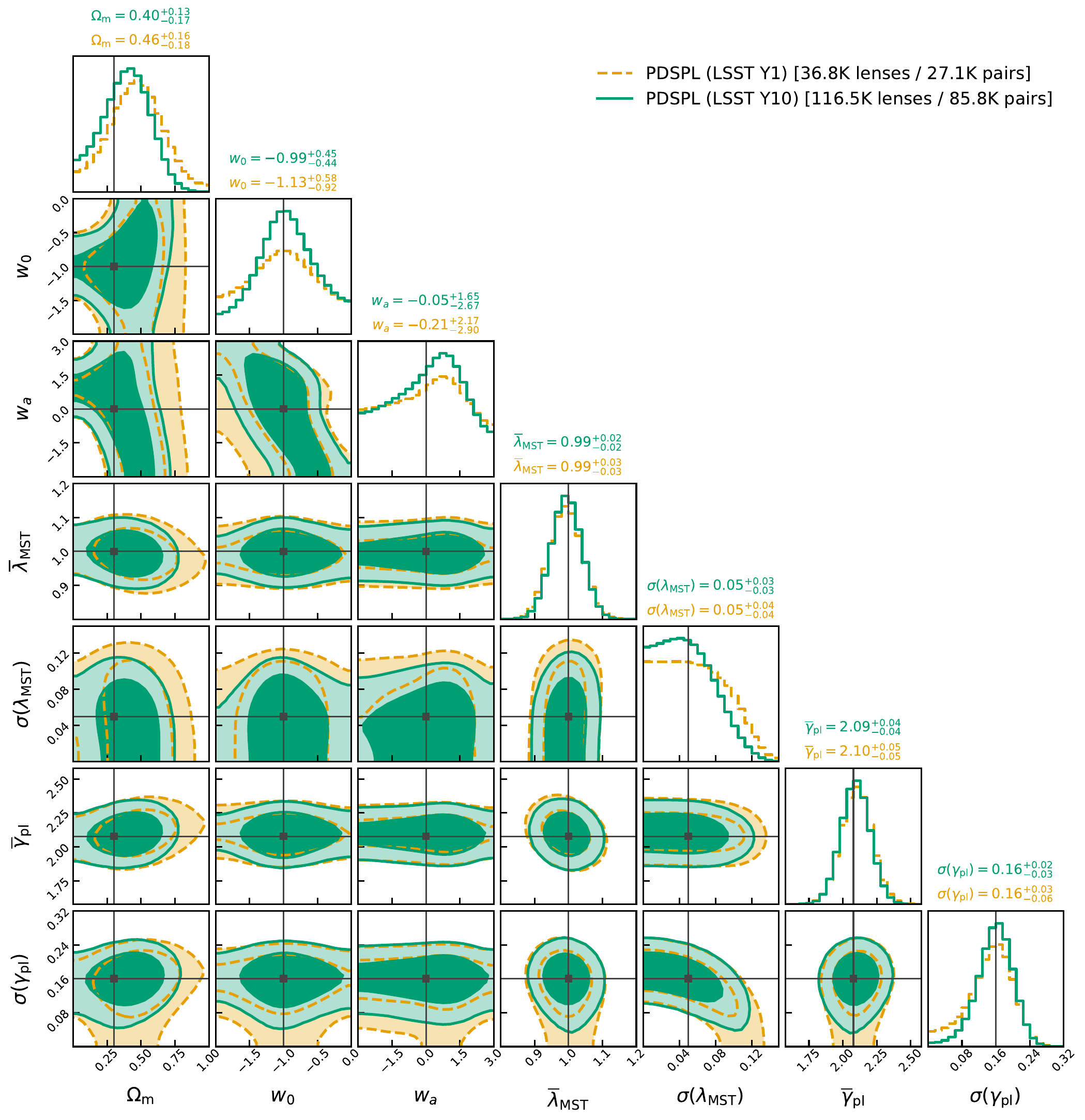} 
  \caption{Comparison of posterior distributions for the LSST Y1 (orange) and LSST Y10 (green) PDSPL samples, illustrating the impact of survey duration on cosmological constraints in a Flat $w_0w_a$CDM cosmology. The transition to the full Y10 sample size ($\sim$86,000 pairs) significantly reduces parameter degeneracies. Contours represent 68\% and 95\% confidence levels, with the ground truth indicated by black lines.
  \label{fig:forecast_w0waCDM_lsst_y1_y10_PDSPLs} 
\vspace{-10pt}
}
\end{figure}

\begin{figure}
\centering
  \includegraphics[width=\textwidth]{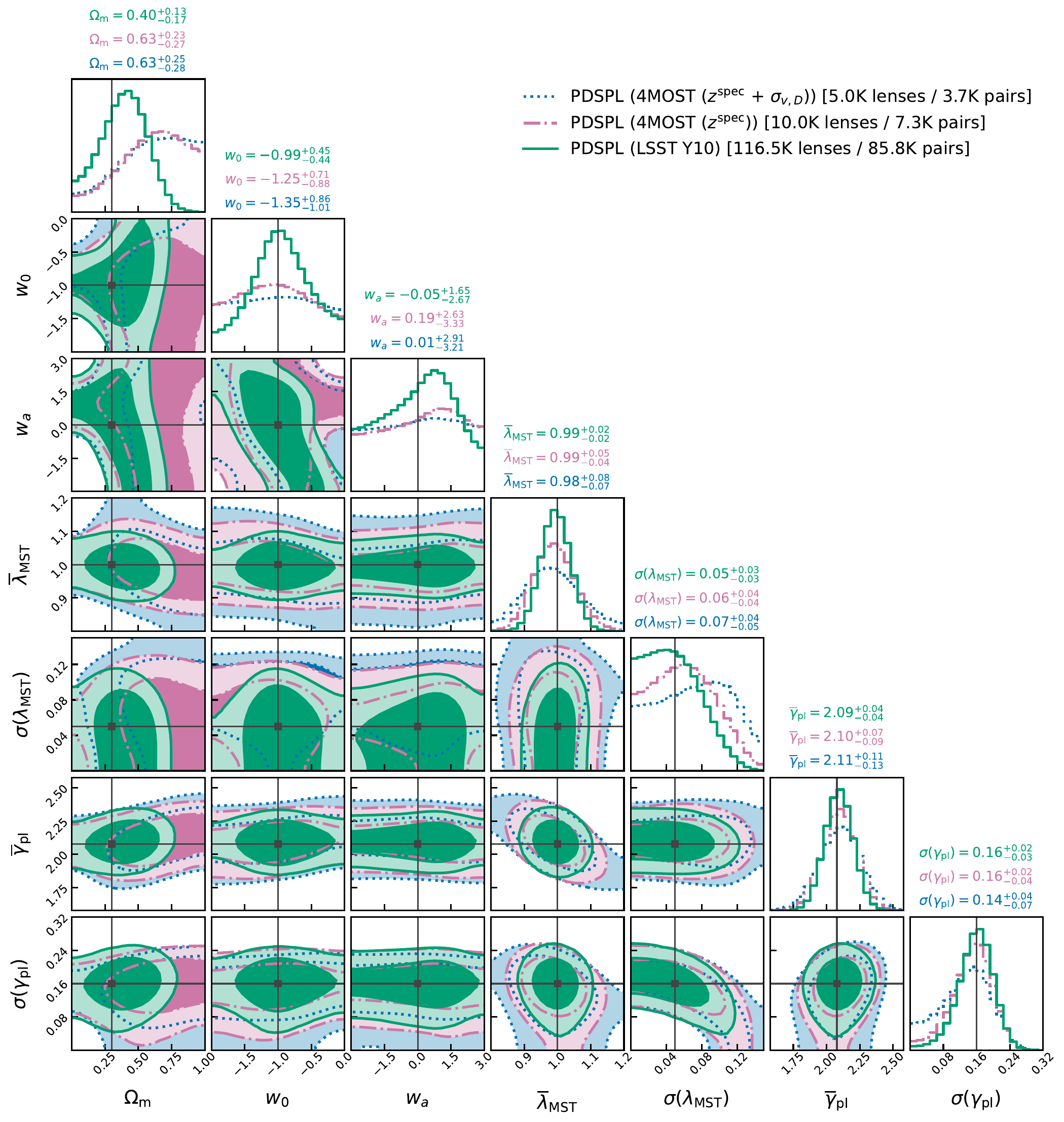} 
  \caption{Comparison of LSST Y10 photometric samples against spectroscopic 4MOST-based PDSPL subsets. Despite the high fidelity pairing with the 4MOST measurements ($z^{\rm spec}$ and $\sigma_{v,D}$), the significantly larger statistical volume of the LSST Y10 photometric sample provides tighter overall constraints on $w_0$ and $w_a$ in a Flat $w_0w_a$CDM cosmology. Contours represent 68\% and 95\% confidence levels, with the ground truth indicated by black lines.
  \label{fig:forecast_w0waCDM_lsst_y10_4MOST_PDSPLs} 
\vspace{-10pt}
}
\end{figure}

We explore the constraining power of the PDSPL samples across two main dimensions: the temporal evolution of the photometric survey (Figure~\ref{fig:forecast_w0waCDM_lsst_y1_y10_PDSPLs}) and the comparative advantage of spectroscopic follow-up (Figure~\ref{fig:forecast_w0waCDM_lsst_y10_4MOST_PDSPLs}) for a Flat $w_0w_a$CDM cosmology. As shown in Figure~\ref{fig:forecast_w0waCDM_lsst_y1_y10_PDSPLs}, the transition from LSST Y1 to Y10 leads to a significant contraction of the cosmological contours. Specifically, the \emph{LSST Y10} sample yields $w_0$ constraints that are highly competitive and perfectly centered on the truth (as mathematically guaranteed by the Asimov formalism). Under our fiducial assumption of unbiased Gaussian photo-z errors, the LSST Y10 sample also provides greater statistical precision than the smaller spectroscopic 4MOST-based PDSPL samples. The 4MOST ($z^{\rm spec}$) and 4MOST ($z^{\rm spec}+\sigma_{v,D}$) samples yield broad standalone constraints on $w_0$, with $\sigma(w_0)\gtrsim0.7$--$1.0$, primarily because of their smaller sample sizes. This comparison concerns statistical precision and does not imply that spectroscopy is unnecessary: representative spectroscopic data remain important for calibrating photo-z biases and controlling systematic accuracy. The effects of spectroscopic redshifts on the full LSST Y10 sample are examined in Appendix~\ref{app:lsst_y10_photo-z_vs_spec-z}, while the effects of degraded photo-z precision and unrecognized bias are presented in Appendix~\ref{app:photoz_precision_bias}.

Figure~\ref{fig:forecast_w0waCDM_DSPL_PDSPL_y10} directly compares the PDSPL constraints from the \emph{LSST Y10} sample against those expected from the 500 DSPL sample. We additionally overlay the constraints obtained when imposing an external Gaussian prior on the matter density, $\Omega_{\rm m}$ of $\mathcal{N}(0.3, 0.05)$. We find that the \emph{LSST Y10} PDSPL sample provides significantly tighter, less degenerate constraints for the dark energy equation of state compared to the 500 DSPLs. When combined with the $\Omega_{\rm m}$ prior, the constraints on $w_0$ narrow considerably for the \emph{LSST Y10} PDSPLs, effectively breaking the remaining degeneracies, whereas the standard DSPL sample remains bottlenecked by the mass-sheet degeneracy.

Beyond cosmology, the hierarchical population model constrains the deflector mass-profile hyperparameters simultaneously. The inference determines the population mean values $\overline\lambda_{\rm MST}$ and $\overline\gamma_{\rm pl}$ to $\sim2\%$ without relying on tight external priors and also constrains the intrinsic scatters $\sigma(\lambda_{\rm MST})$ and $\sigma(\gamma_{\rm pl})$. A comprehensive summary of the inferred posteriors---detailing the median and 1-sigma percentiles for both the cosmological and deflector parameters across all tested scenarios---is provided in Table~\ref{tab:forecast_results}.

\begin{table}[h!]
\centering
\caption{Sampled global parameters and their priors in the hierarchical population model.}\label{tab:forecast_priors}
\begin{tabular}{lll}
\toprule
\textbf{Parameters} & \textbf{Prior} & \textbf{Description} \\
\midrule
\multicolumn{3}{l}{\textbf{Cosmology}} \\\\
$H_0$ [ km s$^{-1}$ Mpc$^{-1}$ ] & Fixed to $70$ & Hubble constant \\
$\Omega_{\rm m}$ & $\mathcal{U}(0.0, 1.0)$ & Matter density \\
$w_0$ & $\mathcal{U}(-3, 0)$ & Dark Energy EoS \\
$w_a$ & $\mathcal{U}(-5, 5)$ & Dark Energy EoS \\
\midrule
\multicolumn{3}{l}{\textbf{Mass profile}} \\\\
$\overline\lambda_{\text{MST}}$ & $\mathcal{U}(0.8, 1.2)$ & Internal MST population mean \\
$\overline\gamma_{\rm pl}$ & $\mathcal{U}(1.0, 3.0)$ & Mean power-law slope of the deflector \\
$\sigma(\gamma_{\rm pl})$ & $\mathcal{U}(0.0, 0.50)$ & Intrinsic scatter in deflector power law slope \\
$\sigma(\lambda_{\text{MST}})$ & $\mathcal{U}(0.0, 0.20)$ & Intrinsic scatter in MST parameter \\
\bottomrule
\end{tabular}
\end{table}

\begin{sidewaystable}[ht]
\centering
\begin{threeparttable}
\caption{Cosmological forecast constraints for the PDSPL samples and 500 DSPLs.}\label{tab:forecast_results}
\vspace{1em}
\begin{tabular}{lccccccc}
\toprule
Sample & $\Omega_{\rm m}$ & $w_0$ & $w_a$ & $\overline\lambda_{\rm MST}$ & $\sigma(\lambda_{\rm MST})$ & $\overline\gamma_{\rm pl}$ & $\sigma(\gamma_{\rm pl})$ \\
\midrule
TRUTH & 0.30 & -1.00 & 0.00 & 1.000 & 0.050 & 2.078 & 0.160 \\
\midrule
\multicolumn{8}{l}{\textbf{Flat $w_0w_a$CDM}} \\
\midrule
DSPL (500 lenses) & $0.53^{+0.26}_{-0.22}$ & $-1.40^{+0.84}_{-0.99}$ & $-0.17^{+2.69}_{-3.08}$ & $0.981^{+0.056}_{-0.046}$ & $0.056^{+0.033}_{-0.033}$ & $2.123^{+0.088}_{-0.104}$ & $0.163^{+0.025}_{-0.036}$ \\
PDSPL (LSST Y1) & $0.46^{+0.16}_{-0.18}$ & $-1.13^{+0.58}_{-0.92}$ & $-0.21^{+2.17}_{-2.90}$ & $0.992^{+0.028}_{-0.026}$ & $0.054^{+0.040}_{-0.037}$ & $2.104^{+0.049}_{-0.051}$ & $0.157^{+0.026}_{-0.058}$ \\
PDSPL (LSST Y10) & $0.40^{+0.13}_{-0.17}$ & $-0.99^{+0.45}_{-0.44}$ & $-0.05^{+1.65}_{-2.67}$ & $0.994^{+0.019}_{-0.019}$ & $0.045^{+0.031}_{-0.030}$ & $2.094^{+0.035}_{-0.035}$ & $0.164^{+0.017}_{-0.032}$ \\
PDSPL (4MOST $z^{\rm spec}$) & $0.63^{+0.23}_{-0.27}$ & $-1.25^{+0.71}_{-0.88}$ & $0.19^{+2.63}_{-3.33}$ & $0.988^{+0.051}_{-0.044}$ & $0.059^{+0.035}_{-0.036}$ & $2.103^{+0.073}_{-0.085}$ & $0.156^{+0.025}_{-0.043}$ \\
PDSPL (4MOST $z^{\rm spec}$ + $\sigma_v$) & $0.63^{+0.25}_{-0.28}$ & $-1.35^{+0.86}_{-1.01}$ & $0.01^{+2.91}_{-3.21}$ & $0.983^{+0.082}_{-0.070}$ & $0.074^{+0.040}_{-0.048}$ & $2.108^{+0.110}_{-0.127}$ & $0.142^{+0.040}_{-0.070}$ \\
\midrule
\multicolumn{8}{l}{\textbf{Flat $w_0w_a$CDM + $\Omega_{m}$ prior$^\dagger$}} \\
\midrule
DSPL (500 lenses) & $0.31^{+0.05}_{-0.05}$ & $-1.39^{+0.73}_{-0.92}$ & $0.11^{+2.23}_{-2.79}$ & $1.013^{+0.056}_{-0.048}$ & $0.056^{+0.035}_{-0.035}$ & $2.036^{+0.083}_{-0.084}$ & $0.150^{+0.024}_{-0.036}$ \\
PDSPL (LSST Y1) & $0.31^{+0.05}_{-0.05}$ & $-1.13^{+0.39}_{-0.61}$ & $0.67^{+1.55}_{-1.72}$ & $1.001^{+0.028}_{-0.027}$ & $0.061^{+0.037}_{-0.040}$ & $2.076^{+0.044}_{-0.045}$ & $0.148^{+0.028}_{-0.061}$ \\
PDSPL (LSST Y10) & $0.31^{+0.05}_{-0.05}$ & $-1.08^{+0.25}_{-0.32}$ & $0.58^{+1.28}_{-1.27}$ & $0.999^{+0.018}_{-0.018}$ & $0.048^{+0.032}_{-0.032}$ & $2.081^{+0.030}_{-0.030}$ & $0.160^{+0.017}_{-0.034}$ \\
PDSPL (4MOST $z^{\rm spec}$) & $0.31^{+0.05}_{-0.05}$ & $-1.05^{+0.57}_{-0.65}$ & $0.76^{+1.58}_{-2.15}$ & $1.016^{+0.056}_{-0.050}$ & $0.059^{+0.037}_{-0.037}$ & $2.043^{+0.090}_{-0.089}$ & $0.147^{+0.025}_{-0.044}$ \\
PDSPL (4MOST $z^{\rm spec}$ + $\sigma_v$) & $0.31^{+0.05}_{-0.05}$ & $-1.17^{+0.74}_{-0.91}$ & $0.45^{+2.24}_{-2.80}$ & $1.026^{+0.087}_{-0.084}$ & $0.075^{+0.038}_{-0.049}$ & $2.020^{+0.145}_{-0.131}$ & $0.129^{+0.040}_{-0.068}$ \\
\bottomrule
\end{tabular}
\begin{tablenotes}
\item[$\dagger$] The prior on $\Omega_{\rm m}$ is $\mathcal{N}(0.3, 0.05)$.
\end{tablenotes}
\end{threeparttable}
\end{sidewaystable}

\begin{figure}
\centering
  \includegraphics[width=\textwidth]{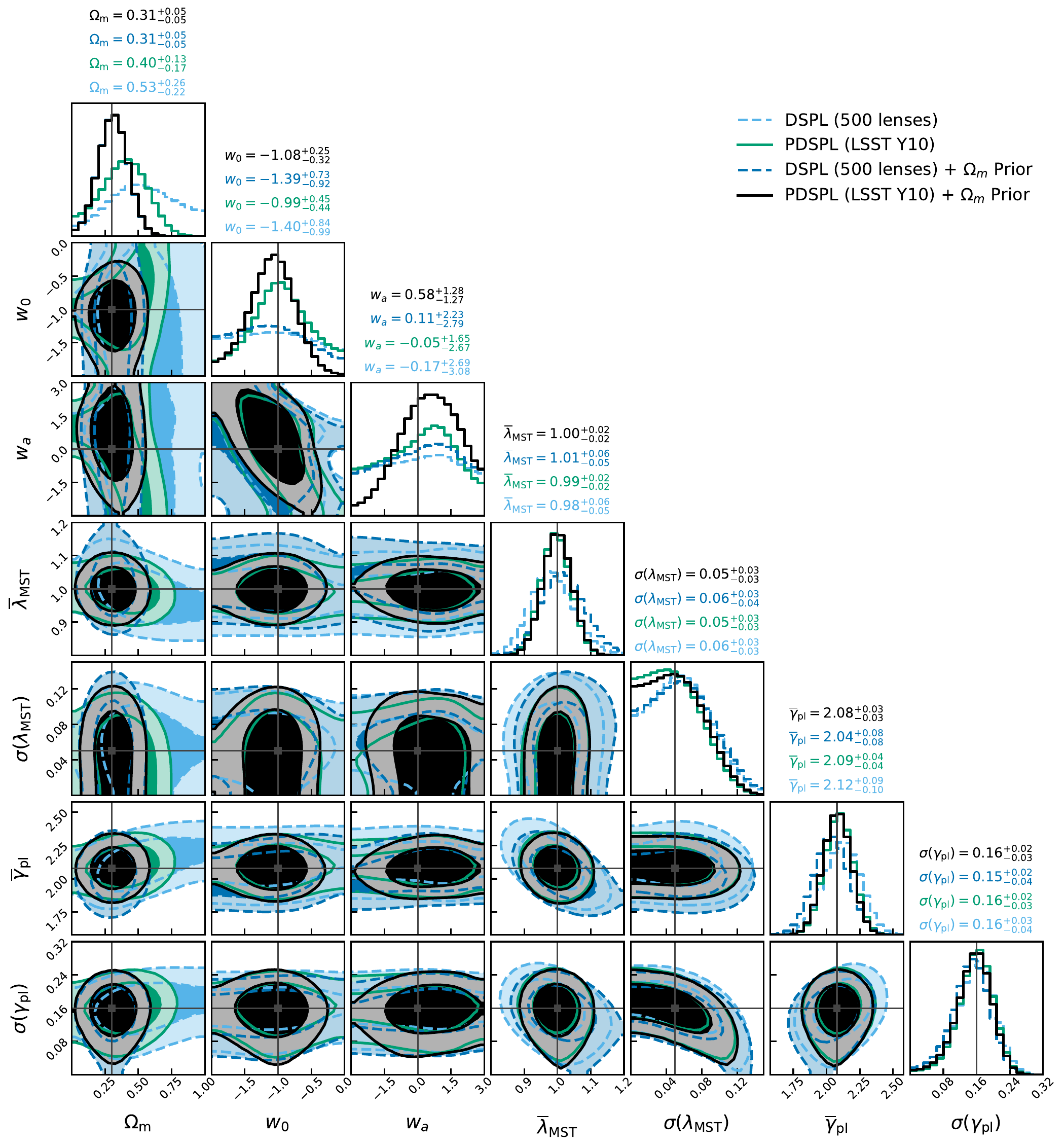} 
  \caption{Comparison of posterior distributions for the DSPL (500 lenses) and LSST Y10 PDSPL samples while inferring Flat $w_0w_a$CDM cosmology. Results are shown both with and without an external prior on $\Omega_{\rm m}$ of $\mathcal{N}(0.3, 0.05)$. The PDSPL sample (solid green/black contours) demonstrates significantly tighter constraints on the dark energy equation of state parameters ($w_0, w_a$) compared to the standard DSPL sample (dashed blue contours). Ground truth is also shown with black horizontal/vertical lines. Note that the DSPL (500 lenses) shows an idealized case where the forecast assumes no mass of the intermediate source.
  \label{fig:forecast_w0waCDM_DSPL_PDSPL_y10} 
\vspace{-10pt}
}
\end{figure}

\section{Discussion}
\label{sec:discussion}
The PDSPL framework proposed in this work transforms strong lensing dark energy cosmography from a regime limited by the stochastic discovery of rare alignments (DSPLs) to one driven by the statistics of wide-field surveys. By leveraging the vast population of single-source plane lenses expected from LSST, Euclid, and Roman, we trade exact geometric alignment for statistical power. In this section, we discuss the methodological advantages, strategies for refining the pair selection, the robustness of this approach against systematic uncertainties, and how this method compares against existing probes of dark energy.

First we note that our current analysis is conservative regarding the information content of the lens models. We have not accounted for the potential to break the mass-sheet degeneracy (MSD) using weak lensing constraints, as demonstrated in recent work on galaxy-scale lenses \citep{Khadka:2024}. Furthermore, high-resolution imaging of lensed arcs would not only constrain the mass density slope, $\gamma_{\rm pl}$, but also allow us to use this parameter to pair deflectors more accurately. Incorporating these priors would likely tighten the constraints on the lens population hyperparameters, further improving the cosmological precision.

\subsection{Geometric Simplicity and Avoiding Intermediate Source MST}

A distinct advantage of the PDSPL formalism is that it bypasses the intermediate source mass, thus avoiding the geometric and modeling complexities of true multi-source plane systems. True DSPLs suffer from ``compound lensing'', where the foreground source acts as a secondary deflector for the background source, in the extreme case creating complex ``Einstein zig-zag'' geometries \cite{Collett:2016, Dux:2025}. Modeling these systems requires computationally expensive multi-plane ray tracing and introduces additional free parameters. More critically, this intermediate source introduces its own independent, unmeasurable mass-sheet degeneracy. Breaking this secondary degeneracy requires virtually impossible spatially resolved kinematics, severely limiting the robustness of derived cosmological constraints. To put the magnitude of this systematic into perspective, literature explicitly modelling genuine DSPLs (such as SDSSJ0946+1006) demonstrates that the intermediate source is not a passive tracer, but a secondary deflector with a non-negligible Einstein radius (e.g., $\theta_{E} \approx 0.15''$; \cite{Collett:2014, Collett:2016}). As shown by these studies, failing to account for this compound lensing effect biases the inferred deflector mass profile and geometric distance ratios by several percent. Given the sensitivity of the dark energy equation of state to these ratios, such an unmodeled systematic bias would severely degrade the cosmological constraining power of any standard DSPL sample. PDSPLs circumvent this bottleneck by design. By pairing two \textit{independent} single-source plane lenses, there is no intermediate mass to model and no secondary mass-sheet degeneracy to break. This cleanly isolates the cosmological signal in the angular diameter distance ratios, free from the confounding effects of lens-lens coupling. Ultimately, avoiding the intermediate source MST ensures more robust constraints and drastically simplifies likelihood evaluation—a crucial optimization for processing the $\mathcal{O}(10^5)$ systems expected from next-generation surveys.

\subsection{Observational Yield and Survey Constraints}
\label{sec:discussion:observational_constraints}
The statistical power of the PDSPL methodology is fundamentally tied to the number of detectable lens systems. Consequently, the constraints forecasted in this work are sensitive to the practical observational realities of upcoming wide-field surveys.

\noindent \underline{\textit{Sensitivity to Survey Seeing Conditions:}}
The baseline forecast assumes optimal stacking with $s_{\rm opt} = 0.5''$, representing an optimistic but physically motivated scenario where strong lenses are discovered from stacks of the best-seeing epochs \citep{Collett:2015}. Under median survey seeing ($s_{\rm med} \approx 0.7''$), the detectable lens sample reduces to $\sim 55,000$ systems (Section~\ref{sec:slsim_lenses}), and the number of deflector pairs scales accordingly. Assuming the constraining power scales as $\sigma(w_0) \propto N_{\rm pairs}^{-1/2}$, this corresponds to a degradation of $\sim 1.4\times$, yielding $\sigma(w_0) \sim 0.62$ without an $\Omega_{\rm m}$ prior. The baseline $s_{\rm opt} = 0.5''$ result should therefore be understood as an upper bound on the statistical performance, with the median-seeing case representing a plausible conservative scenario.

\noindent \underline{\textit{Detectability of Compact Lenses and Space-Based Synergy:}}
While our theoretical detectability criteria in Section~\ref{sec:slsim_lenses} allow for inclusion of lenses down to $\theta_E \sim 0.25''$ (Figure~\ref{fig:slsim_corner_GGL}), identifying such compact systems solely from ground-based imaging is highly ambitious. Automated search algorithms typically degrade near the seeing limit, making $\theta_E \lesssim 0.5''$ lenses exceptionally challenging to confidently identify \cite{Herle:2024, Canameras:2024}. However, morphological verification of these smaller systems will become feasible using optimally drizzled data and, crucially, high-resolution space-based imaging \cite{Silver:2025}. Since the Roman High Latitude Wide Area Survey (HLWAS) and Euclid Wide Survey heavily overlap with the LSST footprint \citep{Capak:2019, ROTAC:2025}, leveraging these space-based observations will be essential to confirm compact lens candidates and preserve the large statistical volume required for the PDSPL methodology. Moreover these space based observatories may provide more precise measurements of the Einstein radii, thus helping with the cosmology.

\subsection{Optimizing Deflector Pairing}\label{sec:discussion:defl_pairing}
The constraining power of the PDSPL method is defined by the "noise floor" introduced by deflector dissimilarity. Our analysis highlights three avenues to suppress this noise and improve the fidelity of the observable $\beta_{\rm E}$.

\noindent \underline{\textit{Limitations of the Simulated Noise Floor:}}
Our forecast relies on the noise floor $\sigma_{\beta_{\rm E},\mathcal D}$ derived from the simulated lens population. This population model may not capture all correlations among the photometric, structural, and total-mass properties of real galaxies. For example, observations of the SLACS lens sample indicate that the inferred total-density power-law slope $\gamma_{\rm pl}$ correlates with galaxy structure, including effective radius, which is one of our pairing observables \citep[see Figure~5 in][]{Auger:2010}. In our simulations, however, $\gamma_{\rm pl}$ is drawn independently of the photometric and structural quantities used for pairing. Consequently, the slopes of the two paired deflectors show little correlation, as illustrated in Figure~\ref{fig:pairing_scatter_quality}. This is an expected consequence of the adopted population model and indicates that our pairing procedure does not preferentially match deflectors in $\gamma_{\rm pl}$.

The individual simulated slopes are not propagated into the Einstein radii used to calibrate $\sigma_{\beta_{\rm E},\mathcal D}$, because those radii are evaluated using the SIS-equivalent convention described in Section~\ref{sec:slsim_lenses}. The present calibration therefore cannot determine how correlations between $\gamma_{\rm pl}$ and the pairing observables would change the pair-differential scatter in real lenses. Such correlations could cause photometrically matched deflectors to have more similar slopes, but additional covariance between photometry, mass normalization, and other total-mass properties could instead increase or otherwise alter the residual scatter. 
Consequently, we cannot conclude that the simulated noise floor is necessarily conservative. The relation between the pairing observables and the residual Einstein-radius-ratio scatter must ultimately be validated or inferred from the observed lens population. Within our simulated samples, selecting the brighter and more massive \textit{4MOST} deflectors reduces the residual scatter, demonstrating that sample selection can trade statistical volume for greater per-pair fidelity.

\noindent \underline{\textit{Refining Matches with Spectroscopy:}}
While the PDSPL method does not fundamentally rely on spectroscopic data, incorporating precise spectroscopic redshifts ($z_{D}^{\rm spec}$) serves as a powerful tool to further suppress the intrinsic pairing scatter. As demonstrated in our analysis (see Figure~\ref{fig:beta_E_vs_D}), replacing photo-z estimates with spectroscopic redshifts significantly improves the matching fidelity. This improvement arises because the Einstein radius ratio ($\beta_{\rm E}$) is highly sensitive to the angular diameter distances of the lenses; eliminating photo-z uncertainties removes a major source of redshift-induced geometric uncertainty. Consequently, obtaining $z_{D}^{\rm spec}$ measurements for the full LSST Y10 deflector population would be highly beneficial—it would simultaneously lower the systematic pairing noise floor and remove the photo-z error budget from the likelihood evaluation, yielding sharper constraints on the dark energy equation of state (as detailed in Appendix~\ref{app:lsst_y10_photo-z_vs_spec-z}). Interestingly, we find that incorporating central velocity dispersion ($\sigma_{v,D}$) constraints alongside $z_{D}^{\rm spec}$ yields no further significant reduction in the scatter. This indicates that, within the simulated 4MOST subsets considered here, precise redshift matching combined with photometric observables may be sufficient to achieve the pairing fidelity found in our analysis. Available kinematic information can nevertheless provide an additional observable for refining and validating the deflector pairing. Achieving this for the full LSST Y10 population will require dedicated wide-field spectroscopic campaigns. Imminent surveys such as WAVES@4MOST \cite{Kaur_WAVES_4MOST:2025} and, on a 10--15 year timescale, proposed facilities like the 12 metre Wide-field Spectroscopic Telescope \citep[WST;][]{WST2024} with $\sim$30,000 fibres represent realistic pathways to delivering spectroscopic redshifts for the full LSST strong lens population within the survey lifetime.

\noindent \underline{\textit{Anisotropic Dissimilarity Weighting:}}
In our current formulation (Equation~\ref{eqn:dissimilarity_parameter}), the dissimilarity parameter $\mathcal{D}_{\rm deflector}$ is calculated as an unweighted Euclidean distance in normalized parameter space. This implicitly treats a $1\sigma$ deviation in redshift as equivalent to a $1\sigma$ deviation in magnitude. However, the sensitivity of $\beta_{\rm E}$ to these parameters is physically anisotropic; variations in deflector redshift directly alter the angular diameter distance ratios, impacting $\beta_{\rm E}$ more severely than photometric fluctuations. While a simple weighted distance metric ($\mathcal{D}^2 = \sum w_k (\Delta P_k)^2$) could account for these differential sensitivities, it ignores the fact that key photometric/spectroscopic observables may possess significant intrinsic covariances \cite{Bernardi:2003}. Consequently, future iterations of this methodology should employ a fully generalized Mahalanobis-like distance metric \cite{Mahalanobis:1936}. This approach would incorporate both the Fisher information content of each parameter relative to $\beta_{\rm E}$ and the physical correlations between the observables themselves, yielding a statistically optimal pairing strategy.

\noindent \underline{\textit{Morphological Matching via Deep Learning:}} 
Finally, our current kD-Tree matching relies on scalar catalog observables (redshift, magnitude, radius, colors or velocity dispersion), discarding the rich morphological information contained in pixel-level data. With the advent of high-resolution imaging from space based surveys like Euclid and Roman, future implementations could employ deep learning techniques, such as contrastive learning, to perform "image-level" matching. These methods can identify morphological analogs by capturing 
higher-order structural features that escape parametric fits, 
increasing the effective sample size of high-fidelity pairs.

\subsection{Inference of $\sigma_{\beta_E,\mathcal D}$}
In the baseline forecasts, the coefficients of the $\sigma_{\beta_E,\mathcal D}$ relation are fixed from the \texttt{SLSim} pipeline. For observational application, these coefficients can instead be inferred jointly with the cosmological and deflector-population parameters. Appendix~\ref{app:inferred_scatter} shows that the LSST Y10 mock PDSPL sample constrains both coefficients with only a modest degradation of the cosmological constraints. This result is conditional on the adopted linear, zero-mean Gaussian scatter model; it does not by itself validate this exact functional form for real lenses. Observational analyses will therefore require posterior-predictive tests and more flexible scatter models where needed. 

\subsection{Synergy with External Datasets}
As shown in Figure~\ref{fig:forecast_w0waCDM_DSPL_PDSPL_y10}, the PDSPL method is most powerful when combined with external probes that constrain the matter density $\Omega_{\rm m}$, such as the Cosmic Microwave Background (e.g., Planck \cite{Planck:2020}), Baryon Acoustic Oscillations (e.g., DESI \citep{DESI:BAO:2025}), or Type Ia Supernovae (e.g., Pantheon+ \citep{Pantheon+:Brout:2022}). The synergy between the geometric probe of PDSPLs (sensitive to expansion history) and the growth-of-structure probes (sensitive to $\Omega_{\rm m}$) highlights the role of strong lensing tomography as a complementary pillar of precision cosmology.

\subsection{Systematic Uncertainties}
While the PDSPL framework bypasses the modeling complexities of true multi-source plane systems, it is susceptible to unique systematic uncertainties that collectively dictate its irreducible ``noise floor'', $\sigma_{\beta_E, \mathcal{D}}$:

\noindent \underline{\textit{Photometric-to-Mass Mapping:}} The assumption that similar photometry implies identical mass profiles carries intrinsic, irreducible scatter (e.g., varying dark matter fractions) that translates directly into uncertainty on $\beta_{\rm E}$.

\noindent \underline{\textit{Environmental Variations:}} Pairing structurally identical galaxies that reside in different local environments (e.g., field vs. cluster) introduces unmodeled differences in external convergence ($\kappa_{\rm ext}$) and shear, elevating the baseline $1\%$ line-of-sight scatter \cite{Johnson:2025}.

\noindent\underline{\textit{Photo-z Errors, Calibration, and Galaxy Evolution:}}
Photo-z uncertainties affect both deflector pairing and the inferred distance ratios. Large uncertainties may pair galaxies across different cosmic epochs, introducing additional mismatch through galaxy evolution, while an unrecognized redshift bias can systematically shift the cosmological inference and does not average down with sample size. Representative spectroscopic data will therefore be required to calibrate the photo-z distributions, particularly for blended lensed sources. We quantify the effects of degraded precision and systematic bias in Appendix~\ref{app:photoz_precision_bias}.

\noindent \underline{\textit{Selection Bias:}} Survey detectability criteria (e.g., minimum $\theta_E$, SNR limits, etc.)~\cite{Collett:2015} preferentially select lenses that scatter upward in deflection angle. Crucially for PDSPLs, this effect is redshift-dependent: an identical deflector requires a more extreme upward mass scatter to clear the detectability threshold if its source is at a lower redshift due to lower geometric lensing efficiency. This Malmquist-like bias systematically skews the inferred population mass parameters ($\overline{\gamma}_{\rm pl}$, $\overline{\lambda}_{\rm MST}$), directly propagating errors into the cosmological constraints unless rigorously corrected via forward-modeling.

\noindent \underline{\textit{Astrophysical Contaminants and Profile Assumptions:}} 
The PDSPL framework relies on the stability of deflector mass profiles and clean photometry. Photometric contamination from luminous Active Galactic Nuclei (AGN) is expected to be minimal in massive early-type galaxies \citep{Kauffmann:2003}, and any such anomalous systems would naturally be rejected during color-based pairing. Structurally, while massive ellipticals may possess flattened inner stellar cores, the \textit{total} enclosed mass at the typical Einstein radius remains robustly well-described by a near-isothermal power law \citep{Koopmans:2009}. Finally, for spectroscopic pairing, typical fiber apertures (e.g., $1.5''$ for 4MOST \cite{4MOST:2012}) subtend kiloparsec scales at typical deflector redshifts. This far exceeds the supermassive black hole (SMBH) sphere of influence, ensuring that measured velocity dispersions trace the global stellar halo and remain unaffected by the intrinsic scatter in SMBH scaling relations.

\subsection{Comparison with Literature and the Impact of Degeneracies}
It is instructive to place the PDSPL methodology in the context of other strong lensing probes of dark energy, particularly those leveraging the statistical power of upcoming surveys like LSST and Euclid. Recent forecasts for ``Lensing + Kinematics'' demonstrate the potential of combining Einstein radius measurements with stellar kinematics for large populations. \cite{Li:2024} show that a sample of 10,000 lenses from Euclid/4MOST can constrain the equation of state to $\sigma(w) \approx 0.11$ in Flat $w$CDM. However, this method relies heavily on the absolute calibration of the kinematic measurements; constraints degrade significantly if the velocity dispersion precision worsens (e.g., from 10 km/s) or if there are systematic uncertainties in the stellar anisotropy models. Similarly, Time Delay Cosmography (TDC) offers robust constraints; \cite{Erickson:2025} forecast that a sample of $\sim800$ lensed AGN from LSST can yield a dark energy Figure of Merit (FoM) of 6.7. While powerful, TDC requires resource-intensive long-term monitoring campaigns to achieve high-precision time delays and is sensitive to the "mass-sheet transformation" degeneracy.

In contrast, the PDSPL framework leverages relative measurements between pairs of deflector galaxies. By comparing carefully matched deflectors, we aim to reduce the sensitivity to differences in their mass normalizations, although these differences do not cancel exactly as they do for a true DSPL. The remaining mass-normalization mismatch is accounted for statistically through the calibrated deflector-pairing scatter. While PDSPLs provide looser constraints per system than TDC or Lensing+Dynamics, the sheer volume of available pairs allows for a self-calibration of the lens population hyperparameters (such as $\lambda_{\rm MST}$ and $\gamma_{\rm pl}$) that is difficult to achieve with smaller samples.

Previous analyses of Double Source Plane Lenses (DSPLs), such as \cite{Divij:2023}, have demonstrated that a sample of 87 DSPLs can constrain the dark energy equation of state to $w_0 = -1.09_{-0.32}^{+0.31}$ within a Flat $w_0w_a$CDM framework, provided that the mass-sheet parameter $\lambda_{\rm MST}$ and density slope $\gamma_{\rm pl}$ are fixed or tight priors are assumed. However, we demonstrate that when the means and intrinsic scatters of these quantities are varied within the hierarchical population model, the constraining power of the limited DSPL sample is substantially degraded due to the mass-sheet degeneracy. We find that the PDSPL methodology---applied to the photometrically selected LSST Y10 sample---recovers constraints of $w_0 = -0.99^{+0.45}_{-0.44}$ with uniform priors. Crucially, this precision is achieved while simultaneously constraining the population hyperparameters.

Finally, compared to weak lensing tomography, our constraints on $w_0$ for the LSST 10-year PDSPL sample with $\Omega_{\rm m}$ prior ($w_0 = -1.08^{+0.25}_{-0.32} \rightarrow \sigma(w_0) \sim 0.29$) are comparable to current Stage III cosmic shear surveys, such as the Dark Energy Survey (DES) Y3 $3\times2$pt analysis which yields $w_0 = -0.98^{+0.32}_{-0.20}$ \citep{DES_Y3:2022}. This demonstrates that the PDSPL technique can extract a cosmological signal from strong lensing that is competitive with established large-scale structure probes.

\section{Conclusion}\label{sec:conclusion}
In this work, we have introduced and characterized the framework of Pseudo Double Source Plane Lenses (PDSPLs) as a novel probe for dark energy. By relaxing the requirement for rare, multi-source plane alignments and instead pairing independent single-source plane lenses with self-similar deflectors, we unlock the statistical potential of the $\sim 10^5$ strong lenses expected from upcoming wide-field surveys.

We derived the transformation of the double-source plane observable (the Einstein radius ratio, $\beta_{\rm E}$) under the Mass-Sheet Degeneracy (MSD). We explicitly showed that $\beta_{\rm E}$ is degenerate with the internal mass sheet parameter $\lambda_{\rm MST}$ and the power-law slope $\gamma_{\rm pl}$. This highlights that precise cosmological inference requires either breaking the MSD or marginalizing over the uncertain deflector properties within a hierarchical population model.

Using simulations from the \texttt{SLSim} pipeline, we quantified the "noise floor" introduced by pairing distinct deflectors. The simulation-derived scatter is conditional on the adopted galaxy population and Einstein-radius prescription. Correlations among photometric observables, mass normalization, density slope, and other mass-profile properties in the real lens population could either reduce or increase the residual scatter. The deflector-mismatch relation must therefore be validated or inferred from observational data. We found that including spectroscopic information when identifying near-identical deflectors significantly reduces the intrinsic scatter in $\beta_{\rm E}$ (from $\sim 19\%$ to $\sim 8\%$). Nevertheless, under the assumed unbiased redshift model, the much larger photometric LSST sample provides tighter statistical constraints than the smaller, high-fidelity spectroscopic subset achievable with campaigns like 4MOST. Spectroscopic calibration remains necessary to validate the redshift distributions and control the accuracy of the cosmological inference.

The cosmological forecasts presented here demonstrate that PDSPLs can yield constraints on the dark energy equation of state parameters that surpass those from current samples of genuine DSPLs. Specifically, we found that:
\begin{itemize}
    \item A limited sample of 500 genuine DSPLs is severely limited by the mass-sheet degeneracy, yielding loose constraints of $w_0 = -1.39^{+0.73}_{-0.92}$ even with an $\Omega_{\rm m}$ prior, when marginalizing over mass profile parameters.
    \item In contrast, the photometrically selected PDSPL sample from LSST Y10 ($\sim 8.6\times10^4$ pairs) provides sufficient statistical leverage to constrain the hyperparameters of the hierarchical population model jointly with cosmology, achieving a precision of $\sigma(w_0) \sim 0.45$ in a Flat $w_0w_a$CDM cosmology assuming unbiased Gaussian photo-z uncertainties of $\sigma_z/(1+z)=0.03$. This constraint further tightens to $\sigma(w_0) \sim 0.29$ when combined with a Gaussian prior of $\mathcal{N}(0.3, 0.05)$ on the matter density ($\Omega_{\rm m}$), demonstrating the strong synergy between PDSPLs and external growth-of-structure probes.
    \item The large PDSPL sample allows the hierarchical population model to constrain the mass-profile population means and intrinsic scatters simultaneously, determining the mean internal mass sheet $\lambda_{\rm MST}$ and power-law slope $\gamma_{\rm pl}$ to a precision of $\sim 2\%$ compared to a precision of $\sim 5\%$ from 500 DSPLs.
\end{itemize}

To contextualise these results, our best constraint of $\sigma(w_0) \sim 0.29$ (with an $\Omega_{\rm m}$ prior) is comparable to current Stage III cosmic shear surveys such as DES Y3 \citep{DES_Y3:2022}. Together with other strong lensing probes of dark energy—such as lensing+kinematics \citep{Li:2024} and time-delay cosmography with lensed quasars \citep{Erickson:2025}—PDSPLs provide a complementary and statistically powerful avenue that exploits the galaxy-galaxy lens catalogs that LSST, Euclid, and Roman will naturally produce. We note that several additional systematic effects discussed in Section~\ref{sec:discussion} are not yet incorporated into these forecasts and will need to be addressed in future work. Ultimately, by converting the abundant population of standard galaxy--galaxy lenses into a tomographic dataset, the PDSPL formalism provides a statistics-driven probe of cosmic expansion in the era of next-generation surveys, although its application to observational data will require calibrated redshift distributions, validated deflector-mismatch models, and explicit control of additional systematic effects.

\acknowledgments

This paper has undergone a joint internal review in the LSST Dark Energy Science Collaboration (DESC) and the Strong Lensing Science Collaboration (SLSC). The internal reviewers were Tian Li and Adam Bolton, and they provided extremely valuable feedback that helped improve the robustness of the analysis and the draft.

\paragraph{Contribution Statement:}SB proposed the pseudo-DSPL concept, developed the initial formalism, and outlined the manuscript. PS generated the simulated lens populations, devised pairing technique, performed the cosmological forecasting and analysis, and wrote the final manuscript. NK contributed to the configuration and implementation of the \texttt{SLSim} pipeline for sample generation. SE, PH, PM, DDN, CT, and BW provided extensive feedback on the initial analysis and the manuscript.

\paragraph{Software:}The source code for this work is publicly available at the github repository \href{https://github.com/timedilatesme/pdspl-analysis}{pdspl-analysis}. This work makes use of the publicly available softwares: \href{https://github.com/LSST-strong-lensing/slsim}{\texttt{SLSim}} \cite{Khadka:2026}, \href{https://github.com/sibirrer/hierArc}{\texttt{hierarc}} \citep{Birrer:2020}, and \href{https://github.com/dfm/emcee}{\texttt{emcee}} \citep{emcee}.

\paragraph{Funding:}PS and SB are supported by DoE Grant DE-SC0026113. SB acknowledges support by Schmidt Futures. SE acknowledges funding from NSF GRFP 2021313357. ASB, SE and PM performed work under DOE Contract DE-AC02-76SF00515. TL is funded by the European Research Council (ERC) under the European Union’s Horizon 2020 research and innovation program (LensEra: grant agreement No 945536). TA acknowledges support from ANID-FONDECYT Regular Project 1240105 and the ANID BASAL project FB210003. GPS acknowledges support from The Royal Society, the Leverhulme Trust and the Science and Technology Facilities Council (grant number ST/X001296/1).

The DESC acknowledges ongoing support from the Institut National de 
Physique Nucl\'eaire et de Physique des Particules in France; the 
Science \& Technology Facilities Council in the United Kingdom; and the
Department of Energy and the LSST Discovery Alliance
in the United States.  DESC uses resources of the IN2P3 
Computing Center (CC-IN2P3--Lyon/Villeurbanne - France) funded by the 
Centre National de la Recherche Scientifique; the National Energy 
Research Scientific Computing Center, a DOE Office of Science User 
Facility supported by the Office of Science of the U.S.\ Department of
Energy under Contract No.\ DE-AC02-05CH11231; STFC DiRAC HPC Facilities, 
funded by UK BEIS National E-infrastructure capital grants; and the UK 
particle physics grid, supported by the GridPP Collaboration.  This 
work was performed in part under DOE Contract DE-AC02-76SF00515.



\bibliography{bibliography}

@ARTICLE{SchneiderSluse:2013,
       author = {{Schneider}, Peter and {Sluse}, Dominique},
        title = "{Mass-sheet degeneracy, power-law models and external convergence: Impact on the determination of the Hubble constant from gravitational lensing}",
      journal = {\aa},
         year = 2013,
        month = nov,
       volume = {559},
          eid = {A37},
        pages = {A37},
          doi = {10.1051/0004-6361/201321882},
archivePrefix = {arXiv},
       eprint = {1306.0901},
 primaryClass = {astro-ph.CO},
       adsurl = {https://ui.adsabs.harvard.edu/abs/2013A&A...559A..37S}
}

@ARTICLE{Birrer:2016,
       author = {{Birrer}, Simon and {Amara}, Adam and {Refregier}, Alexandre},
        title = "{The mass-sheet degeneracy and time-delay cosmography: analysis of the strong lens RXJ1131-1231}",
      journal = {\jcap},
         year = 2016,
        month = aug,
       volume = {2016},
       number = {8},
          eid = {020},
        pages = {020},
          doi = {10.1088/1475-7516/2016/08/020},
archivePrefix = {arXiv},
       eprint = {1511.03662},
 primaryClass = {astro-ph.CO},
       adsurl = {https://ui.adsabs.harvard.edu/abs/2016JCAP...08..020B}
}

@ARTICLE{Sonnenfeld:2018,
       author = {{Sonnenfeld}, Alessandro},
        title = "{On the choice of lens density profile in time delay cosmography}",
      journal = {\mnras},
         year = 2018,
        month = mar,
       volume = {474},
       number = {4},
        pages = {4648-4659},
          doi = {10.1093/mnras/stx3105},
archivePrefix = {arXiv},
       eprint = {1710.05925},
 primaryClass = {astro-ph.CO},
       adsurl = {https://ui.adsabs.harvard.edu/abs/2018MNRAS.474.4648S}
}

@ARTICLE{Kochanek:2020,
       author = {{Kochanek}, C.~S.},
        title = "{Overconstrained gravitational lens models and the Hubble constant}",
      journal = {\mnras},
         year = 2020,
        month = apr,
       volume = {493},
       number = {2},
        pages = {1725-1735},
          doi = {10.1093/mnras/staa344},
archivePrefix = {arXiv},
       eprint = {1911.05083},
 primaryClass = {astro-ph.CO},
       adsurl = {https://ui.adsabs.harvard.edu/abs/2020MNRAS.493.1725K}
}

@ARTICLE{DES+KiDS:2023,
       author = {{Dark Energy Survey and Kilo-Degree Survey Collaboration} and {Abbott}, T.~M.~C. and {Aguena}, M. and {Alarcon}, A. and {Alves}, O. and {Amon}, A. and {Andrade-Oliveira}, F. and {Asgari}, M. and {Avila}, S. and {Bacon}, D. and {Bechtol}, K. and {Becker}, M.~R. and {Bernstein}, G.~M. and {Bertin}, E. and {Bilicki}, M. and {Blazek}, J. and {Bocquet}, S. and {Brooks}, D. and {Burger}, P. and {Burke}, D.~L. and {Camacho}, H. and {Campos}, A. and {Carnero Rosell}, A. and {Carrasco Kind}, M. and {Carretero}, J. and {Castander}, F.~J. and {Cawthon}, R. and {Chang}, C. and {Chen}, R. and {Choi}, A. and {Conselice}, C. and {Cordero}, J. and {Crocce}, M. and {da Costa}, L.~N. and {da Silva Pereira}, M.~E. and {Dalal}, R. and {Davis}, C. and {de Jong}, J.~T.~A. and {DeRose}, J. and {Desai}, S. and {Diehl}, H.~T. and {Dodelson}, S. and {Doel}, P. and {Doux}, C. and {Drlica-Wagner}, A. and {Dvornik}, A. and {Eckert}, K. and {Eifler}, T.~F. and {Elvin-Poole}, J. and {Everett}, S. and {Fang}, X. and {Ferrero}, I. and {Fert{\'e}}, A. and {Flaugher}, B. and {Friedrich}, O. and {Frieman}, J. and {Garc{\'\i}a-Bellido}, J. and {Gatti}, M. and {Giannini}, G. and {Giblin}, B. and {Gruen}, D. and {Gruendl}, R.~A. and {Gutierrez}, G. and {Harrison}, I. and {Hartley}, W.~G. and {Herner}, K. and {Heymans}, C. and {Hildebrandt}, H. and {Hinton}, S.~R. and {Hoekstra}, H. and {Hollowood}, D.~L. and {Honscheid}, K. and {Huang}, H. and {Huff}, E.~M. and {Huterer}, D. and {James}, D.~J. and {Jarvis}, M. and {Jeffrey}, N. and {Jeltema}, T. and {Joachimi}, B. and {Joudaki}, S. and {Kannawadi}, A. and {Krause}, E. and {Kuehn}, K. and {Kuijken}, K. and {Kuropatkin}, N. and {Lahav}, O. and {Leget}, P. -F. and {Lemos}, P. and {Li}, S. -S. and {Li}, X. and {Liddle}, A.~R. and {Lima}, M. and {Lin}, C. -A. and {Lin}, H. and {MacCrann}, N. and {Mahony}, C. and {Marshall}, J.~L. and {McCullough}, J. and {Mena-Fern{\'a}ndez}, J. and {Menanteau}, F. and {Miquel}, R. and {Mohr}, J.~J. and {Muir}, J. and {Myles}, J. and {Napolitano}, N. and {Navarro-Alsina}, A. and {Ogando}, R.~L.~C. and {Palmese}, A. and {Pandey}, S. and {Park}, Y. and {Paterno}, M. and {Peacock}, J.~A. and {Petravick}, D. and {Pieres}, A. and {Plazas Malag{\'o}n}, A.~A. and {Porredon}, A. and {Prat}, J. and {Radovich}, M. and {Raveri}, M. and {Reischke}, R. and {Robertson}, N.~C. and {Rollins}, R.~P. and {Romer}, A.~K. and {Roodman}, A. and {Rykoff}, E.~S. and {Samuroff}, S. and {S{\'a}nchez}, C. and {Sanchez}, E. and {Sanchez}, J. and {Schneider}, P. and {Secco}, L.~F. and {Sevilla-Noarbe}, I. and {Shan}, H. -Y. and {Sheldon}, E. and {Shin}, T. and {Sif{\'o}n}, C. and {Smith}, M. and {Soares-Santos}, M. and {St{\"o}lzner}, B. and {Suchyta}, E. and {Swanson}, M.~E.~C. and {Tarle}, G. and {Thomas}, D. and {To}, C. and {Troxel}, M.~A. and {Tr{\"o}ster}, T. and {Tutusaus}, I. and {van den Busch}, J.~L. and {Varga}, T.~N. and {Walker}, A.~R. and {Weaverdyck}, N. and {Wechsler}, R.~H. and {Weller}, J. and {Wiseman}, P. and {Wright}, A.~H. and {Yanny}, B. and {Yin}, B. and {Yoon}, M. and {Zhang}, Y. and {Zuntz}, J.},
        title = "{DES Y3 + KiDS-1000: Consistent cosmology combining cosmic shear surveys}",
      journal = {The Open Journal of Astrophysics},
         year = 2023,
        month = oct,
       volume = {6},
          eid = {36},
        pages = {36},
          doi = {10.21105/astro.2305.17173},
archivePrefix = {arXiv},
       eprint = {2305.17173},
 primaryClass = {astro-ph.CO},
       adsurl = {https://ui.adsabs.harvard.edu/abs/2023OJAp....6E..36D}
}

@ARTICLE{Johnson:2025,
       author = {{Johnson}, Daniel and {Collett}, Thomas and {Li}, Tian and {Fleury}, Pierre},
        title = "{Line-of-sight effects on double source plane lenses}",
      journal = {arXiv e-prints},
         year = 2025,
        month = jan,
          eid = {arXiv:2501.17153},
        pages = {arXiv:2501.17153},
          doi = {10.48550/arXiv.2501.17153},
archivePrefix = {arXiv},
       eprint = {2501.17153},
 primaryClass = {astro-ph.CO},
       adsurl = {https://ui.adsabs.harvard.edu/abs/2025arXiv250117153J}
}

@ARTICLE{Dux:2025,
       author = {{Dux}, F. and {Millon}, M. and {Lemon}, C. and {Schmidt}, T. and {Courbin}, F. and {Shajib}, A.~J. and {Treu}, T. and {Birrer}, S. and {Wong}, K.~C. and {Agnello}, A. and {Andrade}, A. and {Galan}, A. and {Hjorth}, J. and {Paic}, E. and {Schuldt}, S. and {Schweinfurth}, A. and {Sluse}, D. and {Smette}, A. and {Suyu}, S.~H.},
        title = "{J1721+8842: The first Einstein zigzag lens}",
      journal = {\aa},
         year = 2025,
        month = feb,
       volume = {694},
          eid = {A300},
        pages = {A300},
          doi = {10.1051/0004-6361/202452970},
archivePrefix = {arXiv},
       eprint = {2411.04177},
 primaryClass = {astro-ph.CO},
       adsurl = {https://ui.adsabs.harvard.edu/abs/2025A&A...694A.300D}
}

@ARTICLE{Collett:2016,
       author = {{Collett}, Thomas E. and {Bacon}, David J.},
        title = "{Compound lensing: Einstein zig-zags and high-multiplicity lensed images}",
      journal = {\mnras},
         year = 2016,
        month = feb,
       volume = {456},
       number = {2},
        pages = {2210-2220},
          doi = {10.1093/mnras/stv2791},
archivePrefix = {arXiv},
       eprint = {1510.00242},
 primaryClass = {astro-ph.CO},
       adsurl = {https://ui.adsabs.harvard.edu/abs/2016MNRAS.456.2210C}
}

@ARTICLE{Collett:2014,
       author = {{Collett}, Thomas E. and {Auger}, Matthew W.},
        title = "{Cosmological constraints from the double source plane lens SDSSJ0946+1006}",
      journal = {\mnras},
         year = 2014,
        month = sep,
       volume = {443},
       number = {2},
        pages = {969-976},
          doi = {10.1093/mnras/stu1190},
archivePrefix = {arXiv},
       eprint = {1403.5278},
 primaryClass = {astro-ph.CO},
       adsurl = {https://ui.adsabs.harvard.edu/abs/2014MNRAS.443..969C}
}

@ARTICLE{Caminha:2022,
       author = {{Caminha}, G.~B. and {Suyu}, S.~H. and {Grillo}, C. and {Rosati}, P.},
        title = "{Galaxy cluster strong lensing cosmography. Cosmological constraints from a sample of regular galaxy clusters}",
      journal = {\aa},
         year = 2022,
        month = jan,
       volume = {657},
          eid = {A83},
        pages = {A83},
          doi = {10.1051/0004-6361/202141994},
archivePrefix = {arXiv},
       eprint = {2110.06232},
 primaryClass = {astro-ph.CO},
       adsurl = {https://ui.adsabs.harvard.edu/abs/2022A&A...657A..83C}
}

@ARTICLE{Birrer:2020,
       author = {{Birrer}, S. and {Shajib}, A.~J. and {Galan}, A. and {Millon}, M. and {Treu}, T. and {Agnello}, A. and {Auger}, M. and {Chen}, G.~C. -F. and {Christensen}, L. and {Collett}, T. and {Courbin}, F. and {Fassnacht}, C.~D. and {Koopmans}, L.~V.~E. and {Marshall}, P.~J. and {Park}, J. -W. and {Rusu}, C.~E. and {Sluse}, D. and {Spiniello}, C. and {Suyu}, S.~H. and {Wagner-Carena}, S. and {Wong}, K.~C. and {Barnab{\`e}}, M. and {Bolton}, A.~S. and {Czoske}, O. and {Ding}, X. and {Frieman}, J.~A. and {Van de Vyvere}, L.},
        title = "{TDCOSMO. IV. Hierarchical time-delay cosmography - joint inference of the Hubble constant and galaxy density profiles}",
      journal = {\aa},
         year = 2020,
        month = nov,
       volume = {643},
          eid = {A165},
        pages = {A165},
          doi = {10.1051/0004-6361/202038861},
archivePrefix = {arXiv},
       eprint = {2007.02941},
 primaryClass = {astro-ph.CO},
       adsurl = {https://ui.adsabs.harvard.edu/abs/2020A&A...643A.165B}
}

@ARTICLE{Falco:1985,
       author = {{Falco}, E.~E. and {Gorenstein}, M.~V. and {Shapiro}, I.~I.},
        title = "{On model-dependent bounds on H 0 from gravitational images : application to Q 0957+561 A, B.}",
      journal = {\apjl},
         year = 1985,
        month = feb,
       volume = {289},
        pages = {L1-L4},
          doi = {10.1086/184422},
       adsurl = {https://ui.adsabs.harvard.edu/abs/1985ApJ...289L...1F}
}

@ARTICLE{Schneider:2014,
       author = {{Schneider}, Peter},
        title = "{Can one determine cosmological parameters from multi-plane strong lens systems?}",
      journal = {\aa},
         year = 2014,
        month = aug,
       volume = {568},
          eid = {L2},
        pages = {L2},
          doi = {10.1051/0004-6361/201424450},
archivePrefix = {arXiv},
       eprint = {1406.6152},
 primaryClass = {astro-ph.CO},
       adsurl = {https://ui.adsabs.harvard.edu/abs/2014A&A...568L...2S}
}

@ARTICLE{Gorenstein:1988,
       author = {{Gorenstein}, M.~V. and {Falco}, E.~E. and {Shapiro}, I.~I.},
        title = "{Degeneracies in Parameter Estimates for Models of Gravitational Lens Systems}",
      journal = {\apj},
         year = 1988,
        month = apr,
       volume = {327},
        pages = {693},
          doi = {10.1086/166226},
       adsurl = {https://ui.adsabs.harvard.edu/abs/1988ApJ...327..693G}
}

@ARTICLE{Shajib:2024,
       author = {{Shajib}, Anowar J. and {Smith}, Graham P. and {Birrer}, Simon and {Verma}, Aprajita and {Arendse}, Nikki and {Collett}, Thomas and {Daylan}, Tansu and {Serjeant}, Stephen and {LSST Strong Lensing Science Collaboration}},
        title = "{Strong gravitational lenses from the Vera C. Rubin Observatory}",
      journal = {Philosophical Transactions of the Royal Society of London Series A},
         year = 2025,
        month = may,
       volume = {383},
       number = {2295},
          eid = {20240117},
        pages = {20240117},
          doi = {10.1098/rsta.2024.0117},
archivePrefix = {arXiv},
       eprint = {2406.08919},
 primaryClass = {astro-ph.CO},
       adsurl = {https://ui.adsabs.harvard.edu/abs/2025RSPTA.38340117S}
}

@misc{Li:2025,
      title={Euclid Quick Data Release (Q1). The Strong Lensing Discovery Engine D -- Double-source-plane lens candidates}, 
      author={Euclid Collaboration and T. Li and T. E. Collett and M. Walmsley and N. E. P. Lines and K. Rojas and J. W. Nightingale and W. J. R. Enzi and L. A. Moustakas and C. Krawczyk and R. Gavazzi and G. Despali and P. Holloway and S. Schuldt and F. Courbin and R. B. Metcalf and D. J. Ballard and A. Verma and B. Clément and H. Degaudenzi and A. Melo and J. A. Acevedo Barroso and L. Leuzzi and A. Manjón-García and R. Pearce-Casey and D. Sluse and C. Tortora and R. Massey and G. Mahler and A. More and N. Aghanim and B. Altieri and A. Amara and S. Andreon and N. Auricchio and H. Aussel and C. Baccigalupi and M. Baldi and A. Balestra and S. Bardelli and P. Battaglia and R. Bender and F. Bernardeau and A. Biviano and A. Bonchi and E. Branchini and M. Brescia and J. Brinchmann and S. Camera and G. Cañas-Herrera and V. Capobianco and C. Carbone and V. F. Cardone and J. Carretero and S. Casas and M. Castellano and G. Castignani and S. Cavuoti and K. C. Chambers and A. Cimatti and C. Colodro-Conde and G. Congedo and C. J. Conselice and L. Conversi and Y. Copin and H. M. Courtois and M. Cropper and A. Da Silva and G. De Lucia and A. M. Di Giorgio and C. Dolding and H. Dole and F. Dubath and C. A. J. Duncan and X. Dupac and S. Escoffier and M. Fabricius and M. Farina and R. Farinelli and F. Faustini and S. Ferriol and F. Finelli and S. Fotopoulou and M. Frailis and E. Franceschi and S. Galeotta and K. George and W. Gillard and B. Gillis and C. Giocoli and P. Gómez-Alvarez and J. Gracia-Carpio and B. R. Granett and A. Grazian and F. Grupp and S. V. H. Haugan and H. Hoekstra and W. Holmes and I. M. Hook and F. Hormuth and A. Hornstrup and P. Hudelot and K. Jahnke and M. Jhabvala and B. Joachimi and E. Keihänen and S. Kermiche and A. Kiessling and B. Kubik and M. Kümmel and M. Kunz and H. Kurki-Suonio and Q. Le Boulc'h and A. M. C. Le Brun and D. Le Mignant and S. Ligori and P. B. Lilje and V. Lindholm and I. Lloro and G. Mainetti and D. Maino and E. Maiorano and O. Mansutti and S. Marcin and O. Marggraf and M. Martinelli and N. Martinet and F. Marulli and S. Maurogordato and E. Medinaceli and S. Mei and Y. Mellier and M. Meneghetti and E. Merlin and G. Meylan and A. Mora and M. Moresco and L. Moscardini and R. Nakajima and C. Neissner and R. C. Nichol and S. -M. Niemi and C. Padilla and S. Paltani and F. Pasian and K. Pedersen and W. J. Percival and V. Pettorino and S. Pires and G. Polenta and M. Poncet and L. A. Popa and L. Pozzetti and F. Raison and R. Rebolo and A. Renzi and J. Rhodes and G. Riccio and E. Romelli and M. Roncarelli and R. Saglia and Z. Sakr and D. Sapone and B. Sartoris and J. A. Schewtschenko and M. Schirmer and P. Schneider and T. Schrabback and A. Secroun and G. Seidel and M. Seiffert and S. Serrano and P. Simon and C. Sirignano and G. Sirri and A. Spurio Mancini and L. Stanco and J. Steinwagner and P. Tallada-Crespí and A. N. Taylor and I. Tereno and N. Tessore and S. Toft and R. Toledo-Moreo and F. Torradeflot and I. Tutusaus and E. A. Valentijn and L. Valenziano and J. Valiviita and T. Vassallo and G. Verdoes Kleijn and A. Veropalumbo and Y. Wang and J. Weller and A. Zacchei and G. Zamorani and F. M. Zerbi and E. Zucca and V. Allevato and M. Ballardini and M. Bolzonella and E. Bozzo and C. Burigana and R. Cabanac and A. Cappi and D. Di Ferdinando and J. A. Escartin Vigo and L. Gabarra and M. Huertas-Company and J. Martín-Fleitas and S. Matthew and N. Mauri and A. Pezzotta and M. Pöntinen and C. Porciani and I. Risso and V. Scottez and M. Sereno and M. Tenti and M. Viel and M. Wiesmann and Y. Akrami and S. Alvi and I. T. Andika and S. Anselmi and M. Archidiacono and F. Atrio-Barandela and K. Benson and P. Bergamini and D. Bertacca and M. Bethermin and A. Blanchard and L. Blot and M. L. Brown and S. Bruton and A. Calabro and F. Caro and C. S. Carvalho and T. Castro and F. Cogato and A. R. Cooray and O. Cucciati and S. Davini and F. De Paolis and G. Desprez and A. Díaz-Sánchez and J. J. Diaz and S. Di Domizio and J. M. Diego and P. -A. Duc and A. Enia and Y. Fang and A. G. Ferrari and P. G. Ferreira and A. Finoguenov and A. Fontana and A. Franco and K. Ganga and J. García-Bellido and T. Gasparetto and V. Gautard and E. Gaztanaga and F. Giacomini and F. Gianotti and G. Gozaliasl and M. Guidi and C. M. Gutierrez and A. Hall and W. G. Hartley and C. Hernández-Monteagudo and H. Hildebrandt and J. Hjorth and M. Jauzac and J. J. E. Kajava and Y. Kang and V. Kansal and D. Karagiannis and K. Kiiveri and C. C. Kirkpatrick and S. Kruk and J. Le Graet and L. Legrand and M. Lembo and F. Lepori and G. Leroy and G. F. Lesci and J. Lesgourgues and T. I. Liaudat and A. Loureiro and J. Macias-Perez and G. Maggio and M. Magliocchetti and F. Mannucci and R. Maoli and C. J. A. P. Martins and L. Maurin and M. Migliaccio and M. Miluzio and P. Monaco and C. Moretti and G. Morgante and S. Nadathur and K. Naidoo and A. Navarro-Alsina and S. Nesseris and F. Passalacqua and K. Paterson and L. Patrizii and A. Pisani and D. Potter and S. Quai and M. Radovich and P. -F. Rocci and S. Sacquegna and M. Sahlén and D. B. Sanders and E. Sarpa and C. Scarlata and J. Schaye and A. Schneider and D. Sciotti and E. Sellentin and L. C. Smith and K. Tanidis and G. Testera and R. Teyssier and S. Tosi and A. Troja and M. Tucci and C. Valieri and A. Venhola and D. Vergani and G. Vernardos and G. Verza and P. Vielzeuf and N. A. Walton and J. Wilde and D. Scott},
      year={2025},
      eprint={2503.15327},
      archivePrefix={arXiv},
      primaryClass={astro-ph.GA},
      url={https://arxiv.org/abs/2503.15327}, 
}

@MISC{Wong:2017,
       author = {{Wong}, Kenneth},
        title = "{HST Imaging of the Eye of Horus, a Double Source Plane Gravitational Lens}",
 howpublished = {HST Proposal id.15325. Cycle 25},
         year = 2017,
        month = aug,
          eid = {15325},
        pages = {15325},
       adsurl = {https://ui.adsabs.harvard.edu/abs/2017hst..prop15325W}
}

@ARTICLE{Tanaka:2020,
       author = {{Tanaka}, Keigo and {Tsuji}, Ayumi and {Akamatsu}, Hiroki and {Chan}, J.~H.~H. and {Coupon}, Jean and {Egami}, Eiichi and {Finet}, Francois and {Fujimoto}, Ryuichi and {Ichinohe}, Yuto and {Jaelani}, Anton T. and {Lee}, Chien-Hsiu and {Mitsuishi}, Ikuyuki and {More}, Anupreeta and {More}, Surhud and {Oguri}, Masamune and {Okabe}, Nobuhiro and {Ota}, Naomi and {Rusu}, Cristian E. and {Sonnenfeld}, Alessandro and {Tanaka}, Masayuki and {Ueda}, Shutaro and {Wong}, Kenneth C.},
        title = "{X-ray study of the double source plane gravitational lens system Eye of Horus observed with XMM-Newton}",
      journal = {\mnras},
         year = 2020,
        month = jan,
       volume = {491},
       number = {3},
        pages = {3411-3418},
          doi = {10.1093/mnras/stz3188},
archivePrefix = {arXiv},
       eprint = {1911.04805},
 primaryClass = {astro-ph.GA},
       adsurl = {https://ui.adsabs.harvard.edu/abs/2020MNRAS.491.3411T}
}

@article{Gavazzi:2008,
       author = {{Gavazzi}, Rapha{\"e}l and {Treu}, Tommaso and {Koopmans}, L{\'e}on V.~E. and {Bolton}, Adam S. and {Moustakas}, Leonidas A. and {Burles}, Scott and {Marshall}, Philip J.},
        title = "{The Sloan Lens ACS Survey. VI. Discovery and Analysis of a Double Einstein Ring}",
      journal = {\apj},
         year = 2008,
        month = apr,
       volume = {677},
       number = {2},
        pages = {1046-1059},
          doi = {10.1086/529541},
archivePrefix = {arXiv},
       eprint = {0801.1555},
 primaryClass = {astro-ph},
       adsurl = {https://ui.adsabs.harvard.edu/abs/2008ApJ...677.1046G}
}

@ARTICLE{Tanaka:2016,
       author = {{Tanaka}, Masayuki and {Wong}, Kenneth C. and {More}, Anupreeta and {Dezuka}, Arsha and {Egami}, Eiichi and {Oguri}, Masamune and {Suyu}, Sherry H. and {Sonnenfeld}, Alessandro and {Higuchi}, Ryo and {Komiyama}, Yutaka and {Miyazaki}, Satoshi and {Onoue}, Masafusa and {Oyamada}, Shuri and {Utsumi}, Yousuke},
        title = "{A Spectroscopically Confirmed Double Source Plane Lens System in the Hyper Suprime-Cam Subaru Strategic Program}",
      journal = {\apjl},
         year = 2016,
        month = aug,
       volume = {826},
       number = {2},
          eid = {L19},
        pages = {L19},
          doi = {10.3847/2041-8205/826/2/L19},
archivePrefix = {arXiv},
       eprint = {1606.09363},
 primaryClass = {astro-ph.CO},
       adsurl = {https://ui.adsabs.harvard.edu/abs/2016ApJ...826L..19T}
}

@ARTICLE{Mozumdar:2025,
       author = {{Mozumdar}, Pritom and {Knabel}, Shawn and {Treu}, Tommaso and {Sonnenfeld}, Alessandro and {Shajib}, Anowar J. and {Cappellari}, Michele and {Nipoti}, Carlo},
        title = "{XXII. Accurate stellar velocity dispersions of the SL2S lens sample and the lensing mass fundamental plane}",
      journal = {arXiv e-prints},
         year = 2025,
        month = may,
          eid = {arXiv:2505.13962},
        pages = {arXiv:2505.13962},
          doi = {10.48550/arXiv.2505.13962},
archivePrefix = {arXiv},
       eprint = {2505.13962},
 primaryClass = {astro-ph.GA},
       adsurl = {https://ui.adsabs.harvard.edu/abs/2025arXiv250513962M}
}

@ARTICLE{Auger:2010,
       author = {{Auger}, M.~W. and {Treu}, T. and {Bolton}, A.~S. and {Gavazzi}, R. and {Koopmans}, L.~V.~E. and {Marshall}, P.~J. and {Moustakas}, L.~A. and {Burles}, S.},
        title = "{The Sloan Lens ACS Survey. X. Stellar, Dynamical, and Total Mass Correlations of Massive Early-type Galaxies}",
      journal = {\apj},
         year = 2010,
        month = nov,
       volume = {724},
       number = {1},
        pages = {511-525},
          doi = {10.1088/0004-637X/724/1/511},
archivePrefix = {arXiv},
       eprint = {1007.2880},
 primaryClass = {astro-ph.CO},
       adsurl = {https://ui.adsabs.harvard.edu/abs/2010ApJ...724..511A}
}

@ARTICLE{Faber:1976,
       author = {{Faber}, S.~M. and {Jackson}, R.~E.},
        title = "{Velocity dispersions and mass-to-light ratios for elliptical galaxies.}",
      journal = {\apj},
         year = 1976,
        month = mar,
       volume = {204},
        pages = {668-683},
          doi = {10.1086/154215},
       adsurl = {https://ui.adsabs.harvard.edu/abs/1976ApJ...204..668F}
}

@ARTICLE{Kormendy:1977,
       author = {{Kormendy}, J.},
        title = "{Brightness distributions in compact and normal galaxies. II. Structure parameters of the spheroidal component.}",
      journal = {\apj},
         year = 1977,
        month = dec,
       volume = {218},
        pages = {333-346},
          doi = {10.1086/155687},
       adsurl = {https://ui.adsabs.harvard.edu/abs/1977ApJ...218..333K}
}

@ARTICLE{Djorgovski:1987,
       author = {{Djorgovski}, S. and {Davis}, Marc},
        title = "{Fundamental Properties of Elliptical Galaxies}",
      journal = {\apj},
         year = 1987,
        month = feb,
       volume = {313},
        pages = {59},
          doi = {10.1086/164948},
       adsurl = {https://ui.adsabs.harvard.edu/abs/1987ApJ...313...59D}
}

@ARTICLE{Dressler:1987,
       author = {{Dressler}, Alan and {Lynden-Bell}, Donald and {Burstein}, David and {Davies}, Roger L. and {Faber}, S.~M. and {Terlevich}, Roberto and {Wegner}, Gary},
        title = "{Spectroscopy and Photometry of Elliptical Galaxies. I. New Distance Estimator}",
      journal = {\apj},
         year = 1987,
        month = feb,
       volume = {313},
        pages = {42},
          doi = {10.1086/164947},
       adsurl = {https://ui.adsabs.harvard.edu/abs/1987ApJ...313...42D}
}

@ARTICLE{skypy:2021,
       author = {{Amara}, Adam and {de la Bella}, Lucia and {Birrer}, Simon and {Bridle}, Sarah and {Cordero}, Juan and {Favole}, Ginevra and {Harrison}, Ian and {Harry}, Ian and {Hartley}, William and {Krawczyk}, Coleman and {Lundgren}, Andrew and {Nord}, Brian and {Nuttall}, Laura and {Rollins}, Richard and {Sudek}, Philipp and {Tam}, Sut-Ieng and {Tessore}, Nicolas and {Tolley}, Arthur and {Umetsu}, Keiichi and {Williamson}, Andrew and {Wolz}, Laura},
        title = "{SkyPy: A package for modelling the Universe}",
      journal = {The Journal of Open Source Software},
         year = 2021,
        month = sep,
       volume = {6},
       number = {65},
          eid = {3056},
        pages = {3056},
          doi = {10.21105/joss.03056},
archivePrefix = {arXiv},
       eprint = {2109.06172},
 primaryClass = {astro-ph.IM},
       adsurl = {https://ui.adsabs.harvard.edu/abs/2021JOSS....6.3056A}
}

@ARTICLE{Schechter:1976,
       author = {{Schechter}, P.},
        title = "{An analytic expression for the luminosity function for galaxies.}",
      journal = {\apj},
         year = 1976,
        month = jan,
       volume = {203},
        pages = {297-306},
          doi = {10.1086/154079},
       adsurl = {https://ui.adsabs.harvard.edu/abs/1976ApJ...203..297S}
}

@ARTICLE{Smith:2021,
       author = {{Smith}, Russell J. and {Collett}, Thomas E.},
        title = "{A fully-spectroscopic triple-source-plane lens: the Jackpot completed}",
      journal = {\mnras},
         year = 2021,
        month = aug,
       volume = {505},
       number = {2},
        pages = {2136-2140},
          doi = {10.1093/mnras/stab1399},
archivePrefix = {arXiv},
       eprint = {2104.12790},
 primaryClass = {astro-ph.CO},
       adsurl = {https://ui.adsabs.harvard.edu/abs/2021MNRAS.505.2136S}
}

@ARTICLE{Collett:2015,
       author = {Collett, Thomas E.},
        title = "{The Population of Galaxy-Galaxy Strong Lenses in Forthcoming Optical Imaging Surveys}",
      journal = {\apj},
         year = 2015,
        month = sep,
       volume = {811},
       number = {1},
          eid = {20},
        pages = {20},
          doi = {10.1088/0004-637X/811/1/20},
archivePrefix = {arXiv},
       eprint = {1507.02657},
 primaryClass = {astro-ph.CO},
       adsurl = {https://ui.adsabs.harvard.edu/abs/2015ApJ...811...20C}
}

@ARTICLE{2020SciPy,
  author  = {Virtanen, Pauli and Gommers, Ralf and Oliphant, Travis E. and
            Haberland, Matt and Reddy, Tyler and Cournapeau, David and
            Burovski, Evgeni and Peterson, Pearu and Weckesser, Warren and
            Bright, Jonathan and {van der Walt}, St{\'e}fan J. and
            Brett, Matthew and Wilson, Joshua and Millman, K. Jarrod and
            Mayorov, Nikolay and Nelson, Andrew R. J. and Jones, Eric and
            Kern, Robert and Larson, Eric and Carey, C J and
            Polat, {\.I}lhan and Feng, Yu and Moore, Eric W. and
            {VanderPlas}, Jake and Laxalde, Denis and Perktold, Josef and
            Cimrman, Robert and Henriksen, Ian and Quintero, E. A. and
            Harris, Charles R. and Archibald, Anne M. and
            Ribeiro, Ant{\^o}nio H. and Pedregosa, Fabian and
            {van Mulbregt}, Paul and {SciPy 1.0 Contributors}},
  title   = {{{SciPy} 1.0: Fundamental Algorithms for Scientific
            Computing in Python}},
  journal = {Nature Methods},
  year    = {2020},
  volume  = {17},
  pages   = {261--272},
  adsurl  = {https://rdcu.be/b08Wh},
  doi     = {10.1038/s41592-019-0686-2},
}

@ARTICLE{Collett_4MOST_SLSLS:2023,
       author = {{Collett}, T.~E. and {Sonnenfeld}, A. and {Frohmaier}, C. and {Glazebrook}, K. and {Sluse}, D. and {Motta}, V. and {Verma}, A. and {Anguita}, T. and {Koopmans}, L. and {Tortora}, C. and {Courbin}, F. and {Cabanac}, R. and {Frye}, B. and {Smith}, G.~P. and {Diego}, J.~M. and {Alteiri}, B. and {Lopez}, S. and {Fassnacht}, C. and {Cooray}, A. and {Goobar}, A. and {Ryczanowski}, D. and {Serjeant}, S. and {Richard}, J. and {Treu}, T. and {Moustakas}, L. and {Li}, R. and {Jacobs}, C. and {Lemon}, C. and {Marchetti}, L. and {Hartley}, P. and {Jullo}, E. and {Lee}, C.-H. and {Birrer}, S. and {Fritz}, A. and {Nightingale}, J. and {Napolitano}, N. and {Plazas}, A.~A. and {Kruk}, S. and {Spiniello}, C. and {Grillo}, C. and {Suyu}, S. and {Shajib}, A. and {Vernardos}, G. and {Dye}, S. and {Daylan}, T. and {Newman}, J. and {Schuldt}, S.},
        title = "{The 4MOST Strong Lensing Spectroscopic Legacy Survey (4SLSLS)}",
      journal = {The Messenger},
         year = 2023,
        month = mar,
       volume = {190},
        pages = {49-52},
          doi = {10.18727/0722-6691/5313},
       adsurl = {https://ui.adsabs.harvard.edu/abs/2023Msngr.190...49C}
}

@ARTICLE{LSSTScienceBook:2009,
       author = {{LSST Science Collaboration} and {Abell}, Paul A. and {Allison}, Julius and {Anderson}, Scott F. and {Andrew}, John R. and {Angel}, J. Roger P. and {Armus}, Lee and {Arnett}, David and {Asztalos}, S.~J. and {Axelrod}, Tim S. and {Bailey}, Stephen and {Ballantyne}, D.~R. and {Bankert}, Justin R. and {Barkhouse}, Wayne A. and {Barr}, Jeffrey D. and {Barrientos}, L. Felipe and {Barth}, Aaron J. and {Bartlett}, James G. and {Becker}, Andrew C. and {Becla}, Jacek and {Beers}, Timothy C. and {Bernstein}, Joseph P. and {Biswas}, Rahul and {Blanton}, Michael R. and {Bloom}, Joshua S. and {Bochanski}, John J. and {Boeshaar}, Pat and {Borne}, Kirk D. and {Bradac}, Marusa and {Brandt}, W.~N. and {Bridge}, Carrie R. and {Brown}, Michael E. and {Brunner}, Robert J. and {Bullock}, James S. and {Burgasser}, Adam J. and {Burge}, James H. and {Burke}, David L. and {Cargile}, Phillip A. and {Chandrasekharan}, Srinivasan and {Chartas}, George and {Chesley}, Steven R. and {Chu}, You-Hua and {Cinabro}, David and {Claire}, Mark W. and {Claver}, Charles F. and {Clowe}, Douglas and {Connolly}, A.~J. and {Cook}, Kem H. and {Cooke}, Jeff and {Cooray}, Asantha and {Covey}, Kevin R. and {Culliton}, Christopher S. and {de Jong}, Roelof and {de Vries}, Willem H. and {Debattista}, Victor P. and {Delgado}, Francisco and {Dell'Antonio}, Ian P. and {Dhital}, Saurav and {Di Stefano}, Rosanne and {Dickinson}, Mark and {Dilday}, Benjamin and {Djorgovski}, S.~G. and {Dobler}, Gregory and {Donalek}, Ciro and {Dubois-Felsmann}, Gregory and {Durech}, Josef and {Eliasdottir}, Ardis and {Eracleous}, Michael and {Eyer}, Laurent and {Falco}, Emilio E. and {Fan}, Xiaohui and {Fassnacht}, Christopher D. and {Ferguson}, Harry C. and {Fernandez}, Yanga R. and {Fields}, Brian D. and {Finkbeiner}, Douglas and {Figueroa}, Eduardo E. and {Fox}, Derek B. and {Francke}, Harold and {Frank}, James S. and {Frieman}, Josh and {Fromenteau}, Sebastien and {Furqan}, Muhammad and {Galaz}, Gaspar and {Gal-Yam}, A. and {Garnavich}, Peter and {Gawiser}, Eric and {Geary}, John and {Gee}, Perry and {Gibson}, Robert R. and {Gilmore}, Kirk and {Grace}, Emily A. and {Green}, Richard F. and {Gressler}, William J. and {Grillmair}, Carl J. and {Habib}, Salman and {Haggerty}, J.~S. and {Hamuy}, Mario and {Harris}, Alan W. and {Hawley}, Suzanne L. and {Heavens}, Alan F. and {Hebb}, Leslie and {Henry}, Todd J. and {Hileman}, Edward and {Hilton}, Eric J. and {Hoadley}, Keri and {Holberg}, J.~B. and {Holman}, Matt J. and {Howell}, Steve B. and {Infante}, Leopoldo and {Ivezic}, Zeljko and {Jacoby}, Suzanne H. and {Jain}, Bhuvnesh and {R} and {Jedicke} and {Jee}, M. James and {Garrett Jernigan}, J. and {Jha}, Saurabh W. and {Johnston}, Kathryn V. and {Jones}, R. Lynne and {Juric}, Mario and {Kaasalainen}, Mikko and {Styliani} and {Kafka} and {Kahn}, Steven M. and {Kaib}, Nathan A. and {Kalirai}, Jason and {Kantor}, Jeff and {Kasliwal}, Mansi M. and {Keeton}, Charles R. and {Kessler}, Richard and {Knezevic}, Zoran and {Kowalski}, Adam and {Krabbendam}, Victor L. and {Krughoff}, K. Simon and {Kulkarni}, Shrinivas and {Kuhlman}, Stephen and {Lacy}, Mark and {Lepine}, Sebastien and {Liang}, Ming and {Lien}, Amy and {Lira}, Paulina and {Long}, Knox S. and {Lorenz}, Suzanne and {Lotz}, Jennifer M. and {Lupton}, R.~H. and {Lutz}, Julie and {Macri}, Lucas M. and {Mahabal}, Ashish A. and {Mandelbaum}, Rachel and {Marshall}, Phil and {May}, Morgan and {McGehee}, Peregrine M. and {Meadows}, Brian T. and {Meert}, Alan and {Milani}, Andrea and {Miller}, Christopher J. and {Miller}, Michelle and {Mills}, David and {Minniti}, Dante and {Monet}, David and {Mukadam}, Anjum S. and {Nakar}, Ehud and {Neill}, Douglas R. and {Newman}, Jeffrey A. and {Nikolaev}, Sergei and {Nordby}, Martin and {O'Connor}, Paul and {Oguri}, Masamune and {Oliver}, John and {Olivier}, Scot S. and {Olsen}, Julia K. and {Olsen}, Knut and {Olszewski}, Edward W. and {Oluseyi}, Hakeem and {Padilla}, Nelson D. and {Parker}, Alex and {Pepper}, Joshua and {Peterson}, John R. and {Petry}, Catherine and {Pinto}, Philip A. and {Pizagno}, James L. and {Popescu}, Bogdan and {Prsa}, Andrej and {Radcka}, Veljko and {Raddick}, M. Jordan and {Rasmussen}, Andrew and {Rau}, Arne and {Rho}, Jeonghee and {Rhoads}, James E. and {Richards}, Gordon T. and {Ridgway}, Stephen T. and {Robertson}, Brant E. and {Roskar}, Rok and {Saha}, Abhijit and {Sarajedini}, Ata and {Scannapieco}, Evan and {Schalk}, Terry and {Schindler}, Rafe and {Schmidt}, Samuel},
        title = "{LSST Science Book, Version 2.0}",
      journal = {arXiv e-prints},
         year = 2009,
        month = dec,
          eid = {arXiv:0912.0201},
        pages = {arXiv:0912.0201},
          doi = {10.48550/arXiv.0912.0201},
archivePrefix = {arXiv},
       eprint = {0912.0201},
 primaryClass = {astro-ph.IM},
       adsurl = {https://ui.adsabs.harvard.edu/abs/2009arXiv0912.0201L}
}

@ARTICLE{Li:2024,
       author = {{Li}, Tian and {Collett}, Thomas E. and {Krawczyk}, Coleman M. and {Enzi}, Wolfgang},
        title = "{Cosmology from large populations of galaxy-galaxy strong gravitational lenses}",
      journal = {\mnras},
         year = 2024,
        month = jan,
       volume = {527},
       number = {3},
        pages = {5311-5323},
          doi = {10.1093/mnras/stad3514},
archivePrefix = {arXiv},
       eprint = {2307.09271},
 primaryClass = {astro-ph.CO},
       adsurl = {https://ui.adsabs.harvard.edu/abs/2024MNRAS.527.5311L}
}

@article{emcee,
   author = {{Foreman-Mackey}, D. and {Hogg}, D.~W. and {Lang}, D. and {Goodman}, J.},
    title = {emcee: The MCMC Hammer},
  journal = {PASP},
     year = 2013,
   volume = 125,
    pages = {306-312},
   eprint = {1202.3665},
      doi = {10.1086/670067}
}

@ARTICLE{Divij:2023,
       author = {{Sharma}, Divij and {Collett}, Thomas E. and {Linder}, Eric V.},
        title = "{Testing cosmology with double source lensing}",
      journal = {\jcap},
         year = 2023,
        month = apr,
       volume = {2023},
       number = {4},
          eid = {001},
        pages = {001},
          doi = {10.1088/1475-7516/2023/04/001},
archivePrefix = {arXiv},
       eprint = {2212.00055},
 primaryClass = {astro-ph.CO},
       adsurl = {https://ui.adsabs.harvard.edu/abs/2023JCAP...04..001S}
}

@ARTICLE{DES_Y3:2022,
       author = {{Abbott}, T.~M.~C. and {Aguena}, M. and {Alarcon}, A. and {Allam}, S. and {Alves}, O. and {Amon}, A. and {Andrade-Oliveira}, F. and {Annis}, J. and {Avila}, S. and {Bacon}, D. and {Baxter}, E. and {Bechtol}, K. and {Becker}, M.~R. and {Bernstein}, G.~M. and {Bhargava}, S. and {Birrer}, S. and {Blazek}, J. and {Brandao-Souza}, A. and {Bridle}, S.~L. and {Brooks}, D. and {Buckley-Geer}, E. and {Burke}, D.~L. and {Camacho}, H. and {Campos}, A. and {Carnero Rosell}, A. and {Carrasco Kind}, M. and {Carretero}, J. and {Castander}, F.~J. and {Cawthon}, R. and {Chang}, C. and {Chen}, A. and {Chen}, R. and {Choi}, A. and {Conselice}, C. and {Cordero}, J. and {Costanzi}, M. and {Crocce}, M. and {da Costa}, L.~N. and {da Silva Pereira}, M.~E. and {Davis}, C. and {Davis}, T.~M. and {De Vicente}, J. and {DeRose}, J. and {Desai}, S. and {Di Valentino}, E. and {Diehl}, H.~T. and {Dietrich}, J.~P. and {Dodelson}, S. and {Doel}, P. and {Doux}, C. and {Drlica-Wagner}, A. and {Eckert}, K. and {Eifler}, T.~F. and {Elsner}, F. and {Elvin-Poole}, J. and {Everett}, S. and {Evrard}, A.~E. and {Fang}, X. and {Farahi}, A. and {Fernandez}, E. and {Ferrero}, I. and {Fert{\'e}}, A. and {Fosalba}, P. and {Friedrich}, O. and {Frieman}, J. and {Garc{\'\i}a-Bellido}, J. and {Gatti}, M. and {Gaztanaga}, E. and {Gerdes}, D.~W. and {Giannantonio}, T. and {Giannini}, G. and {Gruen}, D. and {Gruendl}, R.~A. and {Gschwend}, J. and {Gutierrez}, G. and {Harrison}, I. and {Hartley}, W.~G. and {Herner}, K. and {Hinton}, S.~R. and {Hollowood}, D.~L. and {Honscheid}, K. and {Hoyle}, B. and {Huff}, E.~M. and {Huterer}, D. and {Jain}, B. and {James}, D.~J. and {Jarvis}, M. and {Jeffrey}, N. and {Jeltema}, T. and {Kovacs}, A. and {Krause}, E. and {Kron}, R. and {Kuehn}, K. and {Kuropatkin}, N. and {Lahav}, O. and {Leget}, P.-F. and {Lemos}, P. and {Liddle}, A.~R. and {Lidman}, C. and {Lima}, M. and {Lin}, H. and {MacCrann}, N. and {Maia}, M.~A.~G. and {Marshall}, J.~L. and {Martini}, P. and {McCullough}, J. and {Melchior}, P. and {Mena-Fern{\'a}ndez}, J. and {Menanteau}, F. and {Miquel}, R. and {Mohr}, J.~J. and {Morgan}, R. and {Muir}, J. and {Myles}, J. and {Nadathur}, S. and {Navarro-Alsina}, A. and {Nichol}, R.~C. and {Ogando}, R.~L.~C. and {Omori}, Y. and {Palmese}, A. and {Pandey}, S. and {Park}, Y. and {Paz-Chinch{\'o}n}, F. and {Petravick}, D. and {Pieres}, A. and {Plazas Malag{\'o}n}, A.~A. and {Porredon}, A. and {Prat}, J. and {Raveri}, M. and {Rodriguez-Monroy}, M. and {Rollins}, R.~P. and {Romer}, A.~K. and {Roodman}, A. and {Rosenfeld}, R. and {Ross}, A.~J. and {Rykoff}, E.~S. and {Samuroff}, S. and {S{\'a}nchez}, C. and {Sanchez}, E. and {Sanchez}, J. and {Sanchez Cid}, D. and {Scarpine}, V. and {Schubnell}, M. and {Scolnic}, D. and {Secco}, L.~F. and {Serrano}, S. and {Sevilla-Noarbe}, I. and {Sheldon}, E. and {Shin}, T. and {Smith}, M. and {Soares-Santos}, M. and {Suchyta}, E. and {Swanson}, M.~E.~C. and {Tabbutt}, M. and {Tarle}, G. and {Thomas}, D. and {To}, C. and {Troja}, A. and {Troxel}, M.~A. and {Tucker}, D.~L. and {Tutusaus}, I. and {Varga}, T.~N. and {Walker}, A.~R. and {Weaverdyck}, N. and {Wechsler}, R. and {Weller}, J. and {Yanny}, B. and {Yin}, B. and {Zhang}, Y. and {Zuntz}, J. and {DES Collaboration}},
        title = "{Dark Energy Survey Year 3 results: Cosmological constraints from galaxy clustering and weak lensing}",
      journal = {\prd},
         year = 2022,
        month = jan,
       volume = {105},
       number = {2},
          eid = {023520},
        pages = {023520},
          doi = {10.1103/PhysRevD.105.023520},
archivePrefix = {arXiv},
       eprint = {2105.13549},
 primaryClass = {astro-ph.CO},
       adsurl = {https://ui.adsabs.harvard.edu/abs/2022PhRvD.105b3520A}
}

@ARTICLE{bernadietal2010,
       author = {{Bernardi}, M. and {Shankar}, F. and {Hyde}, J.~B. and {Mei}, S. and {Marulli}, F. and {Sheth}, R.~K.},
        title = "{Galaxy luminosities, stellar masses, sizes, velocity dispersions as a function of morphological type}",
      journal = {\mnras},
         year = 2010,
        month = jun,
       volume = {404},
       number = {4},
        pages = {2087-2122},
          doi = {10.1111/j.1365-2966.2010.16425.x},
archivePrefix = {arXiv},
       eprint = {0910.1093},
 primaryClass = {astro-ph.CO},
       adsurl = {https://ui.adsabs.harvard.edu/abs/2010MNRAS.404.2087B}
}

@ARTICLE{Bowden:2025,
       author = {{Bowden}, Duncan J. and {Sahu}, Nandini and {Shajib}, Anowar J. and {Tran}, Kim-Vy and {Barone}, Tania M. and {G.~C.}, Keerthi Vasan and {Ballard}, Daniel J. and {Collett}, Thomas E. and {Dalessandro}, Faith and {Ferrami}, Giovanni and {Glazebrook}, Karl and {Gottemoller}, William J. and {Iwamoto}, Leena and {Jones}, Tucker and {Kacprzak}, Glenn G. and {Lewis}, Geraint F. and {McIntosh-Lombardo}, Haven and {Skobe}, Hannah and {Suyu}, Sherry H. and {Sweet}, Sarah M.},
        title = "{Constraining Cosmology with Double-source-plane Strong Gravitational Lenses from the AGEL Survey}",
      journal = {\apj},
         year = 2025,
        month = nov,
       volume = {993},
       number = {1},
          eid = {124},
        pages = {124},
          doi = {10.3847/1538-4357/ae092e},
archivePrefix = {arXiv},
       eprint = {2509.15012},
 primaryClass = {astro-ph.CO},
       adsurl = {https://ui.adsabs.harvard.edu/abs/2025ApJ...993..124B}
}

@ARTICLE{Sahu:2025,
       author = {{Sahu}, Nandini and {Shajib}, Anowar J. and {Tran}, Kim-Vy and {Skobe}, Hannah and {Rhoades}, Sunny and {Jones}, Tucker and {Glazebrook}, Karl and {Collett}, Thomas E. and {Suyu}, Sherry H. and {Vasan G.~C.}, Keerthi and {Barone}, Tania M. and {Bowden}, Duncan J. and {Ballard}, Daniel and {Kacprzak}, Glenn G. and {Sweet}, Sarah M. and {Lewis}, Geraint F. and {Nanayakkara}, Themiya},
        title = "{Cosmography with the Double-source-plane Strong Gravitational Lens AGEL150745+052256}",
      journal = {\apj},
         year = 2025,
        month = sep,
       volume = {991},
       number = {1},
          eid = {72},
        pages = {72},
          doi = {10.3847/1538-4357/adf442},
archivePrefix = {arXiv},
       eprint = {2504.00656},
 primaryClass = {astro-ph.CO},
       adsurl = {https://ui.adsabs.harvard.edu/abs/2025ApJ...991...72S}
}

@ARTICLE{Jain:2003,
       author = {{Jain}, Bhuvnesh and {Taylor}, Andy},
        title = "{Cross-Correlation Tomography: Measuring Dark Energy Evolution with Weak Lensing}",
      journal = {PRL},
         year = 2003,
        month = oct,
       volume = {91},
       number = {14},
          eid = {141302},
        pages = {141302},
          doi = {10.1103/PhysRevLett.91.141302},
archivePrefix = {arXiv},
       eprint = {astro-ph/0306046},
 primaryClass = {astro-ph},
       adsurl = {https://ui.adsabs.harvard.edu/abs/2003PhRvL..91n1302J}
}

@ARTICLE{Hu:2002,
       author = {{Hu}, Wayne},
        title = "{Dark energy and matter evolution from lensing tomography}",
      journal = {\prd},
         year = 2002,
        month = oct,
       volume = {66},
       number = {8},
          eid = {083515},
        pages = {083515},
          doi = {10.1103/PhysRevD.66.083515},
archivePrefix = {arXiv},
       eprint = {astro-ph/0208093},
 primaryClass = {astro-ph},
       adsurl = {https://ui.adsabs.harvard.edu/abs/2002PhRvD..66h3515H}
}

@ARTICLE{Rubin:Ivezic:2019,
       author = {{Ivezi{\'c}}, {\v{Z}}eljko and {Kahn}, Steven M. and {Tyson}, J. Anthony and {Abel}, Bob and {Acosta}, Emily and {Allsman}, Robyn and {Alonso}, David and {AlSayyad}, Yusra and {Anderson}, Scott F. and {Andrew}, John and {Angel}, James Roger P. and {Angeli}, George Z. and {Ansari}, Reza and {Antilogus}, Pierre and {Araujo}, Constanza and {Armstrong}, Robert and {Arndt}, Kirk T. and {Astier}, Pierre and {Aubourg}, {\'E}ric and {Auza}, Nicole and {Axelrod}, Tim S. and {Bard}, Deborah J. and {Barr}, Jeff D. and {Barrau}, Aurelian and {Bartlett}, James G. and {Bauer}, Amanda E. and {Bauman}, Brian J. and {Baumont}, Sylvain and {Bechtol}, Ellen and {Bechtol}, Keith and {Becker}, Andrew C. and {Becla}, Jacek and {Beldica}, Cristina and {Bellavia}, Steve and {Bianco}, Federica B. and {Biswas}, Rahul and {Blanc}, Guillaume and {Blazek}, Jonathan and {Blandford}, Roger D. and {Bloom}, Josh S. and {Bogart}, Joanne and {Bond}, Tim W. and {Booth}, Michael T. and {Borgland}, Anders W. and {Borne}, Kirk and {Bosch}, James F. and {Boutigny}, Dominique and {Brackett}, Craig A. and {Bradshaw}, Andrew and {Brandt}, William Nielsen and {Brown}, Michael E. and {Bullock}, James S. and {Burchat}, Patricia and {Burke}, David L. and {Cagnoli}, Gianpietro and {Calabrese}, Daniel and {Callahan}, Shawn and {Callen}, Alice L. and {Carlin}, Jeffrey L. and {Carlson}, Erin L. and {Chandrasekharan}, Srinivasan and {Charles-Emerson}, Glenaver and {Chesley}, Steve and {Cheu}, Elliott C. and {Chiang}, Hsin-Fang and {Chiang}, James and {Chirino}, Carol and {Chow}, Derek and {Ciardi}, David R. and {Claver}, Charles F. and {Cohen-Tanugi}, Johann and {Cockrum}, Joseph J. and {Coles}, Rebecca and {Connolly}, Andrew J. and {Cook}, Kem H. and {Cooray}, Asantha and {Covey}, Kevin R. and {Cribbs}, Chris and {Cui}, Wei and {Cutri}, Roc and {Daly}, Philip N. and {Daniel}, Scott F. and {Daruich}, Felipe and {Daubard}, Guillaume and {Daues}, Greg and {Dawson}, William and {Delgado}, Francisco and {Dellapenna}, Alfred and {de Peyster}, Robert and {de Val-Borro}, Miguel and {Digel}, Seth W. and {Doherty}, Peter and {Dubois}, Richard and {Dubois-Felsmann}, Gregory P. and {Durech}, Josef and {Economou}, Frossie and {Eifler}, Tim and {Eracleous}, Michael and {Emmons}, Benjamin L. and {Fausti Neto}, Angelo and {Ferguson}, Henry and {Figueroa}, Enrique and {Fisher-Levine}, Merlin and {Focke}, Warren and {Foss}, Michael D. and {Frank}, James and {Freemon}, Michael D. and {Gangler}, Emmanuel and {Gawiser}, Eric and {Geary}, John C. and {Gee}, Perry and {Geha}, Marla and {Gessner}, Charles J.~B. and {Gibson}, Robert R. and {Gilmore}, D. Kirk and {Glanzman}, Thomas and {Glick}, William and {Goldina}, Tatiana and {Goldstein}, Daniel A. and {Goodenow}, Iain and {Graham}, Melissa L. and {Gressler}, William J. and {Gris}, Philippe and {Guy}, Leanne P. and {Guyonnet}, Augustin and {Haller}, Gunther and {Harris}, Ron and {Hascall}, Patrick A. and {Haupt}, Justine and {Hernandez}, Fabio and {Herrmann}, Sven and {Hileman}, Edward and {Hoblitt}, Joshua and {Hodgson}, John A. and {Hogan}, Craig and {Howard}, James D. and {Huang}, Dajun and {Huffer}, Michael E. and {Ingraham}, Patrick and {Innes}, Walter R. and {Jacoby}, Suzanne H. and {Jain}, Bhuvnesh and {Jammes}, Fabrice and {Jee}, M. James and {Jenness}, Tim and {Jernigan}, Garrett and {Jevremovi{\'c}}, Darko and {Johns}, Kenneth and {Johnson}, Anthony S. and {Johnson}, Margaret W.~G. and {Jones}, R. Lynne and {Juramy-Gilles}, Claire and {Juri{\'c}}, Mario and {Kalirai}, Jason S. and {Kallivayalil}, Nitya J. and {Kalmbach}, Bryce and {Kantor}, Jeffrey P. and {Karst}, Pierre and {Kasliwal}, Mansi M. and {Kelly}, Heather and {Kessler}, Richard and {Kinnison}, Veronica and {Kirkby}, David and {Knox}, Lloyd and {Kotov}, Ivan V. and {Krabbendam}, Victor L. and {Krughoff}, K. Simon and {Kub{\'a}nek}, Petr and {Kuczewski}, John and {Kulkarni}, Shri and {Ku}, John and {Kurita}, Nadine R. and {Lage}, Craig S. and {Lambert}, Ron and {Lange}, Travis and {Langton}, J. Brian and {Le Guillou}, Laurent and {Levine}, Deborah and {Liang}, Ming and {Lim}, Kian-Tat and {Lintott}, Chris J. and {Long}, Kevin E. and {Lopez}, Margaux and {Lotz}, Paul J. and {Lupton}, Robert H. and {Lust}, Nate B. and {MacArthur}, Lauren A. and {Mahabal}, Ashish and {Mandelbaum}, Rachel and {Markiewicz}, Thomas W. and {Marsh}, Darren S. and {Marshall}, Philip J. and {Marshall}, Stuart and {May}, Morgan and {McKercher}, Robert and {McQueen}, Michelle and {Meyers}, Joshua and {Migliore}, Myriam and {Miller}, Michelle and {Mills}, David J.},
        title = "{LSST: From Science Drivers to Reference Design and Anticipated Data Products}",
      journal = {\apj},
         year = 2019,
        month = mar,
       volume = {873},
       number = {2},
          eid = {111},
        pages = {111},
          doi = {10.3847/1538-4357/ab042c},
archivePrefix = {arXiv},
       eprint = {0805.2366},
 primaryClass = {astro-ph},
       adsurl = {https://ui.adsabs.harvard.edu/abs/2019ApJ...873..111I}
}

@ARTICLE{Euclid:Laureijs:2011,
       author = {{Laureijs}, R. and {Amiaux}, J. and {Arduini}, S. and {Augu{\`e}res}, J. -L. and {Brinchmann}, J. and {Cole}, R. and {Cropper}, M. and {Dabin}, C. and {Duvet}, L. and {Ealet}, A. and {Garilli}, B. and {Gondoin}, P. and {Guzzo}, L. and {Hoar}, J. and {Hoekstra}, H. and {Holmes}, R. and {Kitching}, T. and {Maciaszek}, T. and {Mellier}, Y. and {Pasian}, F. and {Percival}, W. and {Rhodes}, J. and {Saavedra Criado}, G. and {Sauvage}, M. and {Scaramella}, R. and {Valenziano}, L. and {Warren}, S. and {Bender}, R. and {Castander}, F. and {Cimatti}, A. and {Le F{\`e}vre}, O. and {Kurki-Suonio}, H. and {Levi}, M. and {Lilje}, P. and {Meylan}, G. and {Nichol}, R. and {Pedersen}, K. and {Popa}, V. and {Rebolo Lopez}, R. and {Rix}, H. -W. and {Rottgering}, H. and {Zeilinger}, W. and {Grupp}, F. and {Hudelot}, P. and {Massey}, R. and {Meneghetti}, M. and {Miller}, L. and {Paltani}, S. and {Paulin-Henriksson}, S. and {Pires}, S. and {Saxton}, C. and {Schrabback}, T. and {Seidel}, G. and {Walsh}, J. and {Aghanim}, N. and {Amendola}, L. and {Bartlett}, J. and {Baccigalupi}, C. and {Beaulieu}, J. -P. and {Benabed}, K. and {Cuby}, J. -G. and {Elbaz}, D. and {Fosalba}, P. and {Gavazzi}, G. and {Helmi}, A. and {Hook}, I. and {Irwin}, M. and {Kneib}, J. -P. and {Kunz}, M. and {Mannucci}, F. and {Moscardini}, L. and {Tao}, C. and {Teyssier}, R. and {Weller}, J. and {Zamorani}, G. and {Zapatero Osorio}, M.~R. and {Boulade}, O. and {Foumond}, J.~J. and {Di Giorgio}, A. and {Guttridge}, P. and {James}, A. and {Kemp}, M. and {Martignac}, J. and {Spencer}, A. and {Walton}, D. and {Bl{\"u}mchen}, T. and {Bonoli}, C. and {Bortoletto}, F. and {Cerna}, C. and {Corcione}, L. and {Fabron}, C. and {Jahnke}, K. and {Ligori}, S. and {Madrid}, F. and {Martin}, L. and {Morgante}, G. and {Pamplona}, T. and {Prieto}, E. and {Riva}, M. and {Toledo}, R. and {Trifoglio}, M. and {Zerbi}, F. and {Abdalla}, F. and {Douspis}, M. and {Grenet}, C. and {Borgani}, S. and {Bouwens}, R. and {Courbin}, F. and {Delouis}, J. -M. and {Dubath}, P. and {Fontana}, A. and {Frailis}, M. and {Grazian}, A. and {Koppenh{\"o}fer}, J. and {Mansutti}, O. and {Melchior}, M. and {Mignoli}, M. and {Mohr}, J. and {Neissner}, C. and {Noddle}, K. and {Poncet}, M. and {Scodeggio}, M. and {Serrano}, S. and {Shane}, N. and {Starck}, J. -L. and {Surace}, C. and {Taylor}, A. and {Verdoes-Kleijn}, G. and {Vuerli}, C. and {Williams}, O.~R. and {Zacchei}, A. and {Altieri}, B. and {Escudero Sanz}, I. and {Kohley}, R. and {Oosterbroek}, T. and {Astier}, P. and {Bacon}, D. and {Bardelli}, S. and {Baugh}, C. and {Bellagamba}, F. and {Benoist}, C. and {Bianchi}, D. and {Biviano}, A. and {Branchini}, E. and {Carbone}, C. and {Cardone}, V. and {Clements}, D. and {Colombi}, S. and {Conselice}, C. and {Cresci}, G. and {Deacon}, N. and {Dunlop}, J. and {Fedeli}, C. and {Fontanot}, F. and {Franzetti}, P. and {Giocoli}, C. and {Garcia-Bellido}, J. and {Gow}, J. and {Heavens}, A. and {Hewett}, P. and {Heymans}, C. and {Holland}, A. and {Huang}, Z. and {Ilbert}, O. and {Joachimi}, B. and {Jennins}, E. and {Kerins}, E. and {Kiessling}, A. and {Kirk}, D. and {Kotak}, R. and {Krause}, O. and {Lahav}, O. and {van Leeuwen}, F. and {Lesgourgues}, J. and {Lombardi}, M. and {Magliocchetti}, M. and {Maguire}, K. and {Majerotto}, E. and {Maoli}, R. and {Marulli}, F. and {Maurogordato}, S. and {McCracken}, H. and {McLure}, R. and {Melchiorri}, A. and {Merson}, A. and {Moresco}, M. and {Nonino}, M. and {Norberg}, P. and {Peacock}, J. and {Pello}, R. and {Penny}, M. and {Pettorino}, V. and {Di Porto}, C. and {Pozzetti}, L. and {Quercellini}, C. and {Radovich}, M. and {Rassat}, A. and {Roche}, N. and {Ronayette}, S. and {Rossetti}, E.},
        title = "{Euclid Definition Study Report}",
      journal = {arXiv e-prints},
         year = 2011,
        month = oct,
          eid = {arXiv:1110.3193},
        pages = {arXiv:1110.3193},
          doi = {10.48550/arXiv.1110.3193},
archivePrefix = {arXiv},
       eprint = {1110.3193},
 primaryClass = {astro-ph.CO},
       adsurl = {https://ui.adsabs.harvard.edu/abs/2011arXiv1110.3193L}
}

@ARTICLE{Roman:Spergel:2015,
       author = {{Spergel}, D. and {Gehrels}, N. and {Baltay}, C. and {Bennett}, D. and {Breckinridge}, J. and {Donahue}, M. and {Dressler}, A. and {Gaudi}, B.~S. and {Greene}, T. and {Guyon}, O. and {Hirata}, C. and {Kalirai}, J. and {Kasdin}, N.~J. and {Macintosh}, B. and {Moos}, W. and {Perlmutter}, S. and {Postman}, M. and {Rauscher}, B. and {Rhodes}, J. and {Wang}, Y. and {Weinberg}, D. and {Benford}, D. and {Hudson}, M. and {Jeong}, W. -S. and {Mellier}, Y. and {Traub}, W. and {Yamada}, T. and {Capak}, P. and {Colbert}, J. and {Masters}, D. and {Penny}, M. and {Savransky}, D. and {Stern}, D. and {Zimmerman}, N. and {Barry}, R. and {Bartusek}, L. and {Carpenter}, K. and {Cheng}, E. and {Content}, D. and {Dekens}, F. and {Demers}, R. and {Grady}, K. and {Jackson}, C. and {Kuan}, G. and {Kruk}, J. and {Melton}, M. and {Nemati}, B. and {Parvin}, B. and {Poberezhskiy}, I. and {Peddie}, C. and {Ruffa}, J. and {Wallace}, J.~K. and {Whipple}, A. and {Wollack}, E. and {Zhao}, F.},
        title = "{Wide-Field InfrarRed Survey Telescope-Astrophysics Focused Telescope Assets WFIRST-AFTA 2015 Report}",
      journal = {arXiv e-prints},
         year = 2015,
        month = mar,
          eid = {arXiv:1503.03757},
        pages = {arXiv:1503.03757},
          doi = {10.48550/arXiv.1503.03757},
archivePrefix = {arXiv},
       eprint = {1503.03757},
 primaryClass = {astro-ph.IM},
       adsurl = {https://ui.adsabs.harvard.edu/abs/2015arXiv150303757S}
}

@ARTICLE{Jullo:2010,
       author = {{Jullo}, E. and {Natarajan}, P. and {Kneib}, J.-P. and {D'Aloisio}, A. and {Limousin}, M. and {Richard}, J. and {Schimd}, C.},
        title = "{Cosmological constraints from strong gravitational lensing in clusters of galaxies.}",
      journal = {Science},
         year = 2010,
        month = aug,
       volume = {329},
        pages = {924-927},
          doi = {10.1126/science.1185759},
archivePrefix = {arXiv},
       eprint = {1008.4802},
 primaryClass = {astro-ph.CO},
       adsurl = {https://ui.adsabs.harvard.edu/abs/2010Sci...329..924J}
}

@ARTICLE{Pantheon+:Brout:2022,
       author = {{Brout}, Dillon and {Scolnic}, Dan and {Popovic}, Brodie and {Riess}, Adam G. and {Carr}, Anthony and {Zuntz}, Joe and {Kessler}, Rick and {Davis}, Tamara M. and {Hinton}, Samuel and {Jones}, David and {Kenworthy}, W. D'Arcy and {Peterson}, Erik R. and {Said}, Khaled and {Taylor}, Georgie and {Ali}, Noor and {Armstrong}, Patrick and {Charvu}, Pranav and {Dwomoh}, Arianna and {Meldorf}, Cole and {Palmese}, Antonella and {Qu}, Helen and {Rose}, Benjamin M. and {Sanchez}, Bruno and {Stubbs}, Christopher W. and {Vincenzi}, Maria and {Wood}, Charlotte M. and {Brown}, Peter J. and {Chen}, Rebecca and {Chambers}, Ken and {Coulter}, David A. and {Dai}, Mi and {Dimitriadis}, Georgios and {Filippenko}, Alexei V. and {Foley}, Ryan J. and {Jha}, Saurabh W. and {Kelsey}, Lisa and {Kirshner}, Robert P. and {M{\"o}ller}, Anais and {Muir}, Jessie and {Nadathur}, Seshadri and {Pan}, Yen-Chen and {Rest}, Armin and {Rojas-Bravo}, Cesar and {Sako}, Masao and {Siebert}, Matthew R. and {Smith}, Mat and {Stahl}, Benjamin E. and {Wiseman}, Phil},
        title = "{The Pantheon+ Analysis: Cosmological Constraints}",
      journal = {\apj},
         year = 2022,
        month = oct,
       volume = {938},
       number = {2},
          eid = {110},
        pages = {110},
          doi = {10.3847/1538-4357/ac8e04},
archivePrefix = {arXiv},
       eprint = {2202.04077},
 primaryClass = {astro-ph.CO},
       adsurl = {https://ui.adsabs.harvard.edu/abs/2022ApJ...938..110B}
}

@ARTICLE{DESI:BAO:2025,
       author = {{Adame}, A.~G. and {Aguilar}, J. and {Ahlen}, S. and {Alam}, S. and {Alexander}, D.~M. and {Alvarez}, M. and {Alves}, O. and {Anand}, A. and {Andrade}, U. and {Armengaud}, E. and {Avila}, S. and {Aviles}, A. and {Awan}, H. and {Bahr-Kalus}, B. and {Bailey}, S. and {Baltay}, C. and {Bault}, A. and {Behera}, J. and {BenZvi}, S. and {Bera}, A. and {Beutler}, F. and {Bianchi}, D. and {Blake}, C. and {Blum}, R. and {Brieden}, S. and {Brodzeller}, A. and {Brooks}, D. and {Buckley-Geer}, E. and {Burtin}, E. and {Calderon}, R. and {Canning}, R. and {Carnero Rosell}, A. and {Cereskaite}, R. and {Cervantes-Cota}, J.~L. and {Chabanier}, S. and {Chaussidon}, E. and {Chaves-Montero}, J. and {Chen}, S. and {Chen}, X. and {Claybaugh}, T. and {Cole}, S. and {Cuceu}, A. and {Davis}, T.~M. and {Dawson}, K. and {de la Macorra}, A. and {de Mattia}, A. and {Deiosso}, N. and {Dey}, A. and {Dey}, B. and {Ding}, Z. and {Doel}, P. and {Edelstein}, J. and {Eftekharzadeh}, S. and {Eisenstein}, D.~J. and {Elliott}, A. and {Fagrelius}, P. and {Fanning}, K. and {Ferraro}, S. and {Ereza}, J. and {Findlay}, N. and {Flaugher}, B. and {Font-Ribera}, A. and {Forero-S{\'a}nchez}, D. and {Forero-Romero}, J.~E. and {Frenk}, C.~S. and {Garcia-Quintero}, C. and {Gazta{\~n}aga}, E. and {Gil-Mar{\'\i}n}, H. and {Gontcho a Gontcho}, S. and {Gonzalez-Morales}, A.~X. and {Gonzalez-Perez}, V. and {Gordon}, C. and {Green}, D. and {Gruen}, D. and {Gsponer}, R. and {Gutierrez}, G. and {Guy}, J. and {Hadzhiyska}, B. and {Hahn}, C. and {Hanif}, M.~M.~S. and {Herrera-Alcantar}, H.~K. and {Honscheid}, K. and {Howlett}, C. and {Huterer}, D. and {Ir{\v{s}}i{\v{c}}}, V. and {Ishak}, M. and {Juneau}, S. and {Kara{\c{c}}ayl{\i}}, N.~G. and {Kehoe}, R. and {Kent}, S. and {Kirkby}, D. and {Kremin}, A. and {Krolewski}, A. and {Lai}, Y. and {Lan}, T.-W. and {Landriau}, M. and {Lang}, D. and {Lasker}, J. and {Le Goff}, J.~M. and {Le Guillou}, L. and {Leauthaud}, A. and {Levi}, M.~E. and {Li}, T.~S. and {Linder}, E. and {Lodha}, K. and {Magneville}, C. and {Manera}, M. and {Margala}, D. and {Martini}, P. and {Maus}, M. and {McDonald}, P. and {Medina-Varela}, L. and {Meisner}, A. and {Mena-Fern{\'a}ndez}, J. and {Miquel}, R. and {Moon}, J. and {Moore}, S. and {Moustakas}, J. and {Mueller}, E. and {Mu{\~n}oz-Guti{\'e}rrez}, A. and {Myers}, A.~D. and {Nadathur}, S. and {Napolitano}, L. and {Neveux}, R. and {Newman}, J.~A. and {Nguyen}, N.~M. and {Nie}, J. and {Niz}, G. and {Noriega}, H.~E. and {Padmanabhan}, N. and {Paillas}, E. and {Palanque-Delabrouille}, N. and {Pan}, J. and {Penmetsa}, S. and {Percival}, W.~J. and {Pieri}, M.~M. and {Pinon}, M. and {Poppett}, C. and {Porredon}, A. and {Prada}, F. and {P{\'e}rez-Fern{\'a}ndez}, A. and {P{\'e}rez-R{\`a}fols}, I. and {Rabinowitz}, D. and {Raichoor}, A. and {Ram{\'\i}rez-P{\'e}rez}, C. and {Ramirez-Solano}, S. and {Rashkovetskyi}, M. and {Ravoux}, C. and {Rezaie}, M. and {Rich}, J. and {Rocher}, A. and {Rockosi}, C. and {Roe}, N.~A. and {Rosado-Marin}, A. and {Ross}, A.~J. and {Rossi}, G. and {Ruggeri}, R. and {Ruhlmann-Kleider}, V. and {Samushia}, L. and {Sanchez}, E. and {Saulder}, C. and {Schlafly}, E.~F. and {Schlegel}, D. and {Schubnell}, M. and {Seo}, H. and {Shafieloo}, A. and {Sharples}, R. and {Silber}, J. and {Slosar}, A. and {Smith}, A. and {Sprayberry}, D. and {Tan}, T. and {Tarl{\'e}}, G. and {Taylor}, P. and {Trusov}, S. and {Ure{\~n}a-L{\'o}pez}, L.~A. and {Vaisakh}, R. and {Valcin}, D. and {Valdes}, F. and {Vargas-Maga{\~n}a}, M. and {Verde}, L. and {Walther}, M. and {Wang}, B. and {Wang}, M.~S. and {Weaver}, B.~A. and {Weaverdyck}, N. and {Wechsler}, R.~H. and {Weinberg}, D.~H. and {White}, M. and {Yu}, J. and {Yu}, Y. and {Yuan}, S. and {Y{\`e}che}, C. and {Zaborowski}, E.~A. and {Zarrouk}, P. and {Zhang}, H. and {Zhao}, C. and {Zhao}, R. and {Zhou}, R. and {Zhuang}, T.},
        title = "{DESI 2024 VI: cosmological constraints from the measurements of baryon acoustic oscillations}",
      journal = {\jcap},
         year = 2025,
        month = feb,
       volume = {2025},
       number = {2},
          eid = {021},
        pages = {021},
          doi = {10.1088/1475-7516/2025/02/021},
archivePrefix = {arXiv},
       eprint = {2404.03002},
 primaryClass = {astro-ph.CO},
       adsurl = {https://ui.adsabs.harvard.edu/abs/2025JCAP...02..021A}
}

@ARTICLE{Khadka:2024,
       author = {{Khadka}, Narayan and {Birrer}, Simon and {Leauthaud}, Alexie and {Nix}, Holden},
        title = "{Breaking the mass-sheet degeneracy in strong lensing mass modelling with weak lensing observations}",
      journal = {\mnras},
         year = 2024,
        month = sep,
       volume = {533},
       number = {1},
        pages = {795-806},
          doi = {10.1093/mnras/stae1832},
archivePrefix = {arXiv},
       eprint = {2404.01513},
 primaryClass = {astro-ph.CO},
       adsurl = {https://ui.adsabs.harvard.edu/abs/2024MNRAS.533..795K}
}

@ARTICLE{Planck:2020,
       author = {{Planck Collaboration} and {Aghanim}, N. and {Akrami}, Y. and {Ashdown}, M. and {Aumont}, J. and {Baccigalupi}, C. and {Ballardini}, M. and {Banday}, A.~J. and {Barreiro}, R.~B. and {Bartolo}, N. and {Basak}, S. and {Battye}, R. and {Benabed}, K. and {Bernard}, J.-P. and {Bersanelli}, M. and {Bielewicz}, P. and {Bock}, J.~J. and {Bond}, J.~R. and {Borrill}, J. and {Bouchet}, F.~R. and {Boulanger}, F. and {Bucher}, M. and {Burigana}, C. and {Butler}, R.~C. and {Calabrese}, E. and {Cardoso}, J.-F. and {Carron}, J. and {Challinor}, A. and {Chiang}, H.~C. and {Chluba}, J. and {Colombo}, L.~P.~L. and {Combet}, C. and {Contreras}, D. and {Crill}, B.~P. and {Cuttaia}, F. and {de Bernardis}, P. and {de Zotti}, G. and {Delabrouille}, J. and {Delouis}, J.-M. and {Di Valentino}, E. and {Diego}, J.~M. and {Dor{\'e}}, O. and {Douspis}, M. and {Ducout}, A. and {Dupac}, X. and {Dusini}, S. and {Efstathiou}, G. and {Elsner}, F. and {En{\ss}lin}, T.~A. and {Eriksen}, H.~K. and {Fantaye}, Y. and {Farhang}, M. and {Fergusson}, J. and {Fernandez-Cobos}, R. and {Finelli}, F. and {Forastieri}, F. and {Frailis}, M. and {Fraisse}, A.~A. and {Franceschi}, E. and {Frolov}, A. and {Galeotta}, S. and {Galli}, S. and {Ganga}, K. and {G{\'e}nova-Santos}, R.~T. and {Gerbino}, M. and {Ghosh}, T. and {Gonz{\'a}lez-Nuevo}, J. and {G{\'o}rski}, K.~M. and {Gratton}, S. and {Gruppuso}, A. and {Gudmundsson}, J.~E. and {Hamann}, J. and {Handley}, W. and {Hansen}, F.~K. and {Herranz}, D. and {Hildebrandt}, S.~R. and {Hivon}, E. and {Huang}, Z. and {Jaffe}, A.~H. and {Jones}, W.~C. and {Karakci}, A. and {Keih{\"a}nen}, E. and {Keskitalo}, R. and {Kiiveri}, K. and {Kim}, J. and {Kisner}, T.~S. and {Knox}, L. and {Krachmalnicoff}, N. and {Kunz}, M. and {Kurki-Suonio}, H. and {Lagache}, G. and {Lamarre}, J.-M. and {Lasenby}, A. and {Lattanzi}, M. and {Lawrence}, C.~R. and {Le Jeune}, M. and {Lemos}, P. and {Lesgourgues}, J. and {Levrier}, F. and {Lewis}, A. and {Liguori}, M. and {Lilje}, P.~B. and {Lilley}, M. and {Lindholm}, V. and {L{\'o}pez-Caniego}, M. and {Lubin}, P.~M. and {Ma}, Y.-Z. and {Mac{\'\i}as-P{\'e}rez}, J.~F. and {Maggio}, G. and {Maino}, D. and {Mandolesi}, N. and {Mangilli}, A. and {Marcos-Caballero}, A. and {Maris}, M. and {Martin}, P.~G. and {Martinelli}, M. and {Mart{\'\i}nez-Gonz{\'a}lez}, E. and {Matarrese}, S. and {Mauri}, N. and {McEwen}, J.~D. and {Meinhold}, P.~R. and {Melchiorri}, A. and {Mennella}, A. and {Migliaccio}, M. and {Millea}, M. and {Mitra}, S. and {Miville-Desch{\^e}nes}, M.-A. and {Molinari}, D. and {Montier}, L. and {Morgante}, G. and {Moss}, A. and {Natoli}, P. and {N{\o}rgaard-Nielsen}, H.~U. and {Pagano}, L. and {Paoletti}, D. and {Partridge}, B. and {Patanchon}, G. and {Peiris}, H.~V. and {Perrotta}, F. and {Pettorino}, V. and {Piacentini}, F. and {Polastri}, L. and {Polenta}, G. and {Puget}, J.-L. and {Rachen}, J.~P. and {Reinecke}, M. and {Remazeilles}, M. and {Renzi}, A. and {Rocha}, G. and {Rosset}, C. and {Roudier}, G. and {Rubi{\~n}o-Mart{\'\i}n}, J.~A. and {Ruiz-Granados}, B. and {Salvati}, L. and {Sandri}, M. and {Savelainen}, M. and {Scott}, D. and {Shellard}, E.~P.~S. and {Sirignano}, C. and {Sirri}, G. and {Spencer}, L.~D. and {Sunyaev}, R. and {Suur-Uski}, A.-S. and {Tauber}, J.~A. and {Tavagnacco}, D. and {Tenti}, M. and {Toffolatti}, L. and {Tomasi}, M. and {Trombetti}, T. and {Valenziano}, L. and {Valiviita}, J. and {Van Tent}, B. and {Vibert}, L. and {Vielva}, P. and {Villa}, F. and {Vittorio}, N. and {Wandelt}, B.~D. and {Wehus}, I.~K. and {White}, M. and {White}, S.~D.~M. and {Zacchei}, A. and {Zonca}, A.},
        title = "{Planck 2018 results. VI. Cosmological parameters}",
      journal = {\aa},
         year = 2020,
        month = sep,
       volume = {641},
          eid = {A6},
        pages = {A6},
          doi = {10.1051/0004-6361/201833910},
archivePrefix = {arXiv},
       eprint = {1807.06209},
 primaryClass = {astro-ph.CO},
       adsurl = {https://ui.adsabs.harvard.edu/abs/2020A&A...641A...6P}
}

@ARTICLE{Erickson:2025,
       author = {{Erickson}, Sydney and {Millon}, Martin and {Venkatraman}, Padmavathi and {Li}, Tian and {Holloway}, Philip and {Marshall}, Phil and {Shajib}, Anowar and {Birrer}, Simon and {Huang}, Xiang-Yu and {Anguita}, Timo and {Dillmann}, Steven and {Khadka}, Narayan and {Napier}, Kate and {Roodman}, Aaron and {The LSST Dark Energy Science Collaboration}},
        title = "{Investigating the Dark Energy Constraint from Strongly Lensed AGN at LSST-Scale}",
      journal = {arXiv e-prints},
         year = 2025,
        month = nov,
          eid = {arXiv:2511.13669},
        pages = {arXiv:2511.13669},
          doi = {10.48550/arXiv.2511.13669},
archivePrefix = {arXiv},
       eprint = {2511.13669},
 primaryClass = {astro-ph.CO},
       adsurl = {https://ui.adsabs.harvard.edu/abs/2025arXiv251113669E}
}

@ARTICLE{Khadka:2026,
       author = {{Khadka}, Narayan and {Birrer}, Simon and {Best}, Henry and {Sharma}, Paras and {Abe}, Katsuya T. and {Tang}, Xianzhe and {Mistick}, Carly and {Urcelay}, Felipe and {Sonmez}, Emrecan M. and {Arendse}, Nikki and {Erickson}, Sydney and {Hjortlund}, Jacob O. and {Holloway}, Phil and {Huang}, Alan and {Karthik}, Rahul and {Lamontagne}, Mia and {Negi}, Vibhore and {Pierel}, Justin R. and {Sanchez}, Bruno and {Ece Saricaoglu}, Aysu and {Shajib}, Anowar and {Shao}, Yixuan and {Venkatraman}, Padma and {Wedig}, Bryce and {Agrawal}, Aadya and {Anguita}, Timo and {Bessa}, Pedro and {Bom}, Clecio R. and {Castillo}, Sofia and {Collett}, Thomas and {Daylan}, Tansu and {Dillmann}, Steven and {Grespan}, Margherita and {Hayes}, Erin E. and {Joseph}, Remy and {Kessler}, Richard and {Li}, Tian and {Marshall}, Phil and {More}, Anupreeta and {Motta}, Veronica and {Narayan}, Gautham and {O'Dowd}, Matt and {Oguri}, Masamune and {Verma}, Aprajita and {Vernardos}, Giorgos and {the Strong Lensing Science Collaboration} and {the LSST Dark Energy Science Collaboration}},
        title = "{SLSim: a strong lensing population simulation package}",
      journal = {arXiv e-prints},
         year = 2026,
        month = mar,
          eid = {arXiv:2603.17138},
        pages = {arXiv:2603.17138},
          doi = {10.48550/arXiv.2603.17138},
archivePrefix = {arXiv},
       eprint = {2603.17138},
 primaryClass = {astro-ph.CO},
       adsurl = {https://ui.adsabs.harvard.edu/abs/2026arXiv260317138K}
}

@ARTICLE{Cowan:2011,
       author = {{Cowan}, Glen and {Cranmer}, Kyle and {Gross}, Eilam and {Vitells}, Ofer},
        title = "{Asymptotic formulae for likelihood-based tests of new physics}",
      journal = {European Physical Journal C},
         year = 2011,
        month = feb,
       volume = {71},
       number = {2},
          eid = {1554},
        pages = {1554},
          doi = {10.1140/epjc/s10052-011-1554-0},
archivePrefix = {arXiv},
       eprint = {1007.1727},
 primaryClass = {physics.data-an},
       adsurl = {https://ui.adsabs.harvard.edu/abs/2011EPJC...71.1554C}
}

@article{Mahalanobis:1936, 
author = {{Mahalanobis}, P.C.},
title =  "{On the Generalised Distance in Statistics.}", 
volume={80}, 
ISSN={0976-8378}, 
url={http://dx.doi.org/10.1007/s13171-019-00164-5}, 
DOI={10.1007/s13171-019-00164-5}, 
number={S1}, 
journal={Sankhya A}, 
publisher={Springer Science and Business Media LLC}, year={2018}, month=dec, pages={1–7} }

@ARTICLE{Bernardi:2003,
       author = {{Bernardi}, Mariangela and {Sheth}, Ravi K. and {Annis}, James and {Burles}, Scott and {Eisenstein}, Daniel J. and {Finkbeiner}, Douglas P. and {Hogg}, David W. and {Lupton}, Robert H. and {Schlegel}, David J. and {SubbaRao}, Mark and {Bahcall}, Neta A. and {Blakeslee}, John P. and {Brinkmann}, J. and {Castander}, Francisco J. and {Connolly}, Andrew J. and {Csabai}, Istv{\'a}n and {Doi}, Mamoru and {Fukugita}, Masataka and {Frieman}, Joshua and {Heckman}, Timothy and {Hennessy}, Gregory S. and {Ivezi{\'c}}, {\v{Z}}eljko and {Knapp}, G.~R. and {Lamb}, Don Q. and {McKay}, Timothy and {Munn}, Jeffrey A. and {Nichol}, Robert and {Okamura}, Sadanori and {Schneider}, Donald P. and {Thakar}, Aniruddha R. and {York}, Donald G.},
        title = "{Early-type Galaxies in the Sloan Digital Sky Survey. II. Correlations between Observables}",
      journal = {AJ},
         year = 2003,
        month = apr,
       volume = {125},
       number = {4},
        pages = {1849-1865},
          doi = {10.1086/374256},
archivePrefix = {arXiv},
       eprint = {astro-ph/0301624},
 primaryClass = {astro-ph},
       adsurl = {https://ui.adsabs.harvard.edu/abs/2003AJ....125.1849B}
}

@ARTICLE{Holloway:2023,
       author = {{Holloway}, Philip and {Verma}, Aprajita and {Marshall}, Philip J. and {More}, Anupreeta and {Tecza}, Matthias},
        title = "{On the detectability of strong lensing in near-infrared surveys}",
      journal = {\mnras},
         year = 2023,
        month = oct,
       volume = {525},
       number = {2},
        pages = {2341-2354},
          doi = {10.1093/mnras/stad2371},
archivePrefix = {arXiv},
       eprint = {2308.00851},
 primaryClass = {astro-ph.GA},
       adsurl = {https://ui.adsabs.harvard.edu/abs/2023MNRAS.525.2341H}
}

@ARTICLE{Wedig:2025,
       author = {{Wedig}, Bryce and {Daylan}, Tansu and {Birrer}, Simon and {Cyr-Racine}, Francis-Yan and {Dvorkin}, Cora and {Finkbeiner}, Douglas P. and {Huang}, Alan and {Huang}, Xiaosheng and {Karthik}, Rahul and {Khadka}, Narayan and {Natarajan}, Priyamvada and {Nierenberg}, Anna M. and {Peter}, Annika H.~G. and {Pierel}, Justin D.~R. and {Tang}, Xianzhe TZ and {Wechsler}, Risa H.},
        title = "{The Roman View of Strong Gravitational Lenses}",
      journal = {\apj},
         year = 2025,
        month = jun,
       volume = {986},
       number = {1},
          eid = {42},
        pages = {42},
          doi = {10.3847/1538-4357/adc24f},
archivePrefix = {arXiv},
       eprint = {2506.03390},
 primaryClass = {astro-ph.CO},
       adsurl = {https://ui.adsabs.harvard.edu/abs/2025ApJ...986...42W}
}

@ARTICLE{Ferrami:2024,
       author = {{Ferrami}, G. and {Wyithe}, J. Stuart B.},
        title = "{A model for galaxy-galaxy strong lensing statistics in surveys}",
      journal = {\mnras},
         year = 2024,
        month = aug,
       volume = {532},
       number = {2},
        pages = {1832-1848},
          doi = {10.1093/mnras/stae1607},
archivePrefix = {arXiv},
       eprint = {2404.03143},
 primaryClass = {astro-ph.CO},
       adsurl = {https://ui.adsabs.harvard.edu/abs/2024MNRAS.532.1832F}
}

@ARTICLE{WST2024,
       author = {{Mainieri}, Vincenzo and {Anderson}, Richard I. and {Brinchmann}, Jarle and {Cimatti}, Andrea and {Ellis}, Richard S. and {Hill}, Vanessa and {Kneib}, Jean-Paul and {McLeod}, Anna F. and {Opitom}, Cyrielle and {Roth}, Martin M. and {Sanchez-Saez}, Paula and {Smiljanic}, Rodolfo and {Tolstoy}, Eline and {Bacon}, Roland and {Randich}, Sofia and {Adamo}, Angela and {Annibali}, Francesca and {Arevalo}, Patricia and {Audard}, Marc and {Barsanti}, Stefania and {Battaglia}, Giuseppina and {Bayo Aran}, Amelia M. and {Belfiore}, Francesco and {Bellazzini}, Michele and {Bellini}, Emilio and {Beltran}, Maria Teresa and {Berni}, Leda and {Bianchi}, Simone and {Biazzo}, Katia and {Bisero}, Sofia and {Bisogni}, Susanna and {Bland-Hawthorn}, Joss and {Blondin}, Stephane and {Bodensteiner}, Julia and {Boffin}, Henri M.~J. and {Bonito}, Rosaria and {Bono}, Giuseppe and {Bouche}, Nicolas F. and {Bowman}, Dominic and {Braga}, Vittorio F. and {Bragaglia}, Angela and {Branchesi}, Marica and {Brucalassi}, Anna and {Bryant}, Julia J. and {Bryson}, Ian and {Busa}, Innocenza and {Camera}, Stefano and {Carbone}, Carmelita and {Casali}, Giada and {Casali}, Mark and {Casasola}, Viviana and {Castro}, Norberto and {Catelan}, Marcio and {Cavallo}, Lorenzo and {Chiappini}, Cristina and {Cioni}, Maria-Rosa and {Colless}, Matthew and {Colzi}, Laura and {Contarini}, Sofia and {Couch}, Warrick and {D'Ammando}, Filippo and {d'Assignies D.}, William and {D'Orazi}, Valentina and {da Silva}, Ronaldo and {Dainotti}, Maria Giovanna and {Damiani}, Francesco and {Danielski}, Camilla and {De Cia}, Annalisa and {de Jong}, Roelof S. and {Dhawan}, Suhail and {Dierickx}, Philippe and {Driver}, Simon P. and {Dupletsa}, Ulyana and {Escoffier}, Stephanie and {Escorza}, Ana and {Fabrizio}, Michele and {Fiorentino}, Giuliana and {Fontana}, Adriano and {Fontani}, Francesco and {Forero Sanchez}, Daniel and {Franois}, Patrick and {Galindo-Guil}, Francisco Jose and {Gallazzi}, Anna Rita and {Galli}, Daniele and {Garcia}, Miriam and {Garcia-Rojas}, Jorge and {Garilli}, Bianca and {Grand}, Robert and {Guarcello}, Mario Giuseppe and {Hazra}, Nandini and {Helmi}, Amina and {Herrero}, Artemio and {Iglesias}, Daniela and {Ilic}, Dragana and {Irsic}, Vid and {Ivanov}, Valentin D. and {Izzo}, Luca and {Jablonka}, Pascale and {Joachimi}, Benjamin and {Kakkad}, Darshan and {Kamann}, Sebastian and {Koposov}, Sergey and {Kordopatis}, Georges and {Kovacevic}, Andjelka B. and {Kraljic}, Katarina and {Kuncarayakti}, Hanindyo and {Kwon}, Yuna and {La Forgia}, Fiorangela and {Lahav}, Ofer and {Laigle}, Clotilde and {Lazzarin}, Monica and {Leaman}, Ryan and {Leclercq}, Floriane and {Lee}, Khee-Gan and {Lee}, David and {Lehnert}, Matt D. and {Lira}, Paulina and {Loffredo}, Eleonora and {Lucatello}, Sara and {Magrini}, Laura and {Maguire}, Kate and {Mahler}, Guillaume and {Zahra Majidi}, Fatemeh and {Malavasi}, Nicola and {Mannucci}, Filippo and {Marconi}, Marcella and {Martin}, Nicolas and {Marulli}, Federico and {Massari}, Davide and {Matsuno}, Tadafumi and {Mattheee}, Jorryt and {McGee}, Sean and {Merc}, Jaroslav and {Merle}, Thibault and {Miglio}, Andrea and {Migliorini}, Alessandra and {Minchev}, Ivan and {Minniti}, Dante and {Miret-Roig}, Nuria and {Monreal Ibero}, Ana and {Montano}, Federico and {Montet}, Ben T. and {Moresco}, Michele and {Moretti}, Chiara and {Moscardini}, Lauro and {Moya}, Andres and {Mueller}, Oliver and {Nanayakkara}, Themiya and {Nicholl}, Matt and {Nordlander}, Thomas and {Onori}, Francesca and {Padovani}, Marco and {Pala}, Anna Francesca and {Panda}, Swayamtrupta and {Pandey-Pommier}, Mamta and {Pasquini}, Luca and {Pawlak}, Michal and {Pessi}, Priscila J. and {Pisani}, Alice and {Popovic}, Lukav C. and {Prisinzano}, Loredana and {Raddi}, Roberto and {Rainer}, Monica and {Rebassa-Mansergas}, Alberto and {Richard}, Johan and {Rigault}, Mickael and {Rocher}, Antoine and {Romano}, Donatella and {Rosati}, Piero and {Sacco}, Germano and {Sanchez-Janssen}, Ruben and {Sander}, Andreas A.~C. and {Sanders}, Jason L. and {Sargent}, Mark and {Sarpa}, Elena and {Schimd}, Carlo and {Schipani}, Pietro and {Sefusatti}, Emiliano and {Smith}, Graham P. and {Spina}, Lorenzo and {Steinmetz}, Matthias and {Tacchella}, Sandro and {Tautvaisiene}, Grazina and {Theissen}, Christopher and {Thomas}, Guillaume and {Ting}, Yuan-Sen and {Travouillon}, Tony and {Tresse}, Laurence and {Trivedi}, Oem and {Tsantaki}, Maria and {Tsedrik}, Maria and {Urrutia}, Tanya and {Valenti}, Elena and {Van der Swaelmen}, Mathieu and {Van Eck}, Sophie and {Verdiani}, Francesco and {Verdier}, Aurelien and {Vergani}, Susanna Diana and {Verhamme}, Anne and {Vernet}, Joel},
        title = "{The Wide-field Spectroscopic Telescope (WST) Science White Paper}",
      journal = {arXiv e-prints},
         year = 2024,
        month = mar,
          eid = {arXiv:2403.05398},
        pages = {arXiv:2403.05398},
          doi = {10.48550/arXiv.2403.05398},
archivePrefix = {arXiv},
       eprint = {2403.05398},
 primaryClass = {astro-ph.IM},
       adsurl = {https://ui.adsabs.harvard.edu/abs/2024arXiv240305398M}
}

@ARTICLE{Kaur_WAVES_4MOST:2025,
       author = {{Kaur}, G. and {Bilicki}, M. and {Bellstedt}, S. and {Tempel}, E. and {Hellwing}, W.~A. and {Baldry}, I. and {Bandi}, B. and {Barsanti}, S. and {Driver}, S. and {Guerra-Varas}, N. and {Holwerda}, B. and {Lagos}, C. and {Loveday}, J. and {Robotham}, A.},
        title = "{Wide Area VISTA Extragalactic Survey (WAVES): Selection of targets for the Wide survey using decision-tree classification}",
      journal = {arXiv e-prints},
         year = 2025,
        month = oct,
          eid = {arXiv:2510.11132},
        pages = {arXiv:2510.11132},
          doi = {10.48550/arXiv.2510.11132},
archivePrefix = {arXiv},
       eprint = {2510.11132},
 primaryClass = {astro-ph.CO},
       adsurl = {https://ui.adsabs.harvard.edu/abs/2025arXiv251011132K}
}

@ARTICLE{ROTAC:2025,
       author = {{Observations Time Allocation Committee}, Roman and {Community Survey Definition Committees}, Core},
        title = "{Roman Observations Time Allocation Committee: Final Report and Recommendations}",
      journal = {arXiv e-prints},
         year = 2025,
        month = may,
          eid = {arXiv:2505.10574},
        pages = {arXiv:2505.10574},
          doi = {10.48550/arXiv.2505.10574},
archivePrefix = {arXiv},
       eprint = {2505.10574},
 primaryClass = {astro-ph.IM},
       adsurl = {https://ui.adsabs.harvard.edu/abs/2025arXiv250510574O}
}

@ARTICLE{Capak:2019,
       author = {{Capak}, P. and {Cuillandre}, J-C. and {Bernardeau}, F. and {Castander}, F. and {Bowler}, R. and {Chang}, C. and {Grillmair}, C. and {Gris}, P. and {Eifler}, T. and {Hirata}, C. and {Hook}, I. and {Jain}, B. and {Kuijken}, K. and {Lochner}, M. and {Oesch}, P. and {Paltani}, S. and {Rhodes}, J. and {Robertson}, B. and {Rubin}, D. and {Scaramella}, R. and {Scarlata}, C. and {Scolnic}, D. and {Silverman}, J. and {Wachter}, S. and {Wang}, Y. and {The Tri-Agency Working Group}},
        title = "{Enhancing LSST Science with Euclid Synergy}",
      journal = {arXiv e-prints},
         year = 2019,
        month = apr,
          eid = {arXiv:1904.10439},
        pages = {arXiv:1904.10439},
          doi = {10.48550/arXiv.1904.10439},
archivePrefix = {arXiv},
       eprint = {1904.10439},
 primaryClass = {astro-ph.IM},
       adsurl = {https://ui.adsabs.harvard.edu/abs/2019arXiv190410439C}
}

@INPROCEEDINGS{4MOST:2012,
       author = {{de Jong}, Roelof S. and {Bellido-Tirado}, Olga and {Chiappini}, Cristina and {Depagne}, {\'E}ric and {Haynes}, Roger and {Johl}, Diana and {Schnurr}, Olivier and {Schwope}, Axel and {Walcher}, Jakob and {Dionies}, Frank and {Haynes}, Dionne and {Kelz}, Andreas and {Kitaura}, Francisco S. and {Lamer}, Georg and {Minchev}, Ivan and {M{\"u}ller}, Volker and {Nuza}, Sebasti{\'a}n. E. and {Olaya}, Jean-Christophe and {Piffl}, Tilmann and {Popow}, Emil and {Steinmetz}, Matthias and {Ural}, Ugur and {Williams}, Mary and {Winkler}, Roland and {Wisotzki}, Lutz and {Ansorge}, Wolfgang R. and {Banerji}, Manda and {Gonzalez Solares}, Eduardo and {Irwin}, Mike and {Kennicutt}, Robert C. and {King}, Dave and {McMahon}, Richard G. and {Koposov}, Sergey and {Parry}, Ian R. and {Sun}, David and {Walton}, Nicholas A. and {Finger}, Gert and {Iwert}, Olaf and {Krumpe}, Mirko and {Lizon}, Jean-Louis and {Vincenzo}, Mainieri and {Amans}, Jean-Philippe and {Bonifacio}, Piercarlo and {Cohen}, Mathieu and {Francois}, Patrick and {Jagourel}, Pascal and {Mignot}, Shan B. and {Royer}, Fr{\'e}d{\'e}ric and {Sartoretti}, Paola and {Bender}, Ralf and {Grupp}, Frank and {Hess}, Hans-Joachim and {Lang-Bardl}, Florian and {Muschielok}, Bernard and {B{\"o}hringer}, Hans and {Boller}, Thomas and {Bongiorno}, Angela and {Brusa}, Marcella and {Dwelly}, Tom and {Merloni}, Andrea and {Nandra}, Kirpal and {Salvato}, Mara and {Pragt}, Johannes H. and {Navarro}, Ram{\'o}n and {Gerlofsma}, Gerrit and {Roelfsema}, Ronald and {Dalton}, Gavin B. and {Middleton}, Kevin F. and {Tosh}, Ian A. and {Boeche}, Corrado and {Caffau}, Elisabetta and {Christlieb}, Norbert and {Grebel}, Eva K. and {Hansen}, Camilla and {Koch}, Andreas and {Ludwig}, Hans-G. and {Quirrenbach}, Andreas and {Sbordone}, Luca and {Seifert}, Walter and {Thimm}, Guido and {Trifonov}, Trifon and {Helmi}, Amina and {Trager}, Scott C. and {Feltzing}, Sofia and {Korn}, Andreas and {Boland}, Wilfried},
        title = "{4MOST: 4-metre multi-object spectroscopic telescope}",
    booktitle = {Ground-based and Airborne Instrumentation for Astronomy IV},
         year = 2012,
       editor = {{McLean}, Ian S. and {Ramsay}, Suzanne K. and {Takami}, Hideki},
       series = {Society of Photo-Optical Instrumentation Engineers (SPIE) Conference Series},
       volume = {8446},
        month = sep,
          eid = {84460T},
        pages = {84460T},
          doi = {10.1117/12.926239},
archivePrefix = {arXiv},
       eprint = {1206.6885},
 primaryClass = {astro-ph.IM},
       adsurl = {https://ui.adsabs.harvard.edu/abs/2012SPIE.8446E..0TD}
}

@ARTICLE{Kauffmann:2003,
       author = {{Kauffmann}, Guinevere and {Heckman}, Timothy M. and {Tremonti}, Christy and {Brinchmann}, Jarle and {Charlot}, St{\'e}phane and {White}, Simon D.~M. and {Ridgway}, Susan E. and {Brinkmann}, Jon and {Fukugita}, Masataka and {Hall}, Patrick B. and {Ivezi{\'c}}, {\v{Z}}eljko and {Richards}, Gordon T. and {Schneider}, Donald P.},
        title = "{The host galaxies of active galactic nuclei}",
      journal = {\mnras},
         year = 2003,
        month = dec,
       volume = {346},
       number = {4},
        pages = {1055-1077},
          doi = {10.1111/j.1365-2966.2003.07154.x},
archivePrefix = {arXiv},
       eprint = {astro-ph/0304239},
 primaryClass = {astro-ph},
       adsurl = {https://ui.adsabs.harvard.edu/abs/2003MNRAS.346.1055K}
}

@ARTICLE{Koopmans:2009,
       author = {{Koopmans}, L.~V.~E. and {Bolton}, A. and {Treu}, T. and {Czoske}, O. and {Auger}, M.~W. and {Barnab{\`e}}, M. and {Vegetti}, S. and {Gavazzi}, R. and {Moustakas}, L.~A. and {Burles}, S.},
        title = "{The Structure and Dynamics of Massive Early-Type Galaxies: On Homology, Isothermality, and Isotropy Inside One Effective Radius}",
      journal = {\apjl},
         year = 2009,
        month = sep,
       volume = {703},
       number = {1},
        pages = {L51-L54},
          doi = {10.1088/0004-637X/703/1/L51},
archivePrefix = {arXiv},
       eprint = {0906.1349},
 primaryClass = {astro-ph.CO},
       adsurl = {https://ui.adsabs.harvard.edu/abs/2009ApJ...703L..51K}
}

@ARTICLE{Herle:2024,
       author = {{Herle}, A. and {O'Riordan}, C.~M. and {Vegetti}, S.},
        title = "{Selection functions of strong lens finding neural networks}",
      journal = {\mnras},
         year = 2024,
        month = oct,
       volume = {534},
       number = {2},
        pages = {1093-1106},
          doi = {10.1093/mnras/stae2106},
archivePrefix = {arXiv},
       eprint = {2307.10355},
 primaryClass = {astro-ph.CO},
       adsurl = {https://ui.adsabs.harvard.edu/abs/2024MNRAS.534.1093H}
}

@ARTICLE{Canameras:2024,
       author = {{Ca{\~n}ameras}, R. and {Schuldt}, S. and {Shu}, Y. and {Suyu}, S.~H. and {Taubenberger}, S. and {Andika}, I.~T. and {Bag}, S. and {Inoue}, K.~T. and {Jaelani}, A.~T. and {Leal-Taix{\'e}}, L. and {Meinhardt}, T. and {Melo}, A. and {More}, A.},
        title = "{HOLISMOKES: XI. Evaluation of supervised neural networks for strong-lens searches in ground-based imaging surveys}",
      journal = {\aa},
         year = 2024,
        month = dec,
       volume = {692},
          eid = {A72},
        pages = {A72},
          doi = {10.1051/0004-6361/202347072},
archivePrefix = {arXiv},
       eprint = {2306.03136},
 primaryClass = {astro-ph.GA},
       adsurl = {https://ui.adsabs.harvard.edu/abs/2024A&A...692A..72C}
}

@ARTICLE{Silver:2025,
       author = {{Silver}, Ethan and {Wang}, R. and {Huang}, Xiaosheng and {Bolton}, Adam S. and {Storfer}, Christopher J. and {Banka}, S.},
        title = "{ML-driven Strong Lens Discoveries: Down to {\ensuremath{\theta}}$_{E}${\ensuremath{\sim}}0.$^{\prime\prime}$03 and M$_{halo}$ $<$ {}10$^{11}$M$_{{\ensuremath{\odot}}}$}",
      journal = {\apj},
         year = 2025,
        month = nov,
       volume = {994},
       number = {1},
          eid = {117},
        pages = {117},
          doi = {10.3847/1538-4357/adf3b0},
archivePrefix = {arXiv},
       eprint = {2507.01943},
 primaryClass = {astro-ph.CO},
       adsurl = {https://ui.adsabs.harvard.edu/abs/2025ApJ...994..117S}
}

@ARTICLE{Birrer:2018,
       author = {{Birrer}, Simon and {Amara}, Adam},
        title = "{lenstronomy: Multi-purpose gravitational lens modelling software package}",
      journal = {Physics of the Dark Universe},
         year = 2018,
        month = dec,
       volume = {22},
        pages = {189-201},
          doi = {10.1016/j.dark.2018.11.002},
archivePrefix = {arXiv},
       eprint = {1803.09746},
 primaryClass = {astro-ph.CO},
       adsurl = {https://ui.adsabs.harvard.edu/abs/2018PDU....22..189B}
}

@ARTICLE{Birrer:2021,
       author = {{Birrer}, Simon and {Shajib}, Anowar and {Gilman}, Daniel and {Galan}, Aymeric and {Aalbers}, Jelle and {Millon}, Martin and {Morgan}, Robert and {Pagano}, Giulia and {Park}, Ji and {Teodori}, Luca and {Tessore}, Nicolas and {Ueland}, Madison and {Van de Vyvere}, Lyne and {Wagner-Carena}, Sebastian and {Wempe}, Ewoud and {Yang}, Lilan and {Ding}, Xuheng and {Schmidt}, Thomas and {Sluse}, Dominique and {Zhang}, Ming and {Amara}, Adam},
        title = "{lenstronomy II: A gravitational lensing software ecosystem}",
      journal = {The Journal of Open Source Software},
         year = 2021,
        month = jun,
       volume = {6},
       number = {62},
          eid = {3283},
        pages = {3283},
          doi = {10.21105/joss.03283},
archivePrefix = {arXiv},
       eprint = {2106.05976},
 primaryClass = {astro-ph.CO},
       adsurl = {https://ui.adsabs.harvard.edu/abs/2021JOSS....6.3283B}
}

@ARTICLE{Caldwell:2005,
       author = {{Caldwell}, R.~R. and {Linder}, Eric V.},
        title = "{Limits of Quintessence}",
      journal = {\prl},
         year = 2005,
        month = sep,
       volume = {95},
       number = {14},
          eid = {141301},
        pages = {141301},
          doi = {10.1103/PhysRevLett.95.141301},
archivePrefix = {arXiv},
       eprint = {astro-ph/0505494},
 primaryClass = {astro-ph},
       adsurl = {https://ui.adsabs.harvard.edu/abs/2005PhRvL..95n1301C}
}

@ARTICLE{DiValentino:2021,
       author = {{Di Valentino}, Eleonora and {Mena}, Olga and {Pan}, Supriya and {Visinelli}, Luca and {Yang}, Weiqiang and {Melchiorri}, Alessandro and {Mota}, David F. and {Riess}, Adam G. and {Silk}, Joseph},
        title = "{In the realm of the Hubble tension-a review of solutions}",
      journal = {Classical and Quantum Gravity},
         year = 2021,
        month = jul,
       volume = {38},
       number = {15},
          eid = {153001},
        pages = {153001},
          doi = {10.1088/1361-6382/ac086d},
archivePrefix = {arXiv},
       eprint = {2103.01183},
 primaryClass = {astro-ph.CO},
       adsurl = {https://ui.adsabs.harvard.edu/abs/2021CQGra..38o3001D}
}

@ARTICLE{Singh:2016,
       author = {{Singh}, Suprit and {Singh}, Parminder},
        title = "{It's a dark, dark world: background evolution of interacting phiCDM models beyond simple exponential potentials}",
      journal = {\jcap},
         year = 2016,
        month = may,
       volume = {2016},
       number = {5},
          eid = {017},
        pages = {017},
          doi = {10.1088/1475-7516/2016/05/017},
archivePrefix = {arXiv},
       eprint = {1507.01535},
 primaryClass = {astro-ph.CO},
       adsurl = {https://ui.adsabs.harvard.edu/abs/2016JCAP...05..017S}
}

@ARTICLE{Zhao:2017,
       author = {{Zhao}, Gong-Bo and {Raveri}, Marco and {Pogosian}, Levon and {Wang}, Yuting and {Crittenden}, Robert G. and {Handley}, Will J. and {Percival}, Will J. and {Beutler}, Florian and {Brinkmann}, Jonathan and {Chuang}, Chia-Hsun and {Cuesta}, Antonio J. and {Eisenstein}, Daniel J. and {Kitaura}, Francisco-Shu and {Koyama}, Kazuya and {L'Huillier}, Benjamin and {Nichol}, Robert C. and {Pieri}, Matthew M. and {Rodriguez-Torres}, Sergio and {Ross}, Ashley J. and {Rossi}, Graziano and {S{\'a}nchez}, Ariel G. and {Shafieloo}, Arman and {Tinker}, Jeremy L. and {Tojeiro}, Rita and {Vazquez}, Jose A. and {Zhang}, Hanyu},
        title = "{Dynamical dark energy in light of the latest observations}",
      journal = {Nature Astronomy},
         year = 2017,
        month = aug,
       volume = {1},
        pages = {627-632},
          doi = {10.1038/s41550-017-0216-z},
archivePrefix = {arXiv},
       eprint = {1701.08165},
 primaryClass = {astro-ph.CO},
       adsurl = {https://ui.adsabs.harvard.edu/abs/2017NatAs...1..627Z}
}

@ARTICLE{Zhang:2026,
       author = {{Zhang}, Zhuoming and {Xu}, Tengpeng and {Chen}, Yun},
        title = "{Dynamical Dark Energy and the Unresolved Hubble Tension: Multi-model Constraints from DESI 2025 and Other Probes}",
      journal = {\apj},
         year = 2026,
        month = mar,
       volume = {999},
       number = {2},
          eid = {248},
        pages = {248},
          doi = {10.3847/1538-4357/ae4738},
archivePrefix = {arXiv},
       eprint = {2512.07281},
 primaryClass = {astro-ph.CO},
       adsurl = {https://ui.adsabs.harvard.edu/abs/2026ApJ...999..248Z}
}

@ARTICLE{Chevallier:2001,
       author = {{Chevallier}, Michel and {Polarski}, David},
        title = "{Accelerating Universes with Scaling Dark Matter}",
      journal = {International Journal of Modern Physics D},
         year = 2001,
        month = jan,
       volume = {10},
       number = {2},
        pages = {213-223},
          doi = {10.1142/S0218271801000822},
archivePrefix = {arXiv},
       eprint = {gr-qc/0009008},
 primaryClass = {gr-qc},
       adsurl = {https://ui.adsabs.harvard.edu/abs/2001IJMPD..10..213C}
}

@ARTICLE{Linder:2003,
       author = {{Linder}, E.~V. and {Jenkins}, A.},
        title = "{Cosmic structure growth and dark energy}",
      journal = {\mnras},
         year = 2003,
        month = dec,
       volume = {346},
       number = {2},
        pages = {573-583},
          doi = {10.1046/j.1365-2966.2003.07112.x},
archivePrefix = {arXiv},
       eprint = {astro-ph/0305286},
 primaryClass = {astro-ph},
       adsurl = {https://ui.adsabs.harvard.edu/abs/2003MNRAS.346..573L}
}

@ARTICLE{deCruz:2026,
       author = {{de Cruz P{\'e}rez}, Javier and {G{\'o}mez-Valent}, Adri{\`a} and {Sol{\`a} Peracaula}, Joan},
        title = "{Dynamical dark energy models in light of the latest observations}",
      journal = {\prd},
         year = 2026,
        month = apr,
       volume = {113},
       number = {8},
          eid = {083521},
        pages = {083521},
          doi = {10.1103/xchk-xlk1},
archivePrefix = {arXiv},
       eprint = {2512.20616},
 primaryClass = {astro-ph.CO},
       adsurl = {https://ui.adsabs.harvard.edu/abs/2026PhRvD.113h3521D}
}

@ARTICLE{Abdalla:2022,
       author = {{Abdalla}, Elcio and {Abell{\'a}n}, Guillermo Franco and {Aboubrahim}, Amin and {Agnello}, Adriano and {Akarsu}, {\"O}zg{\"u}r and {Akrami}, Yashar and {Alestas}, George and {Aloni}, Daniel and {Amendola}, Luca and {Anchordoqui}, Luis A. and {Anderson}, Richard I. and {Arendse}, Nikki and {Asgari}, Marika and {Ballardini}, Mario and {Barger}, Vernon and {Basilakos}, Spyros and {Batista}, Ronaldo C. and {Battistelli}, Elia S. and {Battye}, Richard and {Benetti}, Micol and {Benisty}, David and {Berlin}, Asher and {de Bernardis}, Paolo and {Berti}, Emanuele and {Bidenko}, Bohdan and {Birrer}, Simon and {Blakeslee}, John P. and {Boddy}, Kimberly K. and {Bom}, Clecio R. and {Bonilla}, Alexander and {Borghi}, Nicola and {Bouchet}, Fran{\c{c}}ois R. and {Braglia}, Matteo and {Buchert}, Thomas and {Buckley-Geer}, Elizabeth and {Calabrese}, Erminia and {Caldwell}, Robert R. and {Camarena}, David and {Capozziello}, Salvatore and {Casertano}, Stefano and {Chen}, Geoff C.-F. and {Chluba}, Jens and {Chen}, Angela and {Chen}, Hsin-Yu and {Chudaykin}, Anton and {Cicoli}, Michele and {Copi}, Craig J. and {Courbin}, Fred and {Cyr-Racine}, Francis-Yan and {Czerny}, Bo{\.z}ena and {Dainotti}, Maria and {D'Amico}, Guido and {Davis}, Anne-Christine and {de Cruz P{\'e}rez}, Javier and {de Haro}, Jaume and {Delabrouille}, Jacques and {Denton}, Peter B. and {Dhawan}, Suhail and {Dienes}, Keith R. and {Di Valentino}, Eleonora and {Du}, Pu and {Eckert}, Dominique and {Escamilla-Rivera}, Celia and {Fert{\'e}}, Agn{\`e}s and {Finelli}, Fabio and {Fosalba}, Pablo and {Freedman}, Wendy L. and {Frusciante}, Noemi and {Gazta{\~n}aga}, Enrique and {Giar{\`e}}, William and {Giusarma}, Elena and {G{\'o}mez-Valent}, Adri{\`a} and {Handley}, Will and {Harrison}, Ian and {Hart}, Luke and {Hazra}, Dhiraj Kumar and {Heavens}, Alan and {Heinesen}, Asta and {Hildebrandt}, Hendrik and {Hill}, J. Colin and {Hogg}, Natalie B. and {Holz}, Daniel E. and {Hooper}, Deanna C. and {Hosseininejad}, Nikoo and {Huterer}, Dragan and {Ishak}, Mustapha and {Ivanov}, Mikhail M. and {Jaffe}, Andrew H. and {Jang}, In Sung and {Jedamzik}, Karsten and {Jimenez}, Raul and {Joseph}, Melissa and {Joudaki}, Shahab and {Kamionkowski}, Marc and {Karwal}, Tanvi and {Kazantzidis}, Lavrentios and {Keeley}, Ryan E. and {Klasen}, Michael and {Komatsu}, Eiichiro and {Koopmans}, L{\'e}on V.~E. and {Kumar}, Suresh and {Lamagna}, Luca and {Lazkoz}, Ruth and {Lee}, Chung-Chi and {Lesgourgues}, Julien and {Levi Said}, Jackson and {Lewis}, Tiffany R. and {L'Huillier}, Benjamin and {Lucca}, Matteo and {Maartens}, Roy and {Macri}, Lucas M. and {Marfatia}, Danny and {Marra}, Valerio and {Martins}, Carlos J.~A.~P. and {Masi}, Silvia and {Matarrese}, Sabino and {Mazumdar}, Arindam and {Melchiorri}, Alessandro and {Mena}, Olga and {Mersini-Houghton}, Laura and {Mertens}, James and {Milakovi{\'c}}, Dinko and {Minami}, Yuto and {Miranda}, Vivian and {Moreno-Pulido}, Cristian and {Moresco}, Michele and {Mota}, David F. and {Mottola}, Emil and {Mozzon}, Simone and {Muir}, Jessica and {Mukherjee}, Ankan and {Mukherjee}, Suvodip and {Naselsky}, Pavel and {Nath}, Pran and {Nesseris}, Savvas and {Niedermann}, Florian and {Notari}, Alessio and {Nunes}, Rafael C. and {{\'O} Colg{\'a}in}, Eoin and {Owens}, Kayla A. and {{\"O}z{\"u}lker}, Emre and {Pace}, Francesco and {Paliathanasis}, Andronikos and {Palmese}, Antonella and {Pan}, Supriya and {Paoletti}, Daniela and {Perez Bergliaffa}, Santiago E. and {Perivolaropoulos}, Leandros and {Pesce}, Dominic W. and {Pettorino}, Valeria and {Philcox}, Oliver H.~E. and {Pogosian}, Levon and {Poulin}, Vivian and {Poulot}, Gaspard and {Raveri}, Marco and {Reid}, Mark J. and {Renzi}, Fabrizio and {Riess}, Adam G. and {Sabla}, Vivian I. and {Salucci}, Paolo and {Salzano}, Vincenzo and {Saridakis}, Emmanuel N. and {Sathyaprakash}, Bangalore S. and {Schmaltz}, Martin and {Sch{\"o}neberg}, Nils and {Scolnic}, Dan and {Sen}, Anjan A. and {Sehgal}, Neelima and {Shafieloo}, Arman and {Sheikh-Jabbari}, M.~M. and {Silk}, Joseph and {Silvestri}, Alessandra and {Skara}, Foteini and {Sloth}, Martin S. and {Soares-Santos}, Marcelle and {Sol{\`a} Peracaula}, Joan and {Songsheng}, Yu-Yang and {Soriano}, Jorge F. and {Staicova}, Denitsa and {Starkman}, Glenn D. and {Szapudi}, Istv{\'a}n and {Teixeira}, Elsa M. and {Thomas}, Brooks and {Treu}, Tommaso and {Trott}, Emery and {van de Bruck}, Carsten and {Vazquez}, J. Alberto and {Verde}, Licia and {Visinelli}, Luca and {Wang}, Deng and {Wang}, Jian-Min and {Wang}, Shao-Jiang and {Watkins}, Richard and {Watson}, Scott and {Webb}, John K. and {Weiner}, Neal and {Weltman}, Amanda and {Witte}, Samuel J. and {Wojtak}, Rados{\l}aw and {Yadav}, Anil Kumar},
        title = "{Cosmology intertwined: A review of the particle physics, astrophysics, and cosmology associated with the cosmological tensions and anomalies}",
      journal = {Journal of High Energy Astrophysics},
         year = 2022,
        month = jun,
       volume = {34},
        pages = {49-211},
          doi = {10.1016/j.jheap.2022.04.002},
archivePrefix = {arXiv},
       eprint = {2203.06142},
 primaryClass = {astro-ph.CO},
       adsurl = {https://ui.adsabs.harvard.edu/abs/2022JHEAp..34...49A}
}

@ARTICLE{Agnello:2013,
       author = {{Agnello}, A. and {Auger}, M.~W. and {Evans}, N.~W.},
        title = "{Lensing and dynamics in two simple steps.}",
      journal = {\mnras},
         year = 2013,
        month = feb,
       volume = {429},
        pages = {L35-L39},
          doi = {10.1093/mnrasl/sls020},
archivePrefix = {arXiv},
       eprint = {1209.2554},
 primaryClass = {astro-ph.CO},
       adsurl = {https://ui.adsabs.harvard.edu/abs/2013MNRAS.429L..35A}
}
\bibliographystyle{JHEP}

\clearpage
\begin{appendix}

\section{Numerical compression and Asimov forecasting}
\label{app:numerical_compression}

To efficiently forecast constraints for populations exceeding $10^5$ systems, we emulate the statistical power of the large PDSPL samples using smaller, computationally tractable representative subsets. For a full sample of size $N_{\rm total}$ represented by a subset of size $N_{\rm subset}$, we define a Down-Sampling Factor (DSF) as $\mathrm{DSF} = N_{\rm total} / N_{\rm subset}$.

To prevent random stochastic fluctuations from shifting or biasing the inferred parameter posteriors, we evaluate the likelihood using Asimov datasets \citep{Cowan:2011}. An Asimov dataset replaces the randomized observational data with the exact theoretical expectation of the true underlying model. 

For a Gaussian likelihood where the variance itself is a function of the proposed model parameters $\Theta$, the expected value of the log-likelihood must account for the true underlying variance of the data. Let $\beta_{\text{E, true}, i}$ and $\sigma_{\text{true}, i}^2$ represent the true Einstein radius ratio and the true total variance injected into the $i$-th mock system, respectively. The expectation value of the squared residuals evaluated at a proposed model $\beta_{\text{E, pl}, i}(\Theta)$ is analytically given by:
\begin{equation}
    \mathbb{E}\left[ \left(\beta_{\text{E, meas.}, i} - \beta_{\text{E, pl}, i}(\Theta)\right)^2 \right] = \sigma_{\text{true}, i}^2 + \left( \beta_{\text{E, true}, i} - \beta_{\text{E, pl}, i}(\Theta) \right)^2
\end{equation}

Substituting this expectation into the standard Gaussian log-likelihood yields the exact Asimov log-likelihood for a single system:
\begin{equation}
    \ln \mathcal{L}_i^{\text{Asimov}}(\Theta) = -\frac{1}{2} \left[ \frac{\sigma_{\text{true}, i}^2 + \left( \beta_{\text{E, true}, i} - \beta_{\text{E, pl}, i}(\Theta) \right)^2}{\sigma_{\beta_{\rm E}, i}^2(\Theta)} + \ln\left(2\pi\sigma_{\beta_{\rm E}, i}^2(\Theta)\right) \right]
\end{equation}
where $\sigma_{\beta_{\rm E}, i}^2(\Theta)$ is the parameter-dependent variance defined in Section~\ref{sec:likelihood_fnc}. Note that when the proposed model parameters perfectly match the truth ($\Theta = \Theta_{\text{true}}$), the residual term vanishes and $\sigma_{\beta_{\rm E}, i}^2(\Theta) = \sigma_{\text{true}, i}^2$, thereby maximizing the likelihood exactly at the injected truth.

To rigorously preserve the total information and statistical weight of the full catalog while evaluating only a subset, we uniformly scale the log-likelihood of each system in the representative subset by the DSF. The total compressed Asimov log-likelihood becomes:
\begin{equation}
    \ln \mathcal{L}_{\text{tot}}^{\text{Asimov}}(\Theta) = \sum_{i=1}^{N_{\text{subset}}} \mathrm{DSF} \times \ln \mathcal{L}_i^{\text{Asimov}}(\Theta)
\end{equation}

This scaling ensures that the likelihood —and thus the width of the resulting marginal posterior distributions—scales identically to that of the full population, which narrows proportionally to $1/\sqrt{N_{\rm total}}$. This approach allows for highly accurate, unbiased parameter constraint estimations while significantly reducing the computational overhead of the Markov Chain Monte Carlo (MCMC) sampling.

\section{The Impact of Spectroscopic Redshifts on PDSPL Cosmology}\label{app:lsst_y10_photo-z_vs_spec-z}

\begin{figure}
\centering
  \includegraphics[width=\textwidth]{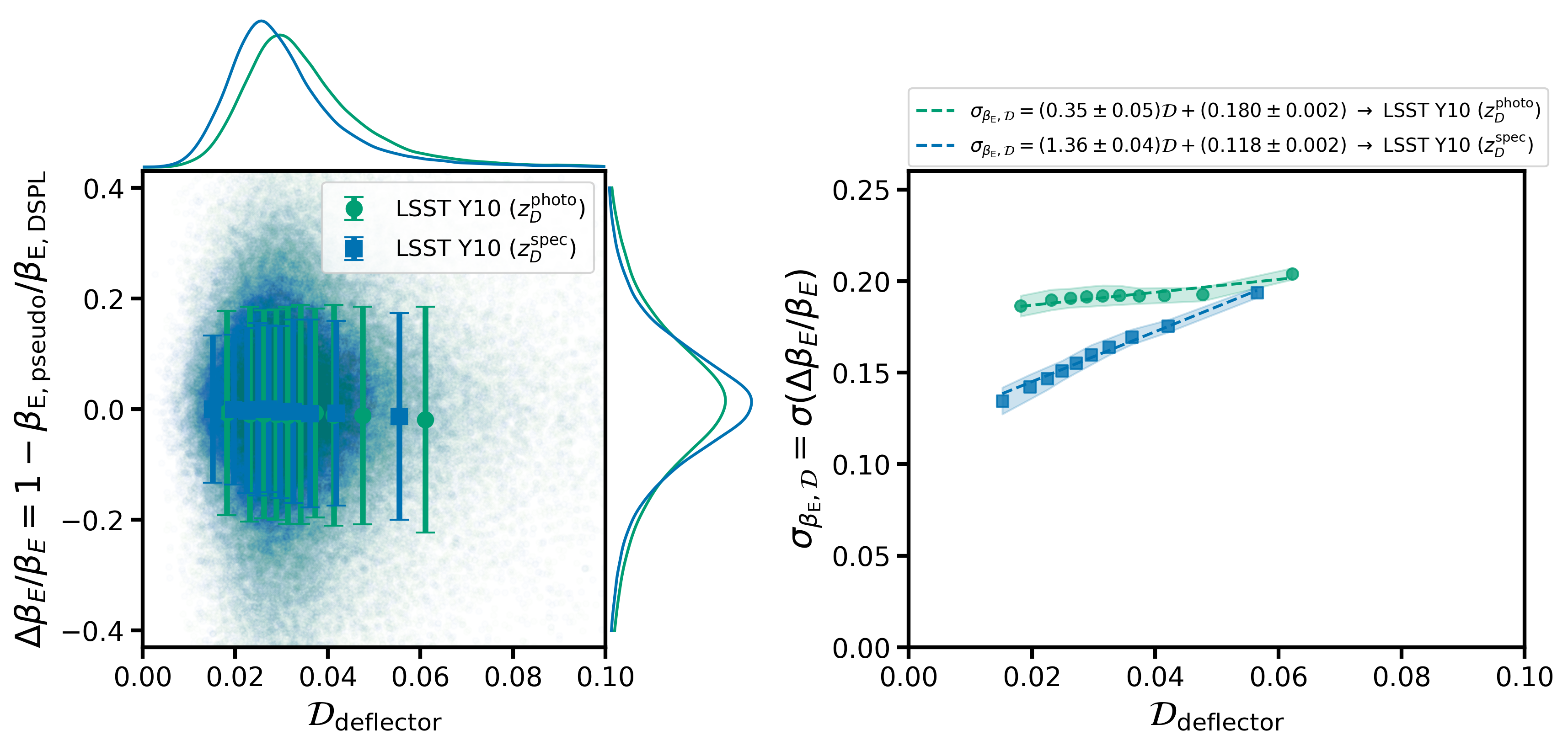} 
  \caption{
  Simulated scatter in $\beta_{\rm E}$ as a function of deflector pair dissimilarity ($\mathcal{D}_{\rm deflector}$) for the LSST Y10 sample (formatting is same as in Figure~\ref{fig:beta_E_vs_D}). Replacing photometric deflector redshifts (green) with exact spectroscopic redshifts (blue) significantly improves pairing fidelity, lowering the baseline noise floor ($\sigma_{\beta_E, \mathcal{D}}$). Because the pairing procedure incorporates random observational errors, the linear fit coefficients for the LSST Y10 ($z_{D}^{\rm phot}$) sample exhibit slight numerical variations across the 500 Monte Carlo realizations compared to the values reported in the main text.
  }
  \label{fig:beta_E_vs_D_LSST_Y10_photo_vs_spec-z} 
\vspace{-10pt}
\end{figure}

\begin{figure}
\centering
  \includegraphics[width=\textwidth]{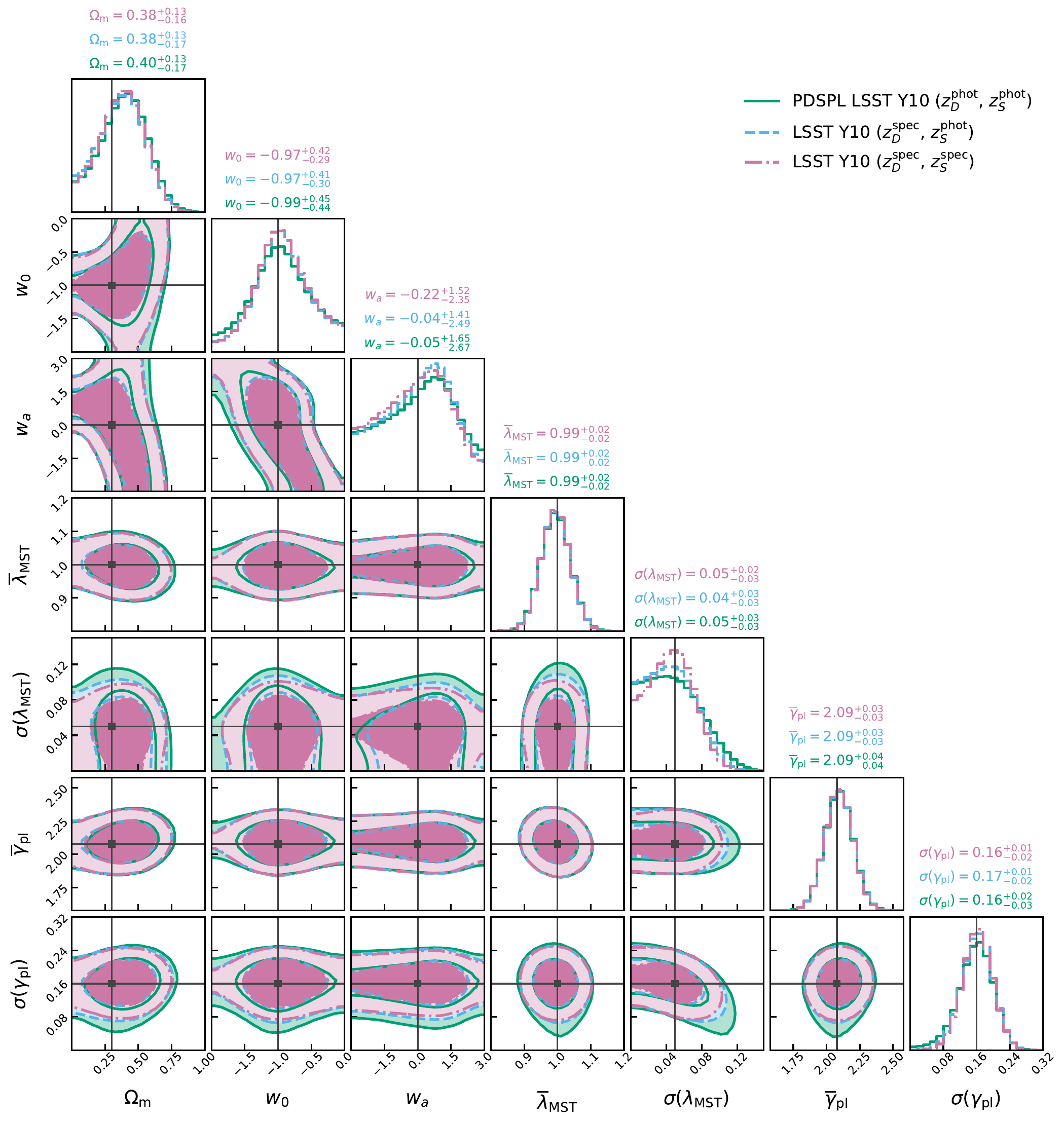} 
  \caption{
  Forecasted posterior distributions for the LSST Y10 PDSPL sample illustrating the impact of spectroscopic redshifts on cosmological constraints. The contours show the 68\% and 95\% confidence levels for a Flat $w_0w_a$CDM cosmology. The baseline photometric configuration (green) is compared against scenarios where exact spectroscopic redshifts are available for the deflectors only (dashed blue) and for both the deflectors and sources (dash dotted pink). The progressive inclusion of spectroscopic data tightens the constraints on $w_0$ and $w_a$ by simultaneously lowering the pairing noise floor and eliminating redshift-induced geometric uncertainties.
  }
  \label{fig:lsst_y10_photoz_vs_specz_forecast} 
\vspace{-10pt}
\end{figure}

While our baseline forecasts demonstrate the viability of PDSPLs using purely photometric data, widespread spectroscopic follow-up represents the theoretical ideal. Incorporating precise spectroscopic redshifts enhances the PDSPL framework through two distinct pathways: First, at the pairing stage, exact deflector redshifts ($z_{D}^{\rm spec}$) drastically improve the fidelity of the kD-Tree algorithm. Removing photo-z errors reduces the intrinsic mismatch between paired lenses, suppressing the average dissimilarity noise floor ($\sigma_{\beta_E, \mathcal{D}}$) from $\sim 19\%$ to $\sim 15\%$, as shown in Figure~\ref{fig:beta_E_vs_D_LSST_Y10_photo_vs_spec-z}. Notably, even with perfect redshift knowledge, this LSST Y10 noise floor remains higher than that of the 4MOST ($z^{\rm spec}$) sample's average noise ($\sim 8\%$, see Figure~\ref{fig:beta_E_vs_D}). This difference partly reflects the distinct selection functions of the two samples. The \textit{4MOST} subset is restricted to the brightest and most massive simulated deflectors and occupies a narrower region of the photometric and structural parameter space (the Fundamental Plane), whereas the full \textit{LSST Y10} catalog includes a broader population of fainter galaxies with greater simulated diversity. Second, at the cosmological inference stage, exact redshifts eliminate the photometric error budget ($\sigma_{\rm photo-z}$, Equation~\ref{eqn:sigma_photo-z}) propagated into the geometric distance ratio ($\beta$) during likelihood evaluation. Figure~\ref{fig:lsst_y10_photoz_vs_specz_forecast} demonstrates this dual impact on our full LSST Y10 forecast ($\sim 86,000$ pairs). Progressing sequentially from a purely photometric baseline ($z_{D}^{\rm phot}$, $z_{S}^{\rm phot}$) to a hybrid sample ($z_{D}^{\rm spec}$, $z_{S}^{\rm phot}$) and finally to an idealized fully spectroscopic sample ($z_{D}^{\rm spec}$, $z_{S}^{\rm spec}$) systematically tightens the dark energy equation of state constraints.

\section{Inferring Dissimilarity Scatter $\sigma_{\beta_E, \mathcal{D}}$}\label{app:inferred_scatter}
In our baseline analysis presented in the main text, the parameters governing the linear dissimilarity scatter relation ($\sigma_{\beta_{\rm E},\mathcal{D}}^{(0)}$ and $\sigma_{\beta_{\rm E},\mathcal{D}}^{(1)}$) were fixed to the values calibrated from our \texttt{SLSim} simulations (Table~\ref{tab:slsim_all_samples_pairing_statistics}). We refer to this baseline configuration as the \textbf{``Fixed Scatter''} model. However, given the vast statistical volume of the LSST Y10 PDSPL sample, these noise hyperparameters can instead be inferred simultaneously with the cosmology and the other hierarchical population parameters. We refer to this self-calibrating configuration as the \textbf{``Free Scatter''} model.

\begin{figure}
\centering
  \includegraphics[width=\textwidth]{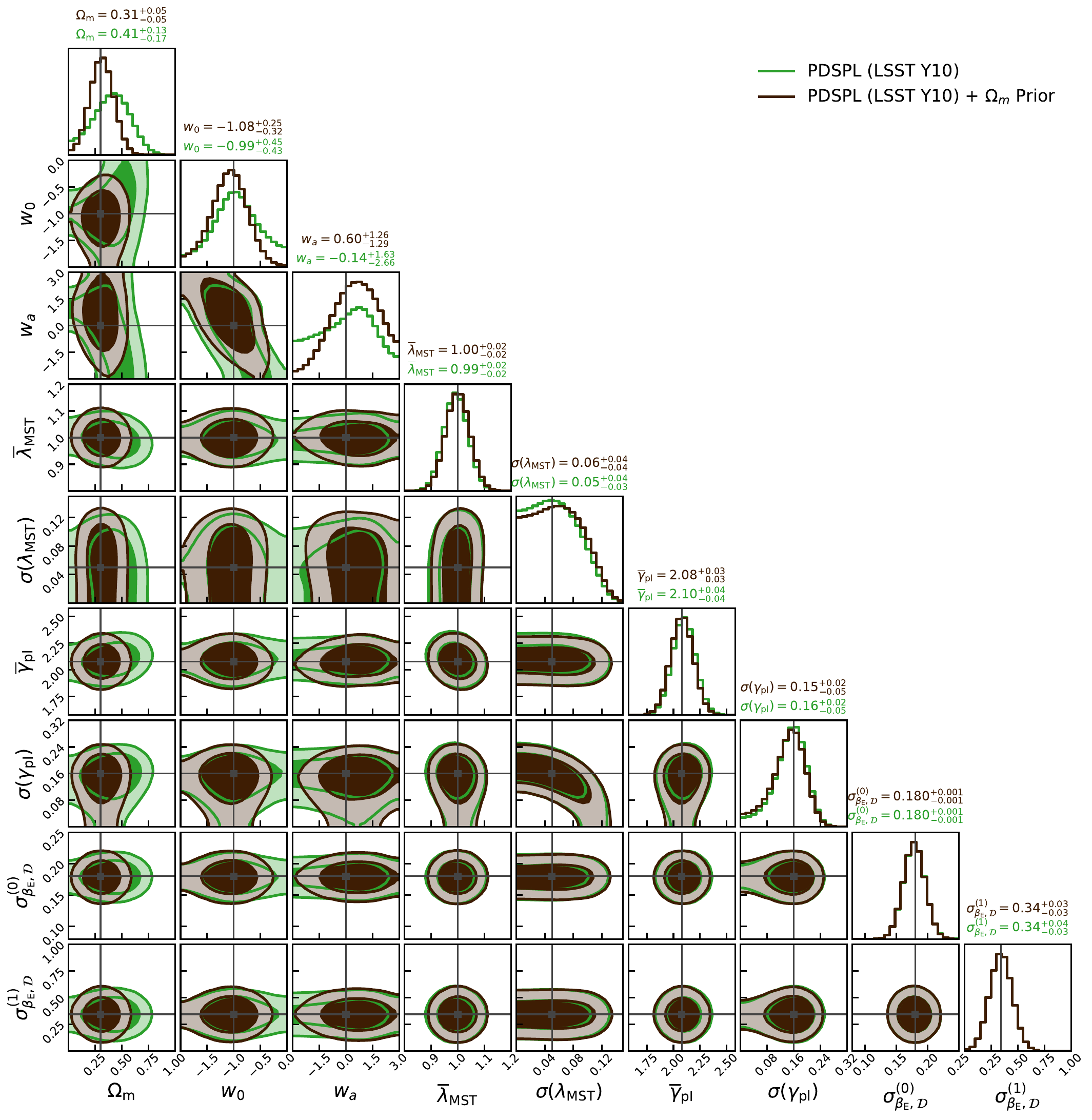} 
  \caption{
  Forecasted posterior distributions for the LSST Y10 PDSPL sample where the intrinsic dissimilarity scatter coefficients ($\sigma_{\beta_{\rm E},\mathcal{D}}^{(0)}$ and $\sigma_{\beta_{\rm E},\mathcal{D}}^{(1)}$) are treated as \emph{free parameters}. Results are shown with (dark brown) and without (green) an external Gaussian prior on $\Omega_{\rm m}$ of $\mathcal{N}(0.3, 0.05)$. The constraints on the dark energy equation of state remain tight, showing that the scatter coefficients can be inferred jointly with the cosmological and population parameters under the adopted linear model.
  \label{fig:free_scatter_lsst_y10_forecast_w0waCDM} 
\vspace{-10pt}
}
\end{figure}

\begin{figure}
\centering
  \includegraphics[width=\textwidth]{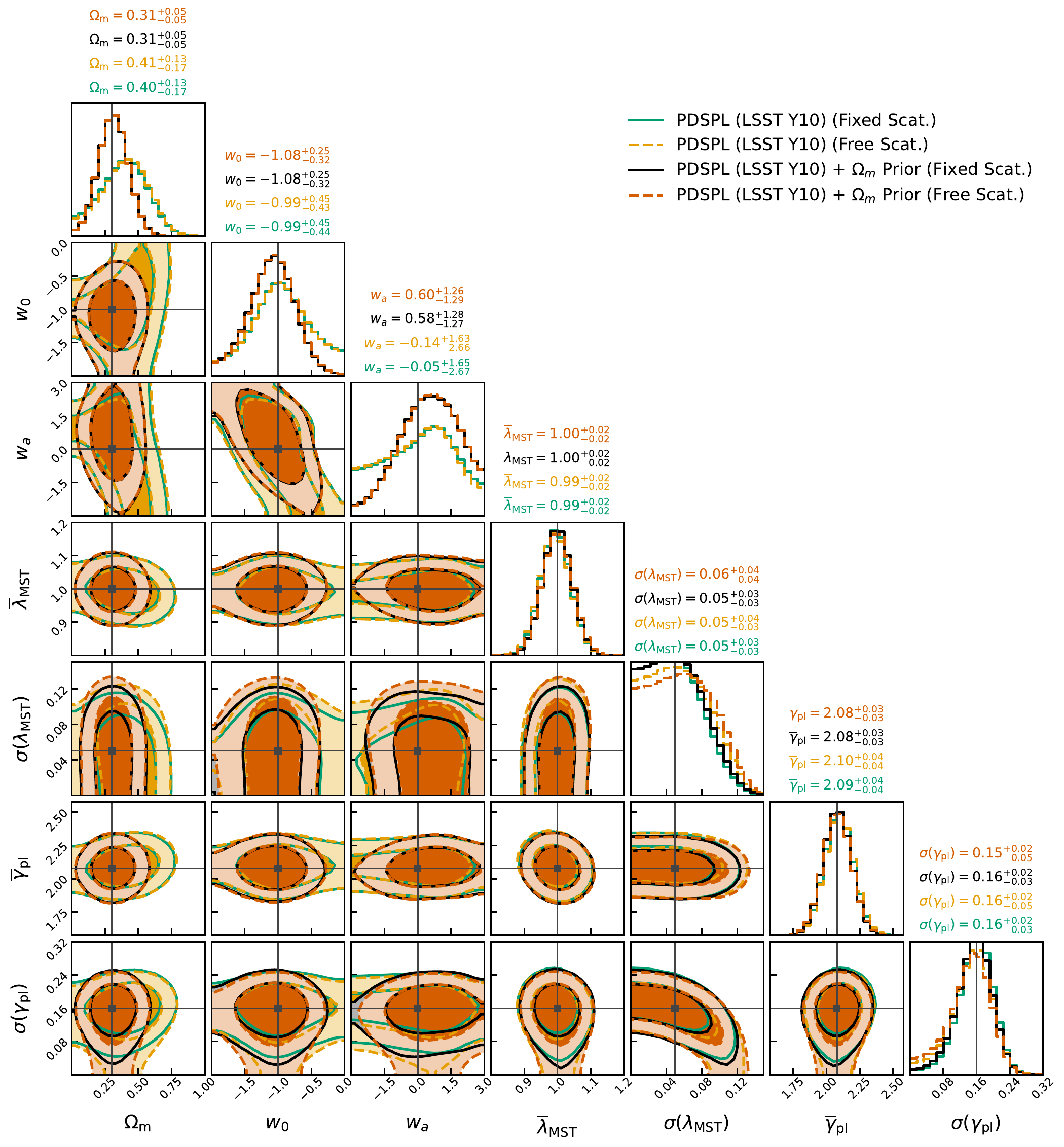} 
  \caption{
  Comparison of the forecasted constraints between the ``Fixed Scatter'' model (where the dissimilarity relation is fixed to its simulation-calibrated values; green) and the ``Free Scatter'' model (where the relation is simultaneously inferred from the data; orange/red) for the LSST Y10 PDSPL sample. Jointly inferring the deflector-mismatch scatter coefficients results in only a minor degradation of the cosmological parameter constraints.
  \label{fig:free_vs_fixed_scatter_lsst_y10_forecast_w0waCDM} 
\vspace{-10pt}
}
\end{figure}

\noindent This ``self-calibration'' approach allows the scatter coefficients to be fitted jointly with cosmology if the variance of the observational sample differs from our simulation-calibrated values. To test this, we applied the extended hierarchical population model to the LSST Y10 mock catalog with broad uniform priors on the scatter coefficients. Figure~\ref{fig:free_scatter_lsst_y10_forecast_w0waCDM} displays the resulting posteriors, demonstrating that the large sample constrains the scatter coefficients without fixing them to the simulation calibration. Figure~\ref{fig:free_vs_fixed_scatter_lsst_y10_forecast_w0waCDM} provides a direct comparison between the fixed-scatter and free-scatter inferences. We find that freeing the scatter parameters produces only a modest degradation of the cosmological constraints within the adopted linear, zero-mean Gaussian model. This result does not establish that the assumed model adequately describes real lenses. For observational data, the conditional mean, functional form, and residual distribution will not be known \textit{a priori}. Future analyses must therefore test alternative parameterizations and perform posterior-predictive checks to validate the scatter model. The simulation-derived calibration is sufficient for the proof-of-concept forecast presented here, but it is not yet sufficient for cosmological inference from real survey data.

\section{Impact of Photo-z Precision and Accuracy}
\label{app:photoz_precision_bias}

Our baseline LSST Y10 forecast assumes unbiased Gaussian photo-z errors of $\sigma_z/(1+z)=0.03$ for the deflector and both sources. Here, we separately examine the effects of photo-z precision and systematic bias. These tests illustrate sensitivity to redshift precision and simple coherent calibration biases; they do not constitute a complete photo-z systematics model and do not include catastrophic outliers, blending-dependent failures, non-Gaussian tails, or correlations between the deflector and source redshift estimates.

\begin{figure}
    \centering
    \includegraphics[width=\textwidth]
    {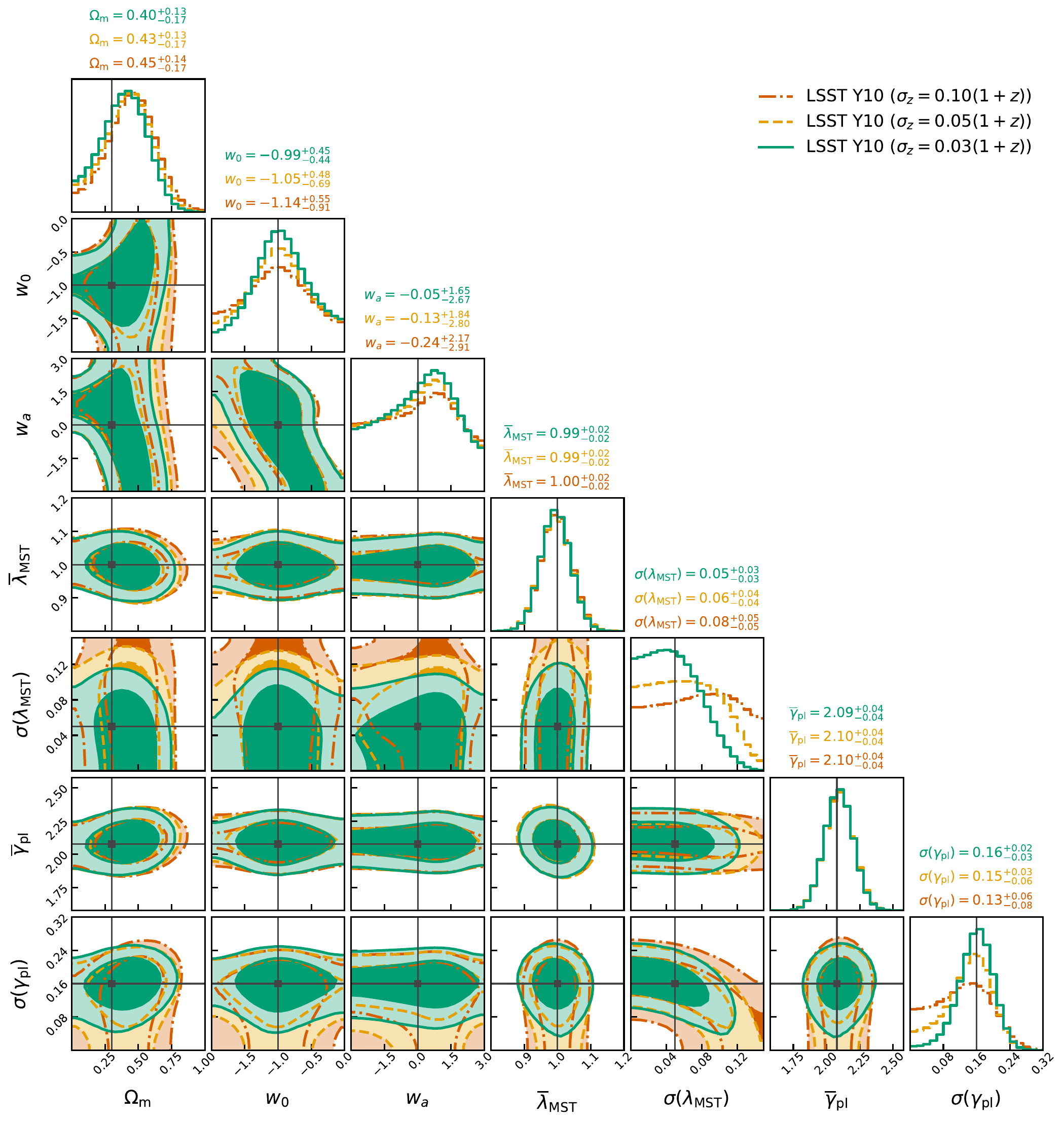}
    \caption{
    Forecast posteriors for the LSST Y10 PDSPL sample assuming photo-z uncertainties of $\sigma_z/(1+z)=0.03$, $0.05$, and $0.10$. The contours show the 68\% and 95\% confidence levels. Increasing the redshift uncertainty broadens the cosmological constraints while preserving consistency with the input cosmology.
    }
    \label{fig:lsst_y10_photoz_precision_forecast_w0waCDM}
\end{figure}

\noindent\textbf{Photo-z precision:}
We repeat the deflector-pairing and cosmological analyses assuming $\sigma_z/(1+z)=0.03$, $0.05$, and $0.10$. For each case, we recalibrate the pairing-scatter relation (Equation~\ref{eqn:model_beta_E_dissimilarity}) and propagate the corresponding redshift uncertainties through the cosmological likelihood. As shown in Figure~\ref{fig:lsst_y10_photoz_precision_forecast_w0waCDM}, degraded photo-z precision broadens the posterior distributions (majorly for $w_0, w_a$ with minor effect on $\Omega_m$) while leaving them consistent with the input cosmology. 


\begin{figure}
    \centering
    \includegraphics[width=\textwidth]
    {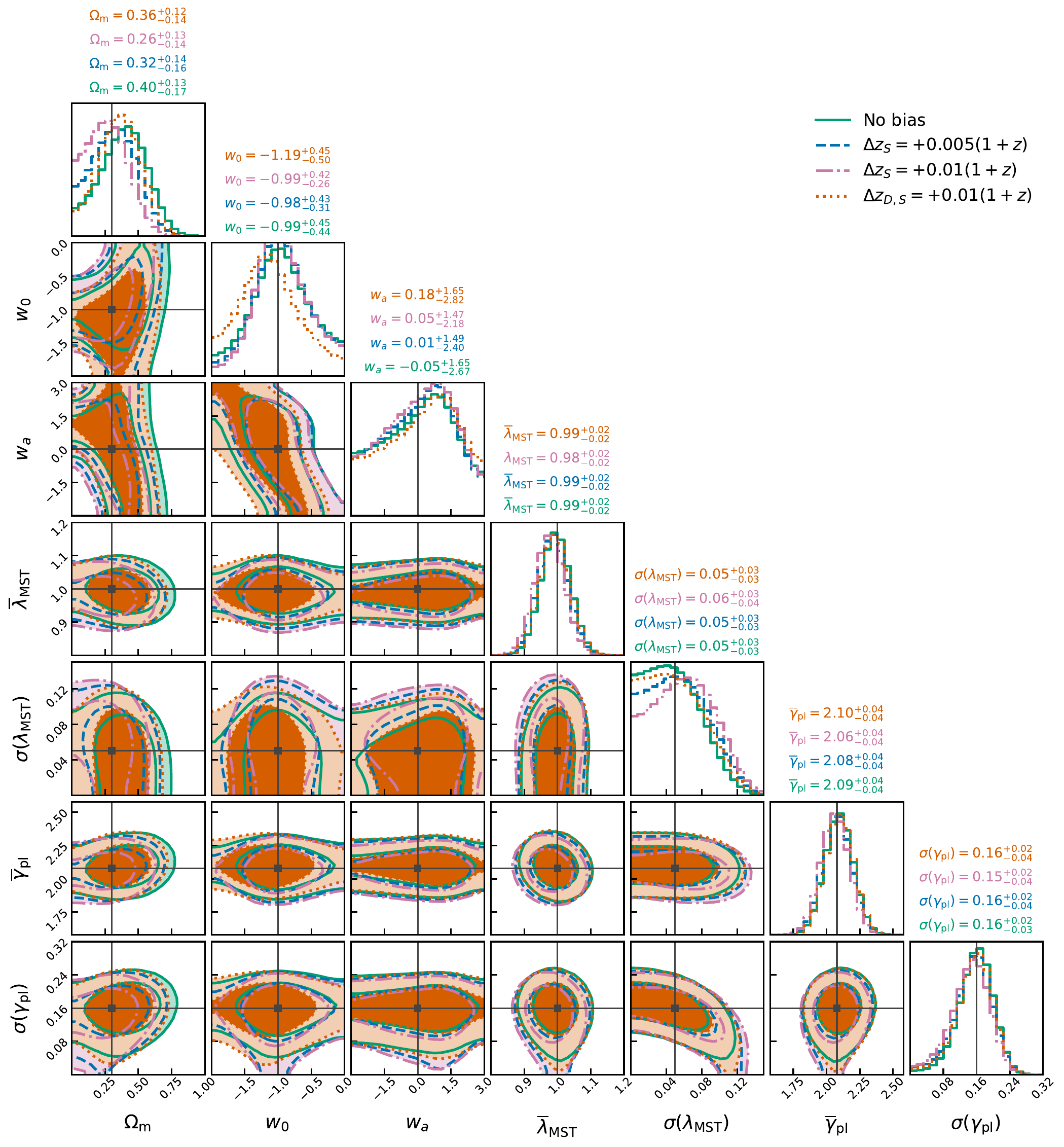}
    \caption{
    Forecast posteriors for the LSST Y10 PDSPL sample with unrecognized photo-z biases. All cases retain the fiducial random uncertainty $\sigma_z/(1+z)=0.03$. We consider biases of $\Delta z_S=0.005(1+z)$ and $0.01(1+z)$ applied to both source redshifts, and a coherent bias of $\Delta z_{D,S}=0.01(1+z)$ applied to the deflector and both sources. The mock observables are generated at the true redshifts, while the likelihood is evaluated using the biased values. The contours show the 68\% and 95\% confidence levels.
    }
    \label{fig:lsst_y10_photoz_bias_forecast_w0waCDM}
\end{figure}

\noindent\textbf{Photo-z accuracy:}
To test systematic accuracy, we retain the fiducial random uncertainty $\sigma_z/(1+z)=0.03$ but introduce an unrecognized redshift bias,
\begin{equation}
    z_{\rm assumed}
    =
    z_{\rm true}
    +
    b(1+z_{\rm true}),
\end{equation}
where the mock observables are generated using the true redshifts and the likelihood is evaluated using the biased values. Figure~\ref{fig:lsst_y10_photoz_bias_forecast_w0waCDM} shows that coherent redshift biases can shift the inferred cosmological parameters rather than merely broaden their posteriors. This test demonstrates the importance of redshift accuracy but does not determine a formal calibration requirement. Representative spectroscopy will be needed to characterize the photo-z distributions and control the systematic accuracy of the photometric sample.

\end{appendix}

\end{document}